# Option Pricing with Greed and Fear Factor: The Rational Finance Approach

Svetlozar Rachev
Texas Tech University

Frank J. Fabozzi
EDHEC Business School

Boryana Racheva-Iotova
BISAM

Abootaleb Shirvani
Texas Tech University

**Abstract:** We explain the main concepts of Prospect Theory and Cumulative Prospect Theory within the framework of rational dynamic asset pricing theory. We derive option pricing formulas when asset returns are altered with a generalized Prospect Theory value function or a modified Prelec's weighting probability function and introduce new parametric classes for Prospect Theory value functions and weighting probability functions consistent with rational dynamic pricing Theory. We study the behavioral finance notion of "greed and fear" from the point of view of rational dynamic asset pricing theory and derive the corresponding option pricing formulas in the case of asset returns that follow continuous diffusions or discrete binomial trees.

**Option Pricing with Greed and Fear Factor: The Rational Finance Approach**

## 1. Introduction

This paper is an attempt to study several behavioral finance (BF) findings with the help of the modern methods of rational dynamic asset pricing theory (RDAPT). In the theory and practice of finance, generally the accepted view is that modern rational finance (RF) (especially with the introduction of high-speed trading) has become very mathematically and computationally advanced. Meanwhile the methods of BF are predominantly based on very sophisticated empirical studies, contributing immensely to a better understanding of important financial markets' empirical phenomena. In this interplay between BF and RF, it is our strong belief that there is no empirical phenomena claimed in mainstream BF that cannot be subject to a successful (while potentially very challenging) RF study. In fact, important findings in BF were or could be quite well explained within the general framework of RF. These principal findings include (1) momentum (long- and short-range dependence observed in asset price time series), (2) non-Gaussian heavy-tailed distributions of asset returns, (3) equity-premium puzzle, (4) volatility puzzle, (5) leverage effect, and (6) asymmetric perception of large returns and large losses. Indeed, we are not the only ones among RF researchers and practitioners who believe that there is no single BF "puzzle" that cannot be reasonably modeled so as to be explained within the RF framework. Looking at the critique of RF by behaviorists, it is clear that some of them have a vague idea of what modern academic and practical RF is all about. Here we describe seven phenomena which modern RF is currently considered standard, meaning that every reasonable RF model should be able to explain those phenomena and deal with real financial industry problems.

First, in the financial industry and academia, risk and reward are rarely measured be the standard variation and the mean of asset, respectively. There are various coherent risk and reward measures used in RF. In modern RF, there are no universal risk and reward measures because there is no universal portfolio problem – thus RF studies and applies classes of risk, reward, and performance measures to capture the specific characteristic of the financial portfolio under consideration. Second, nothing in financial markets is static: gains and losses are measured in milliseconds and often in nanosecond. Third a standard average size traded portfolio consists of many risks factors. Fourth, correlation as a measure for dependence is practically meaningless. The RF applies large dimensional non-Gaussian copulas to capture the dependences in the portfolios returns. The fifth reason is that RF models and monitors asset bubbles and "crowding" effects. Among the practitioners and academics in RF it is not a secret that sufficient conditions for a financial bubble are a persistent large dislocation in the market copula dependence from its equilibrium state and a near critical phase transition of the market viewed as dynamical system. Sixth, there is no major structural break in the financial system that can come overnight, with the exception of catastrophically events of operational, natural or political nature. The market is a huge living thing and specialists in RF are monitoring its health continuously in time. Finally, RF, those large (tens and hundreds of thousands) dimensional financial problems with time-varying internal dependences should be free of arbitrage opportunities and tackled with very advanced methods of RDAPT and financial econometrics.

As we show in this paper, modern RF methods when properly applied can be used by behaviorists to explain many of the so-called puzzles identified in the BF literature. Moreover, some of the models proposed by BF proponents admit arbitrage opportunities that could lead to serious losses to those who use them. The option pricing formulas suggested by some behaviorists is an example. As we show in this paper, some of the premises of BF seen through modern RF are questionable. Our goal in this paper is to raise awareness that BF should be put on solid theoretical quantitative framework embracing the finding of modern RF and avoid relying predominantly on modeling human behavior with samples from people who have no understanding of the theory and practice of investing.[1] Failing to do that will result in BF ultimately serving only technical analysts.

To accomplish this we show using a few examples how some important concepts of BF can be embedded within RF. The paper is structured as follows. In Section 2, we study Kahneman and Tversky (1979) and Tversky and Kahneman (1992) Prospect Theory from the viewpoint of RDAPT. We show the need to modify the Prospect Theory value function (PTVF) to make it consistent with RF and then derive the corresponding option pricing formula. In doing so, we introduce new PTVFs consistent with RDAPT. In Section 3, we study the Cumulative Prospect Theory and derive option pricing formula under a minor modification of Prelec's probability weighting function (PWF). We derive a new PWF consistent with DAPT. In Sections 4 and 5 we study the concept of "greed and fear" in the context of financial markets with risky assets priced based on continuous diffusions and binomial trees. In Section 6 we provide our concluding remark.

## 2. Generalized Prospect Theory Weighting Function and Option Pricing with Logistic -Lévy Asset Return Process

In their seminal papers Kahneman and Tversky (1979) and Tversky and Kahneman (1992) introduced Prospect Theory (PT) and Cumulative Prospect Theory (CPT), critiquing the expected utility theory (EUT). They claim that the EUT cannot be a satisfactory model for the empirically observed in market participants' decision making under risk. From the point view of RF, the main thesis in PT and CPT is that (1) a typical investor (designated from now on as ב) is risk-averse when positive (log-) returns on investment are observed and/or predicted, then the probability for significant returns is very low, as ב becomes risk seeking; and, (2) when ב observes and/or predicts negative returns, ב is generally risk-averse, then the probability for significant losses is very low, ב reduces the risk-aversion level.

---

[1] In regression models, such studies find that the $R^2$ is 0.02. See, for example, the results reported in Tables 3 and 4 in Wang, Rieger, and Hens (2016) and Tables 3, 4, and 5 in Lewellen and Warner (2006).

**Remark 1.** In the PT and the CPT literature, notions of "gains" and "losses" are used to mean different things such as: dollar return, percentage return, log-return, prices, functionals of prices, derivative values, in real or "risk-neutral world". For example, in Barberis and Thaler (2005, p, 17), the reference to utility based on "gains" and "losses is done without specifying whether the utility is defined on asset price or asset return and what is the reward function and what is the loss function. In Tversky (1995, p. 3) losses and gains are in terms of dollar returns, while the conclusions are expressed in percentages. Barberis and Huang (2008, Section 3) state that the utilities defined over the dollar return and the percentage return are "equivalent"; that is, losing 1% on a $1 billion investment and 1% on a $1 investment is the same. We find only one place where clearly the losses and the gains are defined in the asset relative (log) return space, see Hens and Rieger (2010, p.57). That definition of a loss as negative log-return and a gain as positive -log return is what we are going to use in this paper.

**Remark 2.** Tversky and Kahneman (1992) stated:" The most distinctive implication of prospect theory is the fourfold pattern of risk attitudes. Specifically, it is predicted that when faced with a risky prospect people will be: (1) risk-seeking over low-probability gains, (2) risk-averse over high-probability gains, (3) risk-averse over low-probability losses, and (4) risk-seeking over high-probability losses." See also Harbaugh, Krause, and Vesterlund (2009), Ackert and Deaves (2010), Barberis, Mukherjee, and Wang (2016), and Abdellaoui et al. (2016) for some empirical studies on those basic premises of PT and CPT.

As we stated, the goal of this paper is to link and explain the main concepts in PT and CPT within RDAPT, preparing the theoretical basis for testing PT. The work on behavioral dynamic asset pricing viewed from the point of RDAPT is unsatisfactory. The over-reaching error in the BF dynamic asset pricing models from the point of view of the RDAPT is that BF asset return processes are often not semimartingales.[2] Semimartingales are the most general stochastic processes used in RDAPT. Ansel and Stricker (1991)[3] show that a suitable formulation of absence of arbitrage implies that security gains must be semimartingales with finite conditional means. The concept of no arbitrage roughly says that it is impossible for $\beth$ to start a trading strategy (i) with zero dollar invested, (ii) with no inflow or outflow of funds, and (iii) to have a positive return with no risk (that is, with probability 1). If in the frictionless market there is an arbitrage opportunity, all traders will stop trading anything else and instead take a long position in this available arbitrage trade. The market will cease to exist. Indeed, assuming transaction costs, RF can deal with fractional market models which are not semimartingales. But such assumptions about the market with frictions are not made in the works in the asset pricing models proposed by behavioralists. The notion of arbitrage is crucial in the modern theory of finance. It is the cornerstone of the asset pricing theory due to Black and Scholes (1973) and Merton (1973).

---

[2] A stochastic process $X(t), t \geq 0$ defined on a probability basis $(\Omega, \mathcal{F}, \mathbb{F}, \mathbb{P})$, where $\mathbb{F} = (\mathcal{F}_t, 0 \leq t \leq T)$ is a right continuous filtration with $\mathcal{F} = \mathcal{F}_T, T \in (0, \infty], \mathcal{F}_0 = \{\emptyset, \Omega\}$ is called a **semimartingale** if: $(i)$ $X(\cdot)$ is $\mathbb{F}$-adapted càdlàg (right continuous with left limits ) process, and $(ii)$ $X(\cdot)$ can be decomposed as $X(t) = M(t) + A(t) - B(t)$, where $M(t)$ is càdlàg $\mathbb{F}$-adapted local martingale, and $A(t)$ and $B(t)$ are increasing càdlàg $\mathbb{F}$-adapted processes. Furthermore, $M(t), t \geq 0$, is a local martingale, if there exists a strictly increasing sequence of stopping times $\tau^{(k)} \uparrow \infty$ as $k \uparrow \infty$ on , such that the stopped processes $M^{(\tau^{(k)})}(t), t \geq 0, k = 1,2, \ldots$ are martingales. For a detailed exposition on the theory of semimartingales we refer to Métivier M. (1992) and He,Wang and Yan (1992). For a general exposition on RDAPT, see Duffie (2001) and Shiryaev (2003).

[3] See also Delbaen and Schachermayer (1994, 2011) and Chapter 6 in Duffie (2001).

**Remark 3.** Shefrin (2005, p. 103) defined the return distribution of the representative investor as a mixture of two different Gaussian distributions, which is indeed not an infinitely divisible distribution (see Steutel and van Harn (2004), Chapter VI, Section 1)[4] and thus the pricing dynamics of the representative investor is not a semimartingale. Shefrin's model could easily be made consistent with the RDAPT assuming that $t = 0$ and there are a random number of traders: for example $N$, where $N - 2$ has a Poisson distribution, or has a geometric distribution. Alternatively, Shefrin's model could be made consistent with RDAPT by assuming that the market participants trade with high enough transaction costs to wipe out the arbitrage gains, which Shefrin's model generates. Behavioral European option pricing formulas provided in Versluis,

---

[4] Random variable $X$ is **infinitely divisible**, if for every $n = 1,2, \ldots,$ there exist $n$ random variables $X^{(1,n)}, \ldots, X^{(n)}$ such that $X$ has the same distribution as $X^{(1,n)} + \cdots + X^{(n)}$. Normal, Poisson, Stable, log-normal, Student-t, Laplace, Gumbel, Double Pareto, Geometric random variables are infinitely divisible. Binomial and any other random variable with bounded support are not infinitely divisible. For a general exposition on infinitely divisible distributions in finance, see o Sato (1990).

Lehnert and Wolff (2010), Pena, Alemanni, and Zanotti (2011), and Nardon and Pianca (2014) allow for arbitrage opportunities, and should not to applied in real trading, except when taking the short position in those contracts.

2.1 Modified Prospect Theory Value Function and Option Pricing with Logistic-Lévy Asset Return Process

We start with an illustration of how to embed the PT within the RDAPT. Tversky and Kahneman (1992) claim that positive return (gains) and negative returns (losses) generated by financial assets are viewed differently as a result of the general "fear" disposition of traders. To quantify this claim they introduce the **Prospect Theory value function** (PTVF) of the following the form:

$$v(x) = \begin{cases} v^{(+)}(x) := x^{\alpha}, \ for\ x \geq 0 \\ v^{(-)}(x) := -\lambda(-x)^{\beta}, for\ x < 0 \end{cases}, \qquad (1)$$

and estimated parameters $\alpha, \beta$ and $\lambda$, as $\alpha = \beta = 0.88$ and $\lambda = 2.25$. Within the scope of BF, the PTWF $v(x), x \in \mathcal{R},$ should be concave for the gains, that is when $x \geq 0$, and convex for the losses, that is when $x < 0$. See Figures 1a and 1b.

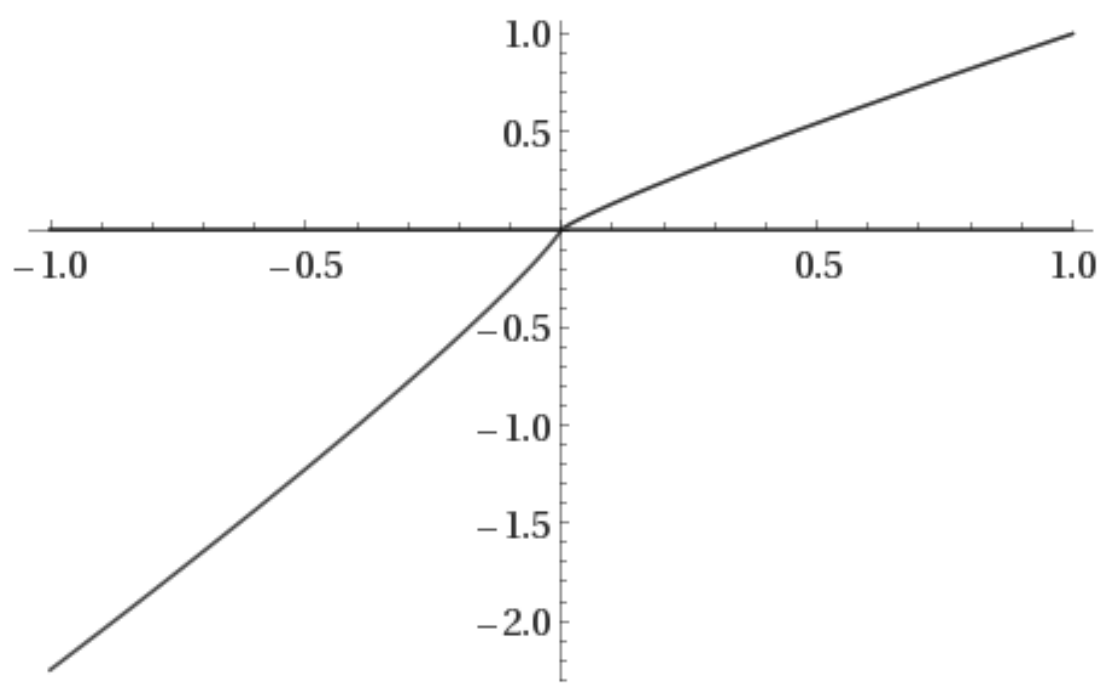

**Figure 1a: Plot of Tversky and Kahneman (1992) PT-value function,** $v(x) = \begin{cases} x^{\alpha}, \alpha = 0.88 \\ -\lambda x^{\beta}, \lambda = 2.25, \beta = 0.88 \end{cases}, \ x \in (-1,1)$

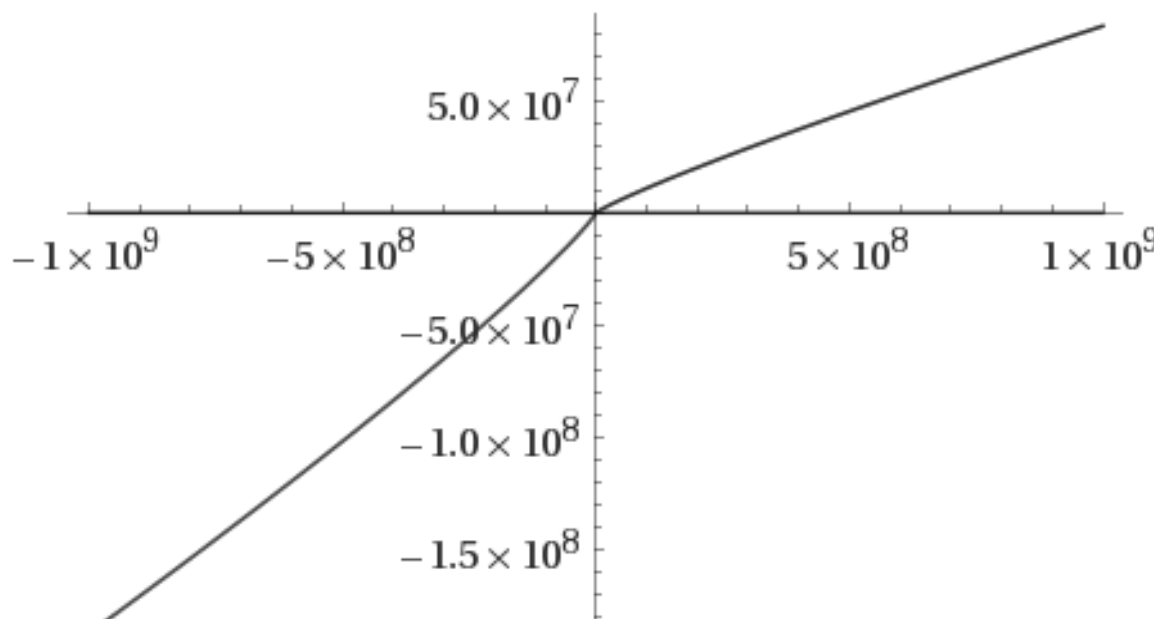

**Figure 1b: Plot of Tversky and Kahneman (1992) PT-value function** $v(x) = \begin{cases} x^{\alpha}, \alpha = 0.88 \\ -\lambda x^{\beta}, \lambda = 2.25, \beta = 0.88 \end{cases}, x \in (-10^9, 10^9)$

Hence it is assumed that in (1), $\alpha \in (0,1)$, $\beta \in (0,1)$ and $\lambda > 1$.

As will be clear from our exposition, in order to make the PT model consistent with RDAPT, the following should be satisfied $(a)$ the prior returns are transformed to posterior return via modification of (1) and $(b)$ the prior and posterior returns are infinitely divisible. We extend the definition of PTWF. We start with a modification of PTWF:

$$w(x) = \begin{cases} w^{(+)}: [0,\infty] \to [-\infty,\infty], continuous\ function\ on\ [0,\infty], \\ \quad with\ \frac{\partial w^{(+)}(x)}{\partial x} > 0, \frac{\partial^2 w^{(+)}(x)}{\partial x^2} < 0,\ for\ all\ x > 0; \\ w^{(-)}: [-\infty,0] \to [-\infty,\infty], continuous\ function\ on [-\infty,0]; \\ \quad with\ \frac{\partial w^{(+)}(x)}{\partial x} > 0, \frac{\partial^2 w^{(+)}(x)}{\partial x^2} > 0,\ for\ all\ x > 0 \end{cases} \qquad (2)$$

which we refer to as the **generalized PTWF**. What distinguishes the generalized PTWF $w(\cdot)$ from the PTWF $v(\cdot)$, is the behavior of the investor ב, when the random asset return, denoted by $\mathfrak{R}$, takes small absolute values $|\mathfrak{R}| < \varepsilon$. Then it could be possible that (i) $w^{(+)}(\mathfrak{R}), 0 < \mathfrak{R} < \varepsilon$ becomes negative, and furthermore, when $\mathfrak{R} \downarrow 0$, it could be that $w^{(+)}(\mathfrak{R}) \downarrow -\infty$; and (ii) $w^{(-)}(\Re), -\varepsilon < \Re < 0$ becomes positive , and furthermore, when $\Re \uparrow 0$, it could be that $w^{(-)}(\Re) \uparrow \infty$, see Figure 2.

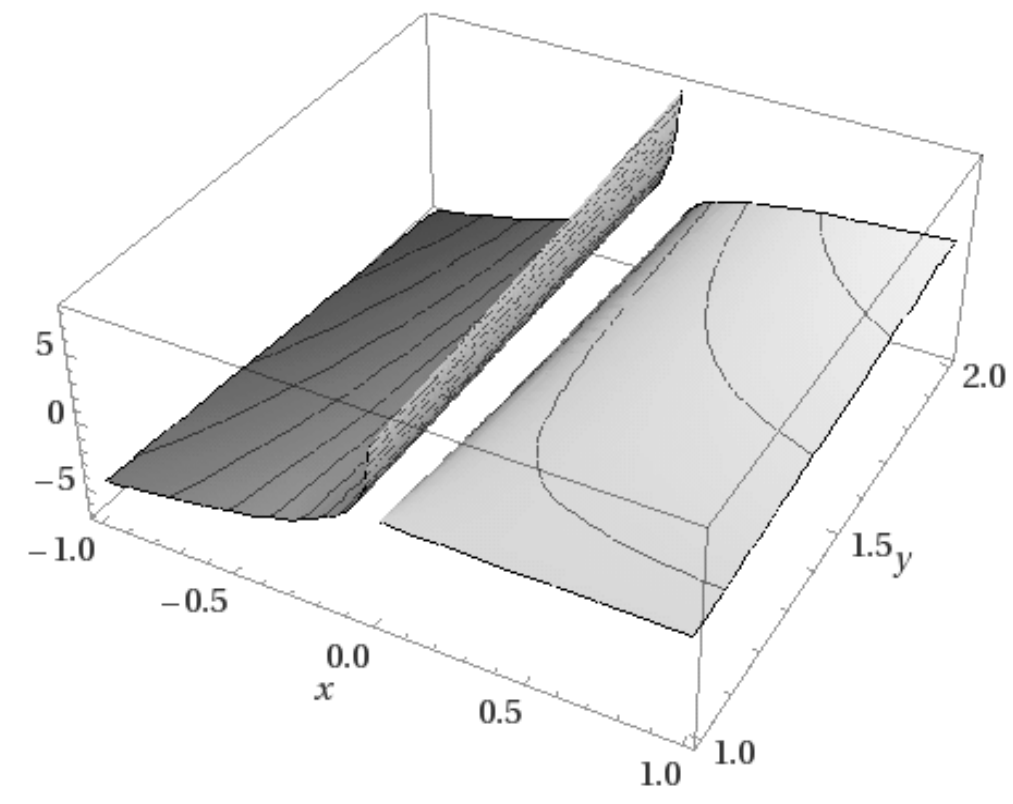

**Figure 2. Plot of PT weighting function**

$$w(x) = \begin{cases} w^{(+)}(x) = (y-1)\ln(10x), 0 < x < 1, 1 < y < 2 \\ w^{(+)}(x) = (y+1)\ln(-10x), -1 < x < 0, 1 < y < 2 \end{cases}$$

Embedding Tversky and Kahneman (1992) approach into RDAPT requires finding an infinitely divisible distribution of an asset return $\Re$, representing the prior views of the investor ב, who based on her "fear-greed profile" chooses a function $v^{(\beth)}(x), x \in R$, of the type (1), decides to alter the distribution of $\Re$, assuming that it is "safer" to use the posterior $\Re^{(\beth)} = v^{(\beth)}(\Re)$. In view of the definition (1), one is tempted to use the two-sided Weibull distribution, that is $\Re = \Re^{(+)} - \Re^{(-)}$, where $\Re^{(+)}$ and $\Re^{(-)}$ are independent identically distributed (iid) with $\Re^{(+)} \triangleq Weibull(\gamma, \delta)$[5], $\gamma > 0, \delta > 0$, that is, its cumulative distribution function (cdf) $F^{(\Re^{(+)})}(x) = \mathbb{P}(\Re^{(+)} \leq x) = 1 - \exp\left(\frac{x}{\delta}\right)^{\gamma}$. Then, define $\Re^{(\beth)} = \Re^{(\beth,+)} - \Re^{(\beth,-)}$ with $\Re^{(\beth,+)} = v^{(+)}(\Re^{(+)}) \triangleq Weibull\left(\frac{\gamma}{\alpha}, \delta^{\alpha}\right)$, $\Re^{(\beth,-)} = v^{(-)}(\Re^{(-)}) \triangleq Weibull\left(\frac{\gamma}{\alpha}, \lambda\delta^{\alpha}\right)$. The problem with this "obvious" approach is that $Weibull(\gamma, \delta)$ is infinitely divisible, if and only if $\gamma < 1$[6]. Because we would like to have $\Re^{(\beth)}$ to be infinitely divisible for every $\alpha \in (0,1)$, the choice of the two-sided Weibull distribution for $\Re$ is not appropriate. A similar problem arises if we choose the two-sided generalized-gamma distribution $\Re = \Re^{(+)} - \Re^{(-)}$, where $\Re^{(+)}$ and $\Re^{(-)}$ are independent identically generalized-gamma distributed $\Re^{(+)} \triangleq GenGamma(\gamma, \delta), \gamma > 0, \delta > 0$, that is, its probability density function (pdf) is given by

$$f^{\left(\Re^{(+)}\right)}(x) = \frac{\partial F^{\left(\Re^{(+)}\right)}(x)}{\partial x} = \frac{|\gamma|}{\Gamma(\delta)} x^{\gamma\delta-1} \exp(-x^{\gamma}), x > 0, \gamma \in \mathcal{R} \setminus \{0\}, \delta > 0. \tag{3}$$

[5] " $\triangleq$ " stands for "equal in distribution".

Unfortunately, $GenGamma(\gamma, \delta)$ is infinitely divisible if and only if $|\gamma| < 1$. A close look at the distributional structure of infinitely divisible non-negative random variables shows that the construction $\Re^{(\beth)} = \Re^{(\beth,+)} - \Re^{(\beth,-)}$, where $\Re^{(\beth,+)}$ and $\Re^{(\beth,-)}$ are iid infinitely divisible random variables (rvs) is not suitable when $\Re^{(\beth)}$ is viewed as the return of the underlying asset in an option contract.

We shall illustrate our approach choosing the Laplace distribution for the prior distribution $\beth$ is dealing with, that is $(i)$ $\Re = \Re^{(0)} + \mathbb{m}$, where $\mathbb{m} = \mathbb{E}\Re$ and $(ii)$ $\Re^{(0)} = \Re^{(+)} - \Re^{(-)}$ has Laplace distribution. That is, $\Re^{(0)} \triangleq Laplace(b), b > 0$, and $\Re^{(+)}$ and $\Re^{(-)}$ are iid exponentially distributed rvs with mean $\mathbb{E}\Re^{(+)} = b$. The pdf $f^{\left(\Re^{(0)}\right)}(x), x \in$ and the cdf $F^{(\Re)}(x)$ of $\Re$, has the form

$$f^{\left(\Re^{(o)}\right)}(x) = f^{(L,\lambda)}(x) = \frac{1}{b} \exp\left(-\frac{|x|}{b}\right), x \in R, b > 0. \tag{4}$$

[6] See Steutel and van Harn (2004), Appendix B, Section3.

$$F^{(L,b)}(x) = \begin{cases} \frac{1}{2}\exp\left(\frac{x}{b}\right) \; if \; x \leq 0 \\ 1 - \frac{1}{2} exp\left(-\frac{x}{b}\right) if \; x \geq 0, \end{cases} \quad (5)$$

see Figures 3a and 3b.

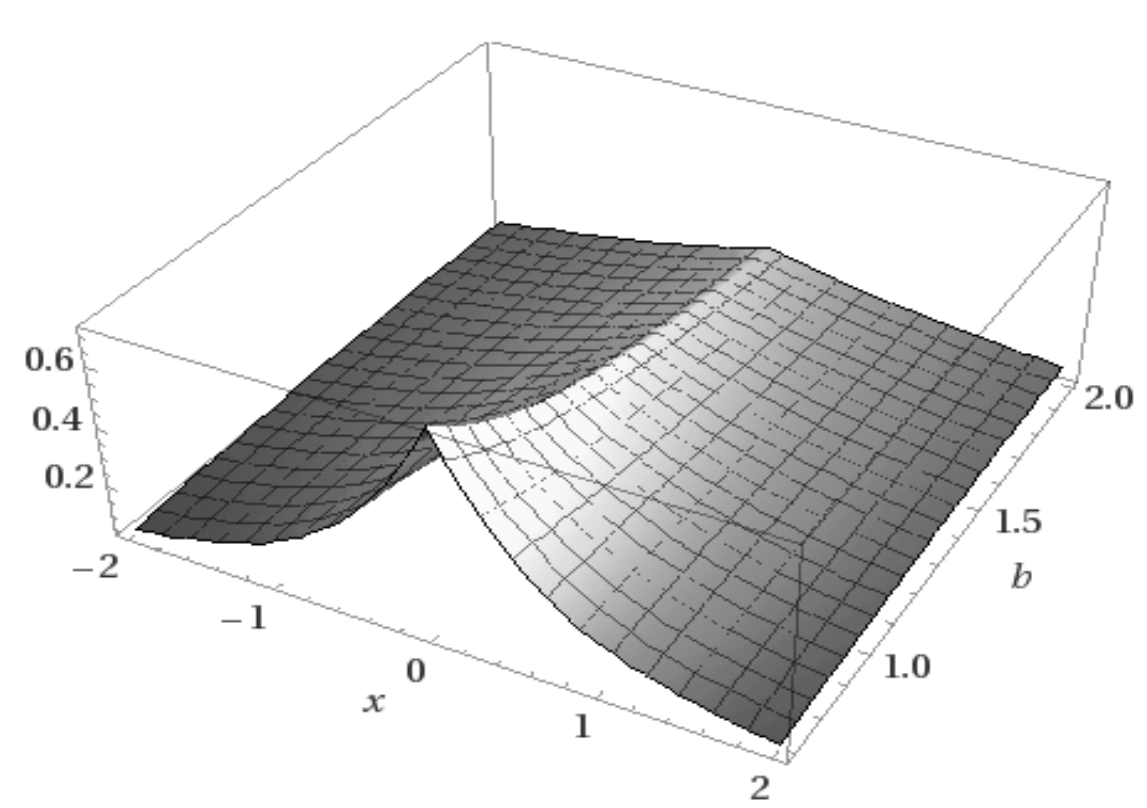

**Figure 3a. Plot of the PDF of Laplace distribution $f^{(L,b)}(x) = \frac{1}{b}\exp\left(-\frac{|x|}{b}\right)$, for $x \in [-2,2], b \in [0.7,2]$.**

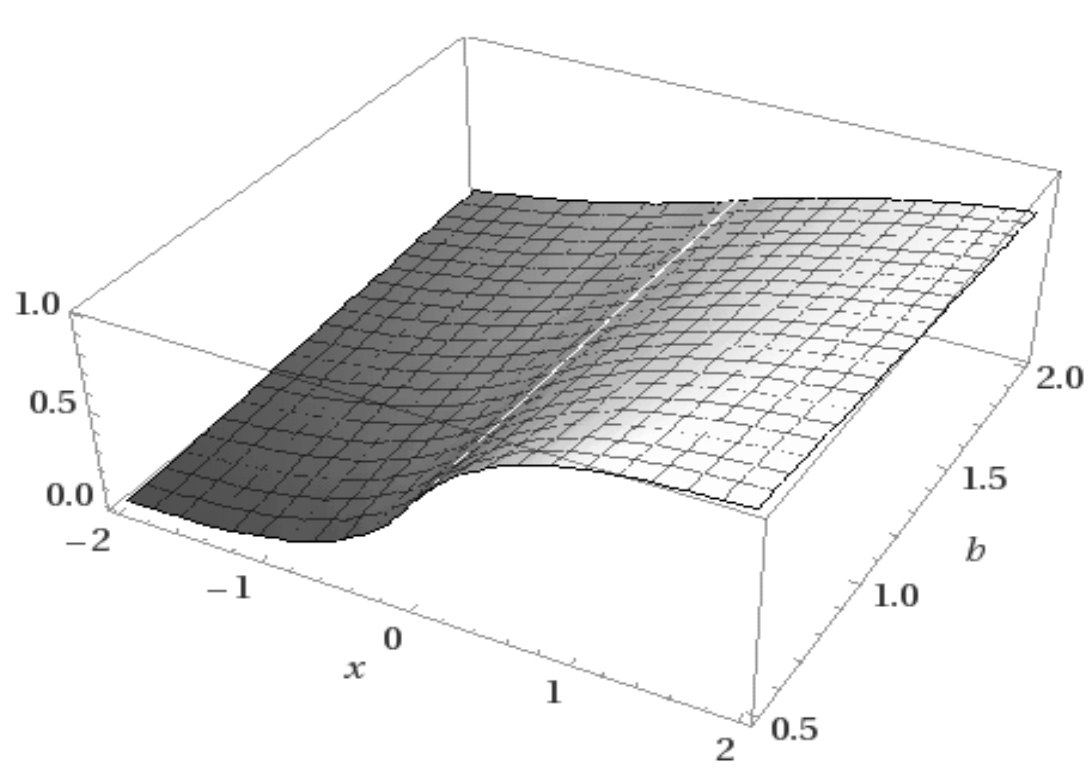

**Figure 3b. Plot of the CDF of Laplace distribution $F^{(L,b)}(x) = \begin{cases} \frac{1}{2}\exp\left(\frac{x}{b}\right) \; if \; x \leq 0 \\ 1 - \frac{1}{2} exp\left(-\frac{x}{b}\right) if \; x \geq 0 \end{cases}$, for $x \in [-2,2], b \in [0.5,2]$.**

$\Re \triangleq Laplace(b)$ has mean $\mathbb{E}\Re^{(0)} = 0$, variance $var\Re = 2b^2$, skewness $\gamma\left(\Re^{(0)}\right) = 0$, excess kurtosis $\kappa\left(\Re^{(0)}\right) = 3$, and characteristic function $\varphi^{\left(\Re^{(0)}\right)}(\theta) = \mathbb{E}e^{i\theta\Re^{(0)}} = \frac{1}{1+b^2\theta^2}, \theta \in \mathcal{R}$.

$\beth$ selects $\mathfrak{R} \triangleq Laplace(b) + \mathbb{m}$ as initial (prior) distribution for $\mathfrak{R}$, for the following four reasons:

$(i)$ $\beth$ has estimated that the stock has mean return $\mathbb{E}\mathfrak{R} = \mathbb{m}$ and variance $var\mathfrak{R} = 2b^2$, and $\beth$ models the distribution of the stock return as $\mathfrak{R} = \mathfrak{R}^{(0)} + \mathbb{m}$, and $\mathfrak{R}^{(0)} = \mathfrak{R}^{(+)} - \mathfrak{R}^{(-)}$, where $\mathfrak{R}^{(+)} > 0$ and $\mathfrak{R}^{(-)} > 0$, are iid infinitely divisible rv's;

$(ii)$ $\mathfrak{R} \triangleq Laplace(b, \mathbb{m}) := Laplace(b) + \mathbb{m}$ is infinitely distributed rv, generating Laplace motion, that is a Lévy process[7] with unit increment distributed as $Laplace(b, \mathbb{m})$. Similar to the Black-Scholes formula pricing formula, when the underlying return process is a Laplace motion with linear drift as provided by Madan, Carr, and Chang (1998)[8];

---

[7] A stochastic process $L(t), t \geq 0$ defined on a probability basis $(\Omega, \mathcal{F}, \mathbb{F}, \mathbb{P})$, where $\mathbb{F} = (\mathcal{F}_t, 0 \leq t \leq T)$ is a right continuous filtration with $\mathcal{F} = \mathcal{F}_T$, $T \in (0, \infty], \mathcal{F}_0 = \{\emptyset, \Omega\}$, is called a **Lévy process** if: $(i)$ $L(\cdot)$ is a càdlàg $\mathbb{F}$-adapted process and $L(0) = 0;$ $(ii)$ $L(\cdot)$ has independent stationary increments, that is, for $0 \leq t^{(0)} < t^{(1)} < \cdots < t^{(n)}$, the rvs $L\left(t^{(i)}\right) - L\left(t^{(i-1)}\right), t = 1, \dots, n,$ are independent rvs; $(ii)$ $L(\cdot)$ has stationary increments, that is, for $0 \leq t < t + s < T$, the distribution of $L(t+s) - L(t)$ does not depend on $t$; $(iv)$ $L(\cdot)$ is stochastically continuous, that is, for every $0 \leq t < T$, and every $\varepsilon > 0$, $\lim_{s \to t} \mathbb{P}(|L(t) - L(s)| > \varepsilon) = 0$. For detailed explosition on Lévy processes in finance, see Sato (1999) and Schoutens (2003) and Applebaum (2009). I WOULD DELETE ALL BUT THE REFERENCE TO THE LAST TWO RFERENCES.

[8] See also Section 8.5 Kotz, Kozubowski, and Podgórski (2001).

$(iii)$ for a suitable choice of a flexible class of generalized PTWF, $w(\cdot)$, $\mathfrak{R}^{(ב,+)} = w^{(+)}\left(\mathfrak{R}^{(+)}\right)$ and $\mathfrak{R}^{(ב,-)} = w^{(-)}\left(\mathfrak{R}^{(-)}\right)$ are independent infinitely divisible rvs, and thus the posterior return $\mathfrak{R}^{(ב)} = \mathfrak{R}^{(ב,+)} - \mathfrak{R}^{(ב,-)}$ is also infinitely divisible rvs.

Consider then the following family of generalized PTWF of logarithmic form:

$$w^{(l)}(x) = \begin{cases} w^{(l,+)}(x) := a\ln x + c,\ for\ x > 0,\ \alpha > 0, c \in \mathcal{R}, \\ w^{(l-)}(x) := -\lambda\ln(-x) - v,\ for\ x < 0,\ \lambda > 0, v \in \mathcal{R}, \end{cases} \tag{6}$$

The parametric family given by (6) is flexible enough to fit a variety of "fear-greed profiles" of ב. To illustrate that let us compare $v^{(+)}(x) := x^{\alpha}$, for $x \in [0.45,0.9], \alpha \in [0.87,0.89]$ (see equation (1)) with suitably fitted $w^{(l,+)}(x) := a\ln x c, x \in [0.45,0.9]$. Matching the first derivatives at point $\frac{1}{2}$, $\frac{\partial^{(j)} y}{\partial x^{(j)}} v^{(+)}\left(\frac{1}{2}\right) = \frac{\partial^{(j)} y}{\partial x^{(j)}} w^{(+)}\left(\frac{1}{2}\right), j = 0,1$, we select $a^{(\alpha)} = \alpha\left(\frac{1}{2}\right)^{\alpha}$, $c^{(\alpha)} = \left(\frac{1}{2}\right)^{\alpha}(1 + \alpha\ln 2)$. Define then $w^{(+,\alpha)}(x) = a^{(\alpha)}\ln x + c^{(\alpha)}$. The graph of the error term $\left(\frac{w^{(+,\alpha)}(x) - v^{(+)}(x)}{v^{(+)}(x)}\right)^2 \leq 0.10, x \in [0.45,0.55], \alpha \in [0.87,0.89]$ is depicted in Figure 4.

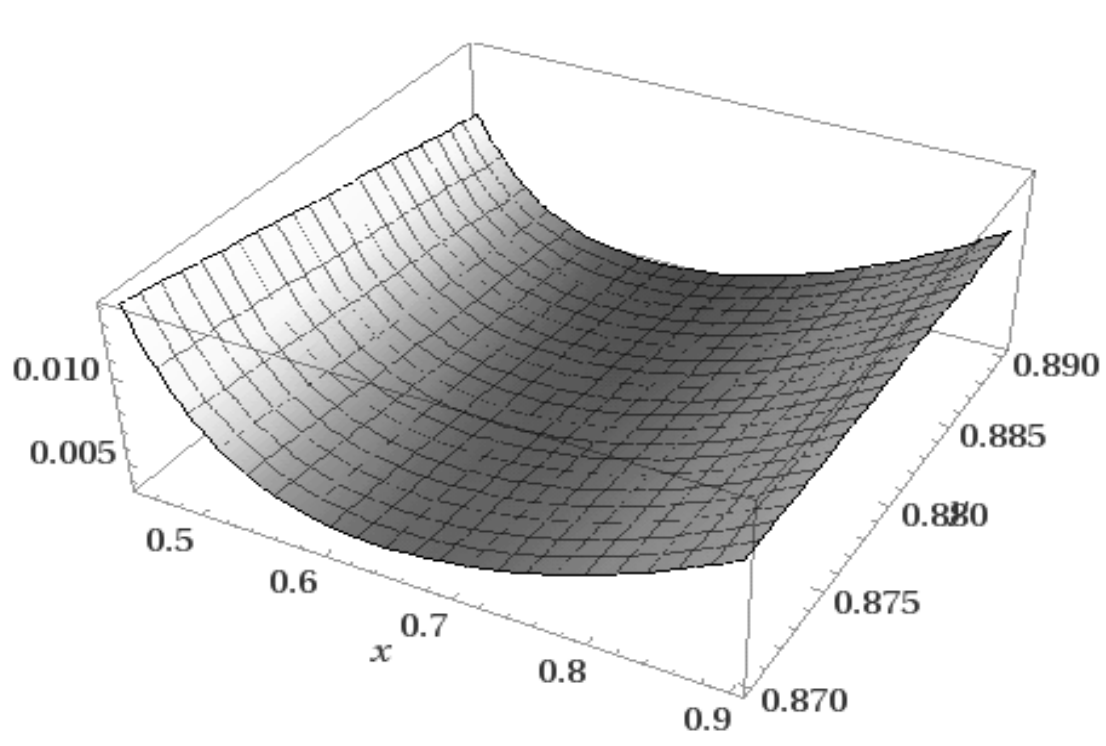

**Figure 4. Plot of the error term** $\left(\frac{w^{(+,\alpha)}(x) - v^{(+)}(x)}{v^{(+)}(x)}\right)^2 \leq \mathbf{0.10}, x \in [\mathbf{0.45}, \mathbf{0.9}], \alpha \in [\mathbf{0.87}, \mathbf{0.89}$

Recall now that a real valued rv $X^{(NG)}$ has a Negative Gumbel distribution if $X^{(NG)} \triangleq NG(\mu,\rho), \mu \in \mathcal{R},\ \rho > 0,$[9] if it has pdf $f^{(NG(\mu,\rho))}$ and cdf $F^{(NG(\mu,a))}$ given by $f^{NG}(x) = \frac{1}{\rho}\exp\left(\frac{x-\mu}{\rho} - e^{\frac{x-\mu}{\rho}}\right), x \in \mathcal{R},$ and $F^{(NG(\mu,a))}(x) = 1 - \exp\left(-e^{\frac{x-\mu}{\rho}}\right), x \in \mathcal{R},$ see Figures 5a and 5b.

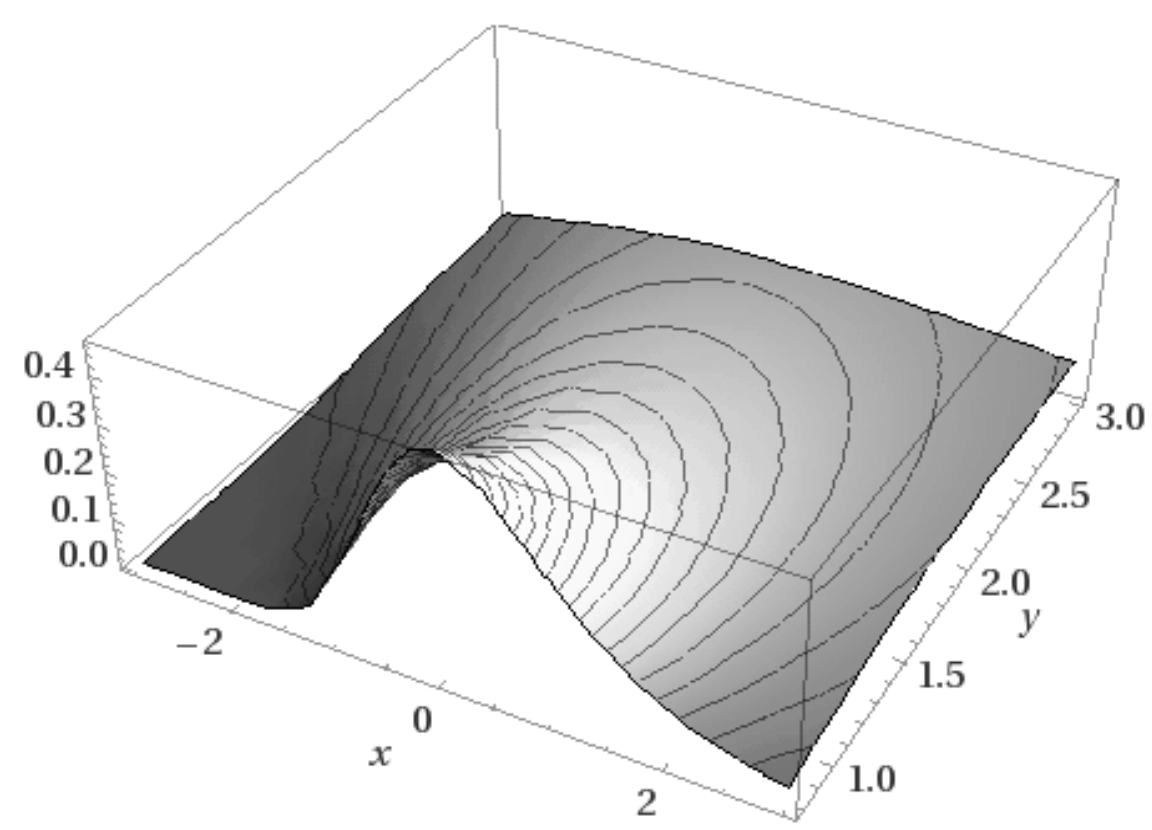

**Figure 5a. Plot of the pdf of Negative Gumbel distribution $f^{(NG(0,\rho))}(x) = \frac{1}{\rho}\exp\left(\frac{x}{\rho} - e^{\frac{x}{\rho}}\right), -3 < x < 3, 0.8 < \rho < 3$**

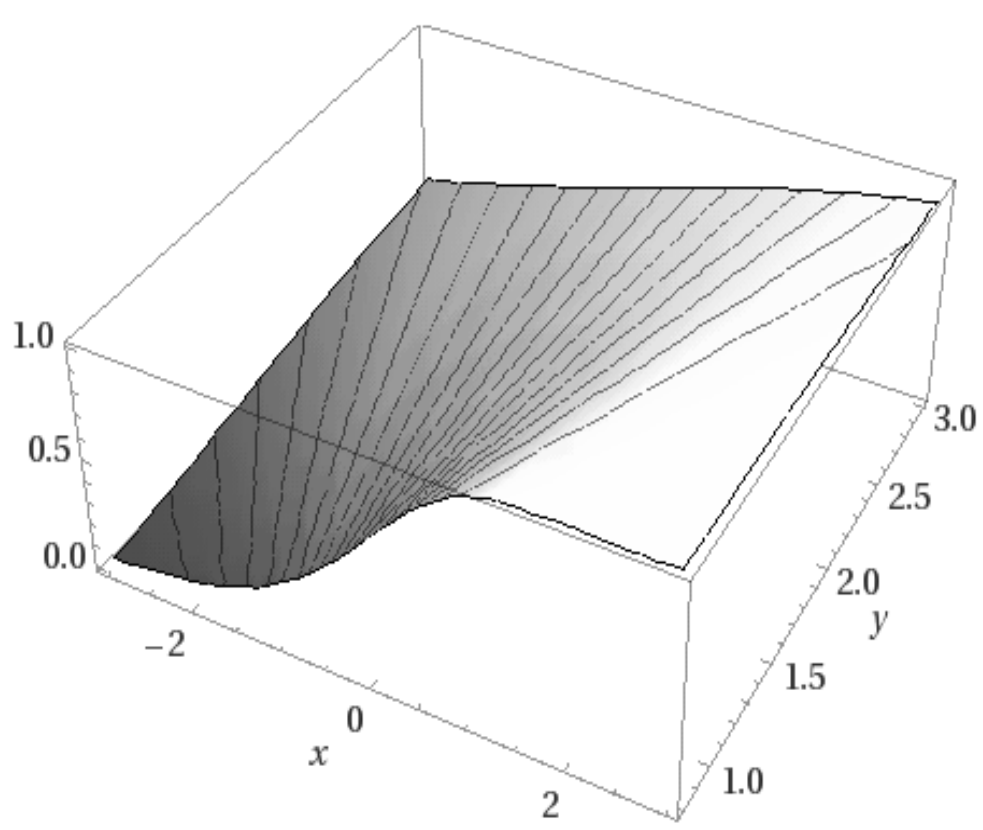

**Figure 5b. Plot of the cdf of Negative Gumbel distribution $F^{(NG(0,\rho))} = 1 - \exp\left(-e^{\frac{x}{\rho}}\right), -4 < x < 4, 0.3 < \rho < 3.$**

[9] See Kotz and Nadarajah (2000. p. 8) and Appendix B, Sectin 2 in Steutel and van Harn (2004).

$\mathbb{E}X^{(NG)} = \mu - \rho\gamma^{(E-M)}$, where $\gamma^{(E-M)} = \lim_{n\uparrow\infty}\left(-ln + \sum_{k=1}^{n}\frac{1}{k}\right)\sim 0.57721$, is the Euler–Mascheroni constant, $varX^{(NG)} = \frac{\pi^2}{6}\rho^2$, skewness $\gamma\left(X^{(NG)}\right) = -\frac{12\sqrt{6}\zeta(3)}{\pi^3}\sim -1.14$, $\zeta(s) = \sum_{n=1}^{\partial} n^{-s}$ is the Riemann zeta function, excess kurtosis $\kappa\left(X^{(NG)}\right) = \frac{1}{12}$ and characteristic function $\varphi^{\left(X^{(NG)}\right)}(\theta) = \mathbb{E}e^{i\theta X^{(NG)}} = e^{i\theta\mu}\Gamma(1 + i\rho\theta), \theta \in \mathcal{R}$. $X^{(NG)}$ is an infinitely divisible rv.

Next, setting $\mathfrak{R}^{(\beth,+)} = w^{(l,+)}\left(\mathfrak{R}^{(+)}\right)$ and $\mathfrak{R}^{(\beth,-)} = -w^{(l-)}\left(-\mathfrak{R}^{(-)}\right)$ leads to $\mathbb{P}\left(\mathfrak{R}^{(\beth,-)} \leq x\right) = 1 - \exp(-\exp\left(\frac{x+v-\lambda lnb}{\lambda}\right)$ and $\mathbb{P}\left(\mathfrak{R}^{(\beth,+)} \leq x\right) = 1 - \exp\left(-exp\left(\frac{x-(c+alnb)}{a}\right)\right), x \in \mathcal{R}$. Thus, $\mathfrak{R}^{(\beth,-)} \triangleq NegGumbel\left(\mu^{(-)}, \rho\right)$, with $\mu^{(-)} = \lambda lnb - v$ , while $\mathfrak{R}^{(\beth,+)} \triangleq NegGumbel\left(\mu^{(+)}, \rho\right)$ =with $\mu^{(+)} = alnb + c$.

As a result, $\mathfrak{R}^{(\beth,0)} = \mathfrak{R}^{(\beth,+)} - \mathfrak{R}^{(\beth,-)}$ has a characteristic function

$$\varphi^{\left(\mathfrak{R}^{(\beth,0)}\right)}(\theta) = e^{(c+v)i\theta}\Gamma(1 + i\rho\theta)\Gamma(1 - i\rho\theta) = e^{(c+v)i\theta}B(1 + i\rho\theta, 1 - i\rho\theta), \quad (7)$$

where $B(x, y) = \frac{\Gamma(x)\Gamma(y)}{\Gamma(x+y)}$ is the Beta function. Thus, $\mathfrak{R}^{(\beth)}$ has a logistic distribution $\mathfrak{R}^{(\beth)} \triangleq Logist(m, \rho)$ [10], $m = c + v$, with pdf

$$f^{\left(\mathfrak{R}^{(\beth,0)}\right)}(x) = f^{(m,\rho)}(x) = \frac{\exp\left(-\frac{x-m}{\rho}\right)}{\rho\left(1+exp\left(-\frac{x-m}{\rho}\right)\right)^2}, x \in \mathcal{R}, \quad (8)$$

[10] See Fisher (1921), McDonald (1991), McDonald and Nelson (1993), Johnson, Kotz, and Balakrishnan (1995), and Ficher (2000a, (200b).

(see Figure 6a) and cdf $F^{(\mathfrak{R}^{(ב,0)})}(x) = F^{(\mathfrak{R},m,\rho)}(x) = \frac{1}{1+exp\left(-\frac{x-m}{\rho}\right)}, x \in \mathcal{R}$ (see Figure 6b). We define posterior return $\mathfrak{R}^{(ב)} := \mathfrak{R}^{(ב,0)} + \mathbb{m}$, where $\mathbb{m} = \mathbb{E}\mathfrak{R}$. Then $\mathfrak{R}^{(ב)}$ is infinitely divisible, with $\mathbb{E}\mathfrak{R}^{(ב)} = \mathbb{m}^{(ב)} := m + \mathbb{m},\ var\ \mathfrak{R}^{(ב)} = \frac{\rho^2\pi^2}{3}, \gamma\left(\mathfrak{R}^{(ב)}\right) = 0$ and $\kappa\left(\mathfrak{R}^{(ב)}\right) = 1.2$.

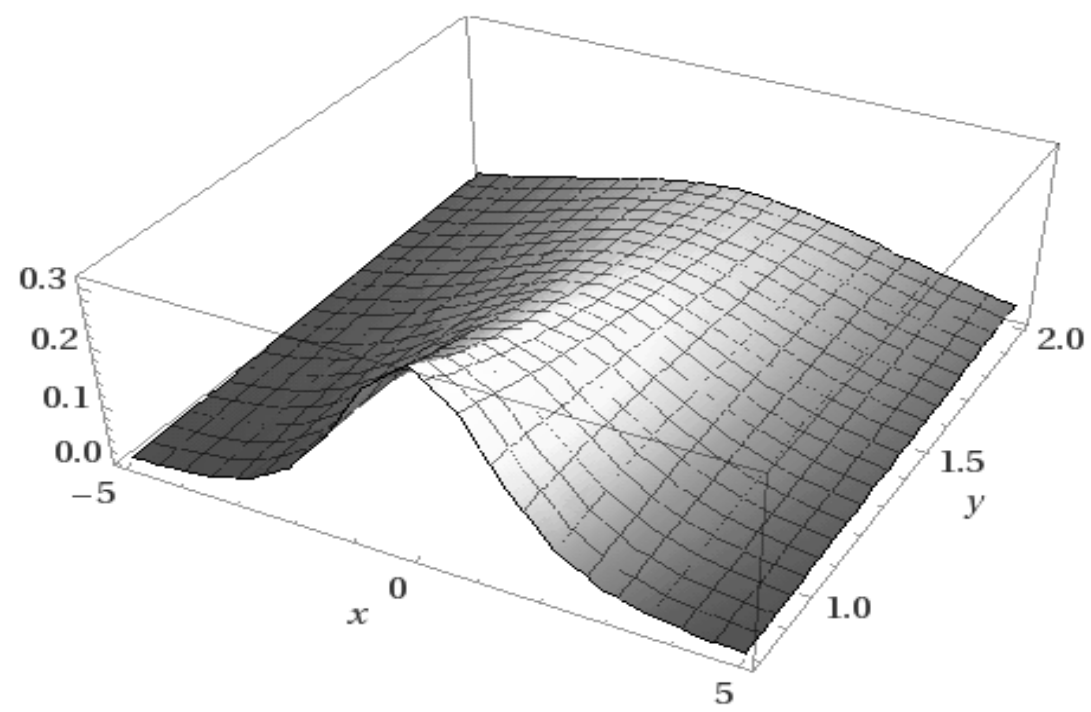

**Figure 6a. Plot of the pdf of logistic distribution** $\boldsymbol{f^{(\mathfrak{R},o,\rho)}(x) = \frac{\exp\left(-\frac{x}{\rho}\right)}{\rho\left(1+exp\left(-\frac{x}{\rho}\right)\right)^2}, -5 < x < 5, 0.8 < \rho < 2}$

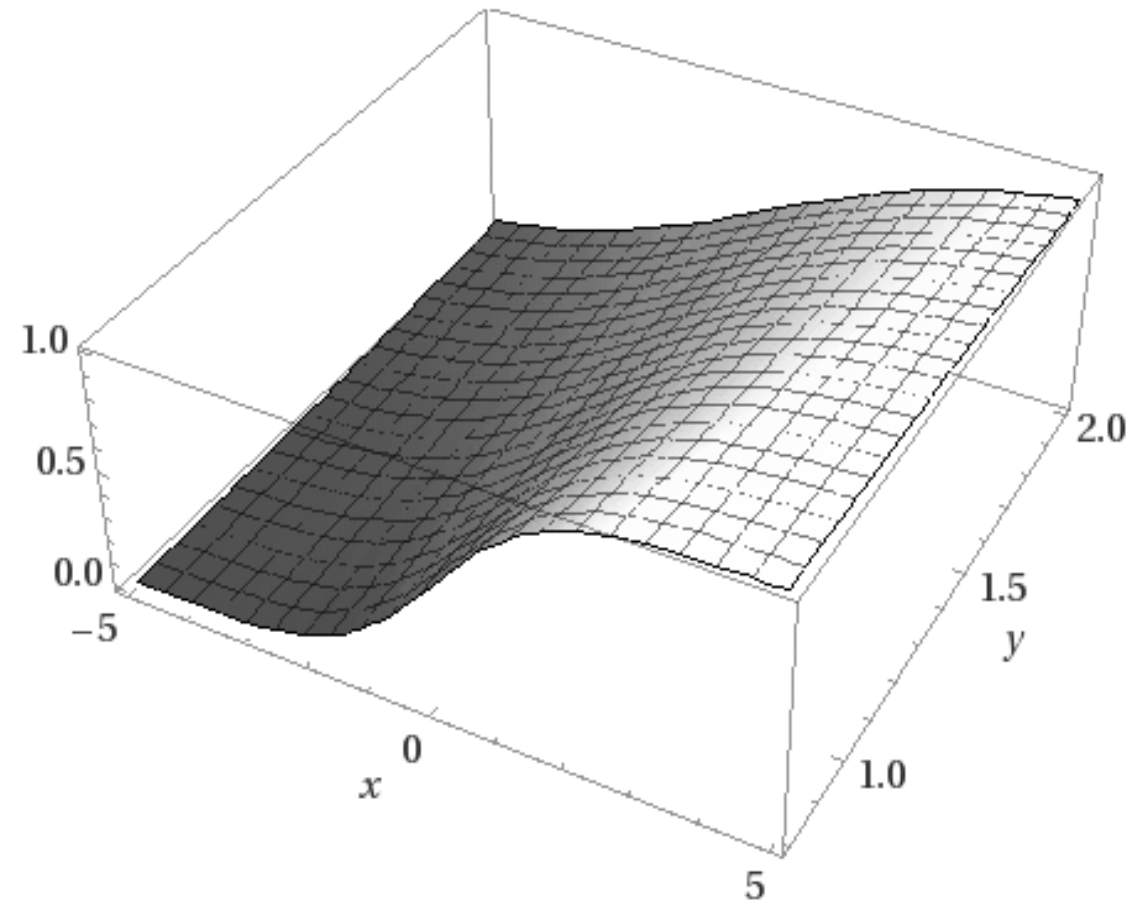

**Figure 6b. Plot of the cdf of logistic distribution** $\boldsymbol{F^{(\mathfrak{R},o,\rho)}(x) = \frac{1}{1+exp\left(-\frac{x}{\rho}\right)}, -5 < x < 5, 0.8 < \rho < 2}$

We pass now to option pricing when the underlying stock-return $\mathfrak{R}^{(\beth)}$ has pdf (8)[11]. Consider a market with

$(i)$ a risky asset (stock) $\mathcal{S}$ with price process $S(t), t \in [0, T]$ following exponential logistic Lévy motion $S(t) = S(0)\mathrm{e}^{\mathcal{L}(\mathrm{t})}, t \geq 0$, where $\mathcal{L}(\cdot)$ is a logistic Lévy motion, that is a Lévy process with $\mathcal{L}(1) \triangleq Logist\left( \mathbb{m}^{(\beth)}, \rho \right)$;

$(ii)$ a riskless asset (bond) $\mathcal{B}$ with price process $B(t) = e^{rt}, t \in [0, T]$, where $r > 0$ is the riskless rate.

Consider a European call option, $\mathcal{C}$, with price process $C(t), t \in [0, T]$ with maturity $T$ and exercise price $K > 0$. To determine $C(0)$ via risk-neutral valuation, we first define the **Esscher density**[12] $f^{(Ess)}(x; t, h), x \in \mathcal{R}, t \geq 0, h > 0$, of $\mathcal{L}(\cdot)$. Let $f^{(\mathcal{L}(\mathrm{t}))}(x), x \in \mathcal{R}$ be the pdf of $\mathcal{L}(\mathrm{t})$, and let $h > 0$ be such that $\mathcal{M}^{(\mathcal{L}(\mathrm{t}))}(\mathrm{h}) = \mathbb{E} \exp\left(h\mathcal{L}(\mathrm{t})\right) = \left(\mathcal{M}^{(\mathcal{L}(1))}(\mathrm{h})\right)^t < \infty$, and define $f^{(Ess)}(x; t, h) := e^{hx} \left(\mathcal{M}^{(\mathcal{L}(1))}(\mathrm{h})\right)^{-t} f^{(\mathcal{L}(\mathrm{t}))}(x)$[13]$, x \in \mathcal{R}, t \geq 0$. Then the characteristic function of $f^{(Ess)}(x; t, h)$ is given by

$$\varphi^{(Ess)}(\theta; t, h) = \int_{-\infty}^{\infty} e^{i\theta x} f^{(Ess)}(x; t, h) dx = \frac{\varphi^{(\mathcal{L}(\mathrm{t}))}(\theta - ih)}{\mathcal{M}^{(\mathcal{L}(\mathrm{t}))}(\mathrm{h})}, \theta \in \mathcal{R}, t \geq 0, \qquad (9)$$

---

[11] See Ficher (2000a) for the detailed proof. Here we only sketch the derivation of the option value at time 0.

[12] See Eberlein, Papapantoleon, and Shiryaev (2009).

[13] $f^{(\mathcal{L}(\mathrm{t}))}(x), x \in \mathcal{R}$ can be obtain from the characteristic function $\varphi^{(\mathcal{L}(\mathrm{t}))}(\theta)$, via standard inversion formula: $f^{(\mathcal{L}(\mathrm{t}))}(x) = \frac{1}{2\pi} \int_{-\infty}^{\infty} e^{-i\theta x} \varphi^{(\mathcal{L}(\mathrm{t}))}(\theta) d\theta$.

where $\varphi^{(\mathcal{L}(\mathrm{t}))}(\theta), \theta \in \mathcal{R}$, is the characteristic function of $\mathcal{L}(\mathrm{t})$,

$$\varphi^{(\mathcal{L}(\mathrm{t}))}(\theta) = \mathbb{E}\exp\left(i\theta\mathcal{L}(\mathrm{t})\right) = \left(e^{\,\mathbb{m}^{(\beth)} i\theta} B(1 + i\rho\theta, 1 - i\rho\theta)\right)^{t}. \qquad (10)$$

Let $\mathcal{M}^{(\mathcal{L}(\mathrm{t}))}(\mathrm{u};\mathrm{h}) := \frac{\mathcal{M}^{(\mathcal{L}(\mathrm{t}))}(\mathrm{u}+\mathrm{h})}{\mathcal{M}^{(\mathcal{L}(\mathrm{t}))}(\mathrm{h})} = \left(\mathcal{M}^{(\mathcal{L}(1))}(\mathrm{u};\mathrm{h})\right)^{t}$. Then the martingale equation of the form

$C(t), S(0) = \mathbb{E}^{(\mathbb{Q})}\left(S(t)\right), t \geq 0, \mathbb{Q} \sim \mathbb{P}$ is equivalent to $r = \ln\left(\mathcal{M}^{(\mathcal{L}(1))}\left(1; h^{(\mathbb{Q})}\right)\right)$, where $h^{(\mathbb{Q})}$ is a root of the martingale function:

$$\mathbb{M}(z) := r - \mathbb{m}^{(\beth)} + \ln\left(B\left(1 + i\rho(z+1), 1 - i\rho(z+1)\right)\right), -\frac{1}{\rho} < z < \frac{1}{\rho} - 1.$$

Letting $\mathbb{k} := lnK - lnS(0)$.then the value of the option $\mathcal{C}$ at time $t = 0$, is given by

$$C(0) = S(0)\int_{\mathbb{k}}^{\infty} f^{(Ess)}\left(x; T, h^{(\mathbb{Q})} + 1\right)dx - e^{-rT}K\int_{\mathbb{k}}^{\infty} f^{(Ess)}\left(x; T, h^{(\mathbb{Q})}\right)dx. \qquad (11)$$

Having formula (11), the goal of a potential empirical study, will be

$(i)$ having daily return data for S&P500 stocks, to calibrate parameters of $\Re \triangleq Laplace(b, \mathbb{m})$ and the generalized PTWF (6) from the market option data on the S&P500 stocks:

$(ii)$ having estimated parameters of the generalized PTWF (6), to make a conclusion about "greed and fear" market sentiment.

We complete this section with a comment on the fact that PTWF (1)[14]. Let $\Re^{(i)}, i = 1,2$ be two returns on two risky assets $\mathcal{S}^{(i)}, i = 1,2$, with cdf $F^{(i)}(x) = \mathbb{P}\left(\Re^{(i)} \leq x\right), x \in \mathcal{R}$. Then $\mathcal{S}^{(1)}$ first-order stochastically dominates $\mathcal{S}^{(2)}$ (denoted by $\mathcal{S}^{(1)} \succ_{(1)} \mathcal{S}^{(2)}$, or equivalently

[14] We refer to Levy (1998) for a general reference on stochastic dominance.

$\mathfrak{R}^{(1)} \succ_{(1)} \mathfrak{R}^{(2)}$, or $F^{(1)} \succ_{(1)} F^{(2)}$), if and only if $F^{(1)}(x) \leq F^{(2)}(x)$ for all $x \in \mathcal{R}$. Consider the class $\mathfrak{U}$ of all non-decreasing functions such $u(x), x \in \mathcal{R}$ with finite $\mathbb{E}\left|u\left(\mathcal{S}^{(1)}\right)\right|$. Then $\mathcal{S}^{(1)} \succ_{(1)} \mathcal{S}^{(2)}$, if and only if $\mathbb{E}\left(u\left(\mathcal{S}^{(1)}\right)\right) \geq \mathbb{E}\left(u\left(\mathcal{S}^{(2)}\right)\right)$ for all $u \in \mathfrak{U}$. The "defect" of the PT is that if in the space of prior returns $\mathfrak{R}^{(1)} \succ_{(1)} \mathfrak{R}^{(2)}$, then, in general, the order will not preserved among the posterior returns $\mathfrak{R}^{(\beth,1)} = v\left(\mathfrak{R}^{(1)}\right)$, $\mathfrak{R}^{(\beth,2)} = v\left(\mathfrak{R}^{(2)}\right)$, see equation (1). With the definitions given by (5), (6) and (8), the first-order stochastically dominant (FOSD) principle is preserved. Indeed suppose that $\mathcal{S}^{(1)} \succ_{(1)} \mathcal{S}^{(2)}$, and the returns $\mathfrak{R}^{(i)}$have are defined by $\mathfrak{R}^{(i)} = \mathfrak{m}^{(i)} + Laplace\left(b^{(i)}\right)$. Because the cdfs of $Laplace\left(b^{(i)}\right)$ with $b^{(1)} \neq b^{(2)}$will always intersect at points $\pm\frac{1}{b^{(2)}-b^{(1)}} ln\frac{b^{(1)}}{b^{(2)}}$, then $\mathfrak{R}^{(1)} \succ_{(1)} \mathfrak{R}^{(2)}$ if and only if $\mathfrak{R}^{(i)} = \mathfrak{m}^{(i)} + Laplace(b), \mathfrak{m}^{(1)} \geq 0$. Thus, $\mathfrak{R}^{(\beth,1)} \succ_{(1)} \mathfrak{R}^{(\beth,2)}$ as required according to the FOSD principle.

**Remark 4**: We believe that the fact that PT fails the FOSD principle, is irrelevant within the scope of BF. In fairness to the PT, we believe that in this instance the critique to PT is ill-founded. First, what is the order (if it exists) between prior $\mathfrak{R}^{(1)}$ and $\mathfrak{R}^{(2)}$, should be irrelevant to the order (if it exists) between the posterior returns $\mathfrak{R}^{(\beth,1)} = v\left(\mathfrak{R}^{(1)}\right)$, $\mathfrak{R}^{(\beth,2)} = v\left(\mathfrak{R}^{(2)}\right)$, as the goal of the transformation $v(\cdot)$ is to change $\beth's$ perception of risk and return, and probably to correct the prior order. Secondly, the very definition of FOSD could be subject to a critique. (We don't claim authorship of the following example, which probably exits in the literature, but we could not find the exact reference). Consider a market where all returns are strictly positively or strictly negatively dependent on the market index return $\mathfrak{R}^{(M)}$, which cdf $F^{\left(\mathfrak{R}^{(M)}\right)}(x), x \in \mathcal{R}$ is strictly increasing and continuous. Suppose that $\mathfrak{R}^{(M)}$, is the positive driving source in the market, in a sense that the stocks following the market index direction always gain, those who do not follow index direction always loose. Let $\mathcal{U} \coloneqq F^{\left(\mathfrak{R}^{(1)}\right)}\left(\mathfrak{R}^{(M)}\right)$. Then every stock $\mathcal{S}$, with return $\mathfrak{R}$ and cdf $F^{(\mathfrak{R})}(x), x \in \mathcal{R}$, has the following representation: either $\mathfrak{R} = F^{(\mathfrak{R},inv)}(\mathcal{U})$, or $\mathfrak{R} = F^{(\mathfrak{R},inv)}(1 - \mathcal{U})$. Next, let $F^{(i)}, i = 1,2$, be two distribution functions, and $F^{(1)} \succ_{(1)} F^{(2)}$. Consider stocks $\mathcal{S}^{(i)}, i = 1,2$ with returns $\mathfrak{R}^{(1)} = F^{(1,inv)}(1 - \mathcal{U})$ and $\mathfrak{R}^{(2)} = F^{(2,inv)}(1 - \mathcal{U})$, where $F^{(i,inv)}(u) = \sup\{x \in \mathcal{R}: F^{(i)}(x) \leq u\}$ is the inverse of $F^{(i)}$. While under the definition of FOSD, $\mathcal{S}^{(1)} \succ_{(1)} \mathcal{S}^{(2)}$, as a matter of fact, $\mathcal{S}^{(1)}$ is a losing stock, while $\mathcal{S}^{(2)}$ is a winning stock.

2.2 General Approach to Prospect Theory Value Function and Option Pricing

In this section, we attempt to define a general approach to prospect theory value function. Our goal is to illustrate that it cannot be a unique class of the PTWP as suggested by equation (1) that should be used for every prior return distribution. As a matter of fact, the shape of the PTWP, should be suggested by the prior return distribution and the "greed and fear" profile of ℶ.

Let $\mathfrak{R}$ and $\mathfrak{R}^{(\beth)}$ be infinitely divisible prior and posterior asset returns. We assume that the moment generating function (mgf) $\mathcal{M}^{(\mathfrak{R}^{(\beth)})}(s) = \mathbb{E}e^{s\mathfrak{R}^{(\beth)}} < \infty, 0 < s < s^{(0)}$[15] .We shall illustrate our point of the need of specific PTWF for a given prior probability distribution $\mathbb{P}^{\mathfrak{R}}$ for $\mathfrak{R}$ and desired by ℶ posterior prior distribution $\mathbb{P}^{\mathfrak{R}^{(\beth)}}$.

**Example 1.** Suppose that ℶ's prior return distribution is $\mathfrak{R} \triangleq Laplace(0, b)$, that is, it has cdf

$$F^{(\mathfrak{R})}(x) = F_{Laplace(0,b)}(x) = \begin{cases} \frac{1}{2}e^{\frac{x}{b}}, x < 0 \\ 1 - \frac{1}{2}e^{-\frac{x}{b}}, x \geq 0 \end{cases}, x \in \mathcal{R}, b > 0, \mathbb{E}(\mathfrak{R}) = 0,$$

$\sigma(\mathfrak{R}) = \sqrt{var\mathfrak{R}} = \sqrt{2}b, \gamma(\mathfrak{R}) = 0, \kappa(\mathfrak{R}) = 3$ (see Figures 3a and 3b). This distribution has exponential tails, and ℶ transform the tails to become thinner, namely Gaussian. ℶ would like to determine the PTVF $\varphi^{(\beth,b,\sigma)}(x), x \in R$, which will transform $\Re$ to $\Re^{(\beth)} \triangleq \mathcal{N}(0,\sigma^2)$, that is, $\Re^{(\beth)} = \varphi^{(\beth,b,\sigma)}(\Re)$. Then, indeed $\varphi^{(\beth,b,\sigma)}(x) = F^{inv}_{\mathcal{N}(0,\sigma^2)} \circ \boldsymbol{F_{Laplace(0,b)}(x)}, x \in \mathcal{R}$, see Figure 7a.

[15] This condition can be relaxed, see Hurst, Platen and Rachev (1999) and Rachev et al (2011).

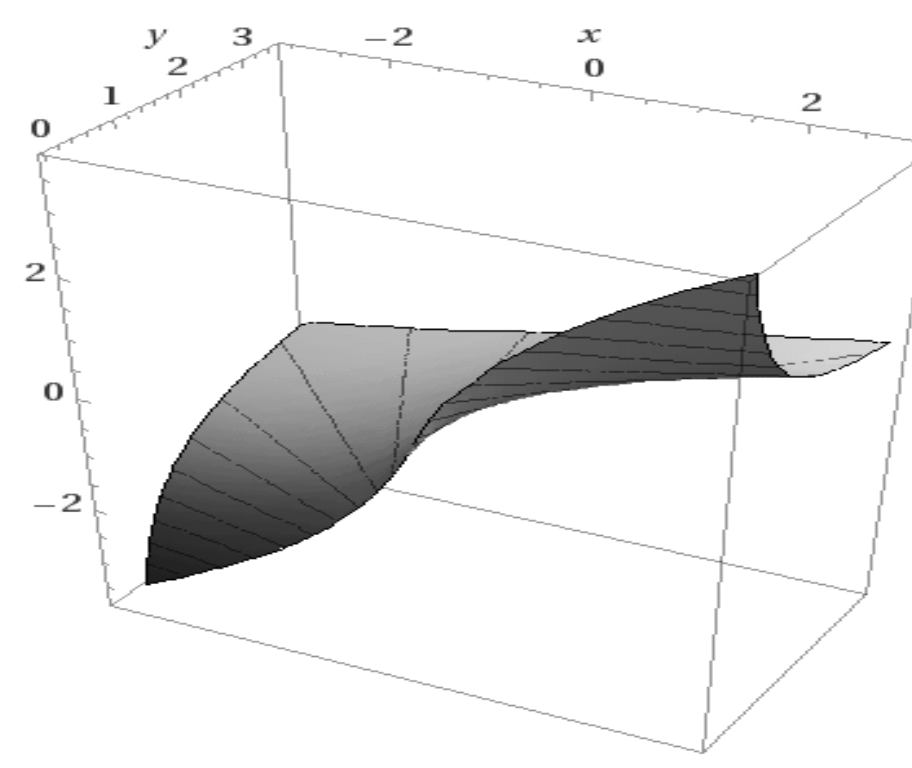

**Figure 7a. Plot of the PTVF $\boldsymbol{\varphi^{(\beth,b,1)}(x)} =$ $\boldsymbol{F^{inv}_{\mathcal{N}(0,1)} \circ F_{Laplace(0,b)}(x), -3 < x < 3, 0.3535 < b < 3.535}$, that is, for $\boldsymbol{0.5 < \sigma(\Re) < 5}$**

As Figure 7a shows, the smaller $\frac{\sigma(\Re)}{\sigma}$ the more pronounced is the convex-concave behavior $\varphi^{(\beth,b,\sigma)}(\cdot)$.

Suppose now ב would like to determine PTVF $\varphi^{(\beth,b,\rho)}(x) \quad x \in R$, which will transform $\Re = \Re \triangleq Laplace(0,b)$ to $\Re^{(\beth)} \triangleq DPareto(\rho), \rho \in (1,\infty)$, (see Example 2 for detailed definition an properties of Double Pareto distribution). Then $\Re^{(\beth)} \triangleq \varphi^{(\beth,b,\rho)}(\Re)$ implies

$$\varphi^{(\beth,b,\rho)}(x) = F^{(inv)}_{DPareto(\rho)}\left(\boldsymbol{F_{Laplace(0,b)}(x)}\right) = \begin{cases} 1 - \mathrm{e}^{\frac{-x}{(\rho-1)b}} & , x < 0 \\ \mathrm{e}^{\frac{x}{(\rho-1)b}} - 1, & x \geq 0 \end{cases}$$

As Figure 7b shows, the smaller $(\rho - 1)b \in (0,1)$ is the more pronounce concave-convex behavior $\varphi^{(\beth,b,\sigma)}(\cdot)$.

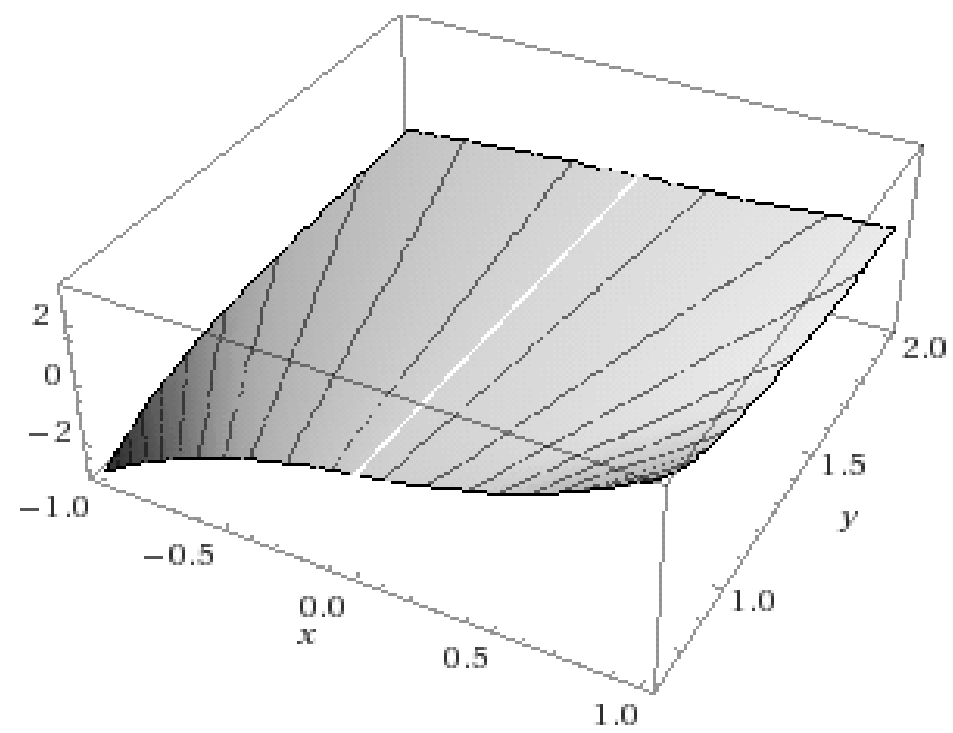

**Figure 7b. Plot of the PTVF**

$$\boldsymbol{\varphi}^{(\beth,\boldsymbol{b},\boldsymbol{\rho})}(\boldsymbol{x}) = \boldsymbol{F}^{(\boldsymbol{inv})}_{\boldsymbol{DPareto}(\boldsymbol{\rho})}\left(\boldsymbol{F}_{\boldsymbol{Laplace}(\mathbf{0},\boldsymbol{b})}(\boldsymbol{x})\right)$$

$$= \begin{cases} \mathbf{1} - \mathbf{e}^{\frac{-\boldsymbol{x}}{(\boldsymbol{\rho}-\mathbf{1})\boldsymbol{b}}}, -\mathbf{1} < \boldsymbol{x} < \mathbf{0} \\ \mathbf{e}^{\frac{\boldsymbol{x}}{(\boldsymbol{\rho}-\mathbf{1})\boldsymbol{b}}} - \mathbf{1}, \mathbf{0} \leq \boldsymbol{x} < \mathbf{1}, \boldsymbol{y} = (\boldsymbol{\rho}-\mathbf{1})\boldsymbol{b} \in \mathbf{0.7} < \mathbf{2} \end{cases}$$

Applying PTWF (1) to ב, is useless from the point of view of RDAPT, as it will lead to $\mathbb{P}^{\mathfrak{R}^{(\beth)}}$ being non-infinitely divisible distribution.

**Example 2.** Suppose that ב's prior return distribution is Double-Pareto[16] $\mathfrak{R} = \mathfrak{R}^{(DP,\rho)} \triangleq DPareto(\rho), \rho \in (1,\infty)$, with pdf $f^{(\mathfrak{R})}(x) = f_{DPareto(b)}(x) = \frac{1}{2}(\rho - 1)(1 + |x|)^{-\rho}$, and cdf

$$F^{(\mathfrak{R})}(x) = F_{DPareto(\rho)}(x) = \begin{cases} \frac{1}{2}(1-x)^{1-\rho}, x < 0 \\ 1 - \frac{1}{2}(1+x)^{1-\rho}, x \geq 0, \end{cases}$$

see Figures 8a and 8b.

[16] See Section B2 in Steutel and van Harn (2004).

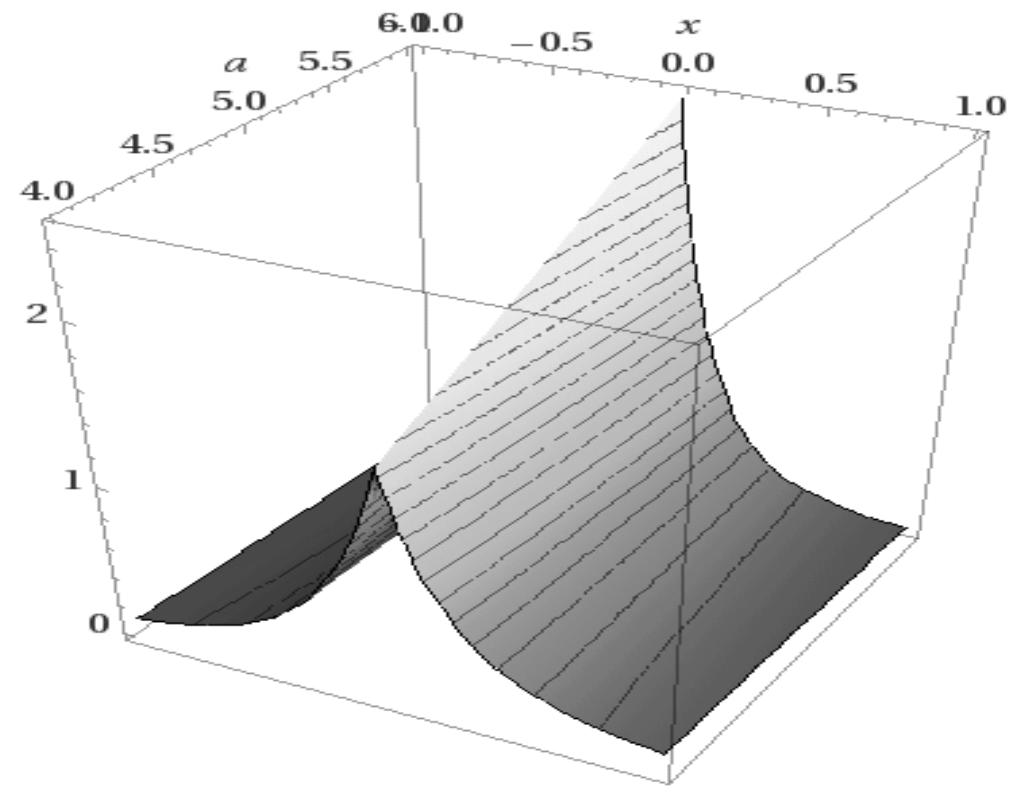

**Figure 8a. Plot of the pdf of Double Pareto Distribution,** $\boldsymbol{f^{(\Re)}(x) = f_{DPareto(b)}(x) = \frac{1}{2}(\rho - 1)(1 + |x|)^{-\rho}, -1 < x < 1, 4 < \rho < 6}$

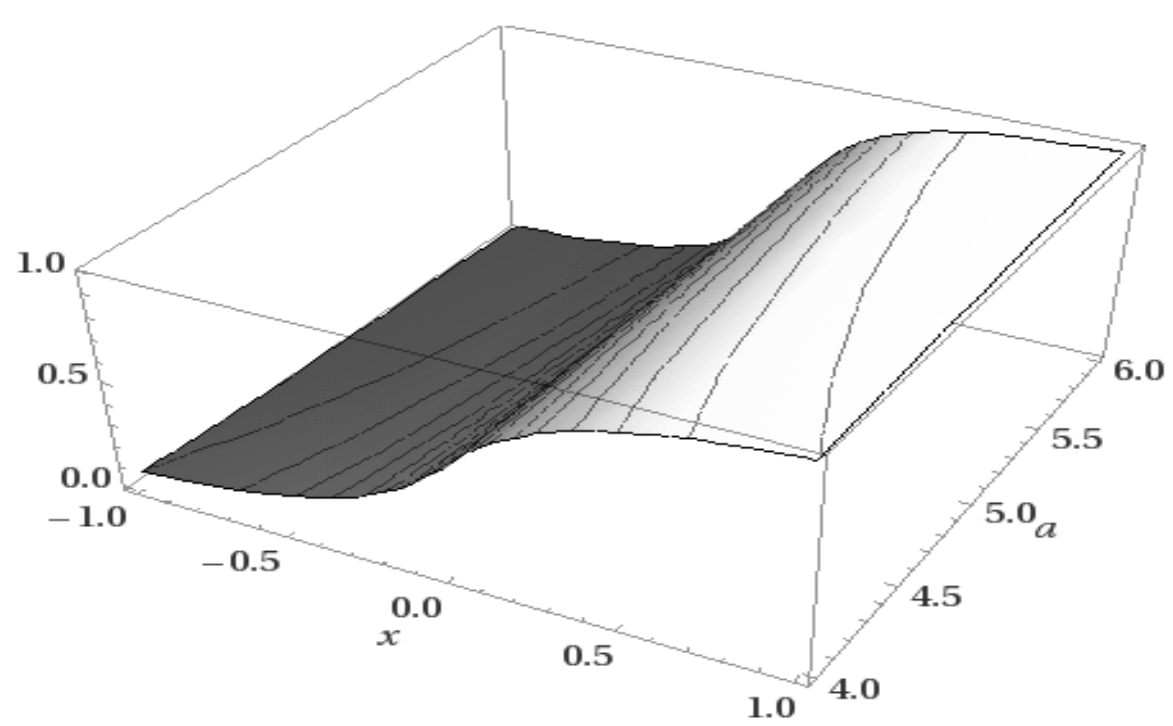

**Figure 8b. Plot of the cdf of Double Pareto Distribution,** $\boldsymbol{F^{(\Re)}(x) = F_{DPareto(\rho)}(x) = \begin{cases} \frac{1}{2}(1 - x)^{1-\rho}, x < 0 \\ 1 - \frac{1}{2}(1 + x)^{1-\rho}, x \geq 0 \end{cases}, -1 < x < 1, 4 < \rho < 6}$

Double-Pareto distribution has heavy Pareto tails and $\mathbb{E}\left|\Re^{(DP,\rho)}\right|^{a} < \infty$ if $\rho \in (1, a)$, and $\mathbb{E}\left|\Re^{(DP,\rho)}\right|^{a} = \infty$ if $\rho \in [a, \infty)$. $\beth$ , like to determine the PTVF $\varphi^{(\beth,b,\rho)}(x), x \in \mathcal{R}$, which will transform $\Re$ to and Laplace distribution $\Re^{(\beth)} \triangleq Laplace(0, b)$ from $\Re^{(\beth)} \triangleq \varphi^{(\beth,b,\rho)}(\Re)$, it follows that $\varphi^{(\beth,b,\rho)}(x) = F^{(inv)}_{Laplace(0,b)} \circ F_{DPareto(\rho)}(x) = \begin{cases} b(1 - \rho)\ln(1 - x), \ x < 0 \\ -b(1 - \rho)\ln(1 + x), \ x \geq 0 \end{cases},$ see Figures 9a and 9b.

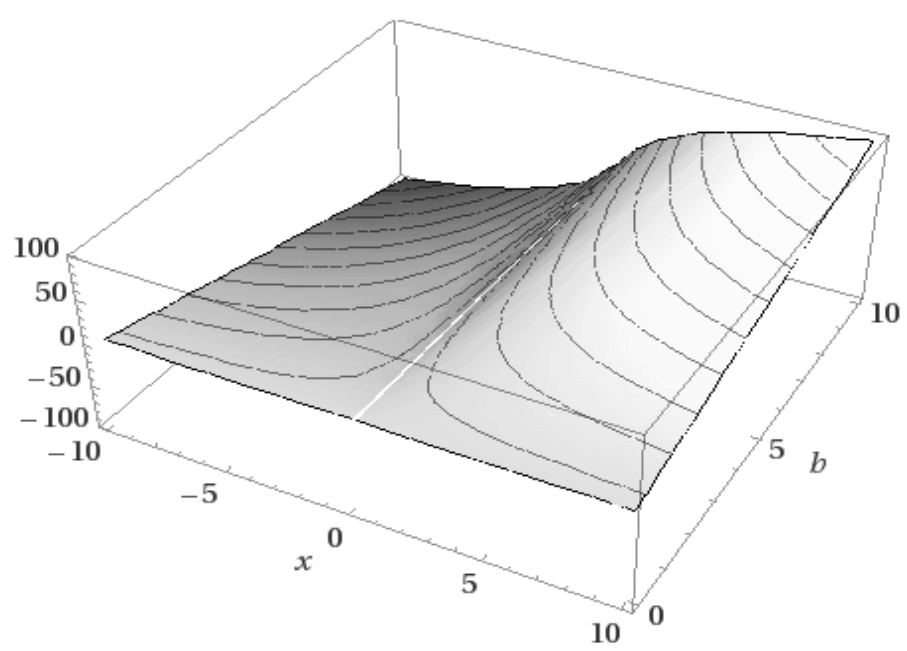

**Figure 9a. Plot of** $\boldsymbol{\varphi^{(\text{ב},b\rho)}(x)} = \begin{cases} \boldsymbol{b(1-\rho)\ln(1-x), 10 < \ x < 0} \\ \boldsymbol{-b(1-\rho)\ln(1+x), 0 \le x \le 10} \end{cases}, \boldsymbol{0.1 < b < 10, \rho = 5}$

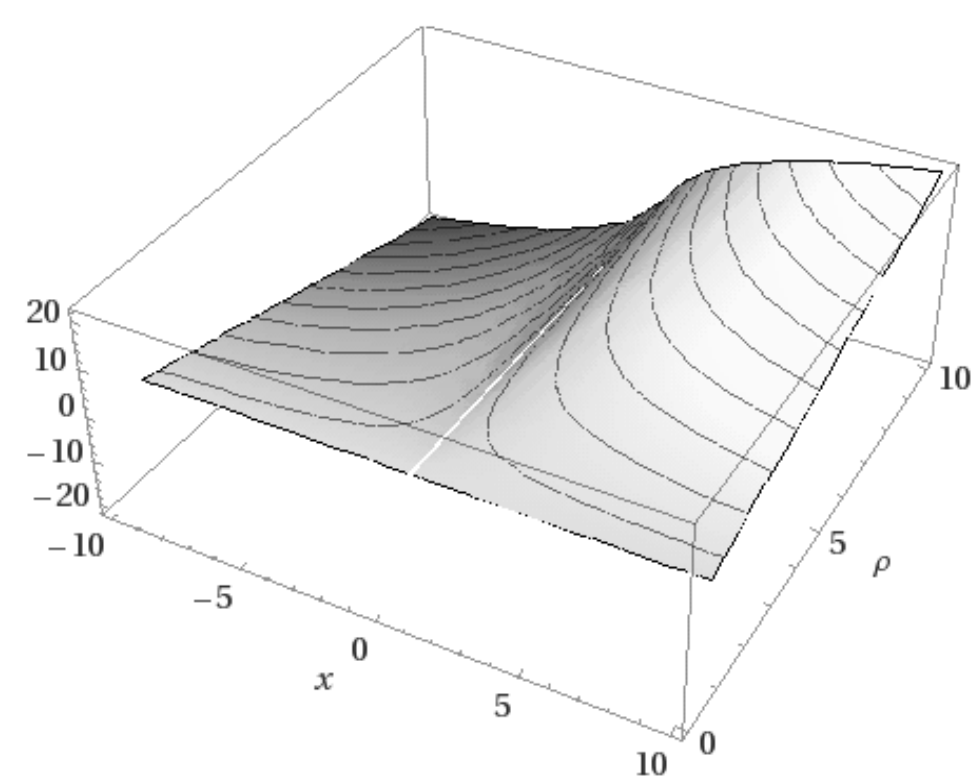

**Figure 9b. Plot of** $\boldsymbol{\varphi^{(\text{ב},b\rho)}(x)} = \begin{cases} \boldsymbol{b(1-\rho)\ln(1-x), 10 < \ x < 0} \\ \boldsymbol{-b(1-\rho)\ln(1+x), 0 \le x \le 10} \end{cases}, \boldsymbol{b = 1, 1 < \rho < 10}$

Suppose that ב's posterior return distribution is again Double-Pareto $\mathfrak{R}^{(\text{ב})} \triangleq DPareto\left(\rho^{(\text{ב})}\right), \rho^{(\text{ב})} \in (1, \infty)$. The corresponding PTWF is given by

$$\varphi^{(\text{ב},\rho,\rho^{(\text{ב})})}(x) = F^{(inv)}_{DPareto\left(\rho^{(\text{ב})}\right)}\left(F_{DPareto(\rho)}(x)\right) =$$

$$= \begin{cases} 1-(1-x)^{\frac{1-\rho}{1-\rho^{(\text{ב})}}}, & x < 0 \\ (1+x)^{\frac{1-\rho}{1-\rho^{(\text{ב})}}} - 1, & x \ge 0 \end{cases}$$

As Figure 9c shows, the larger $\frac{1-\rho}{1-\rho^{(\beth)}}$ is the more pronounce concave-convex behavior $\varphi^{(\beth,\rho,\rho^{(\beth)})}(\cdot)$.

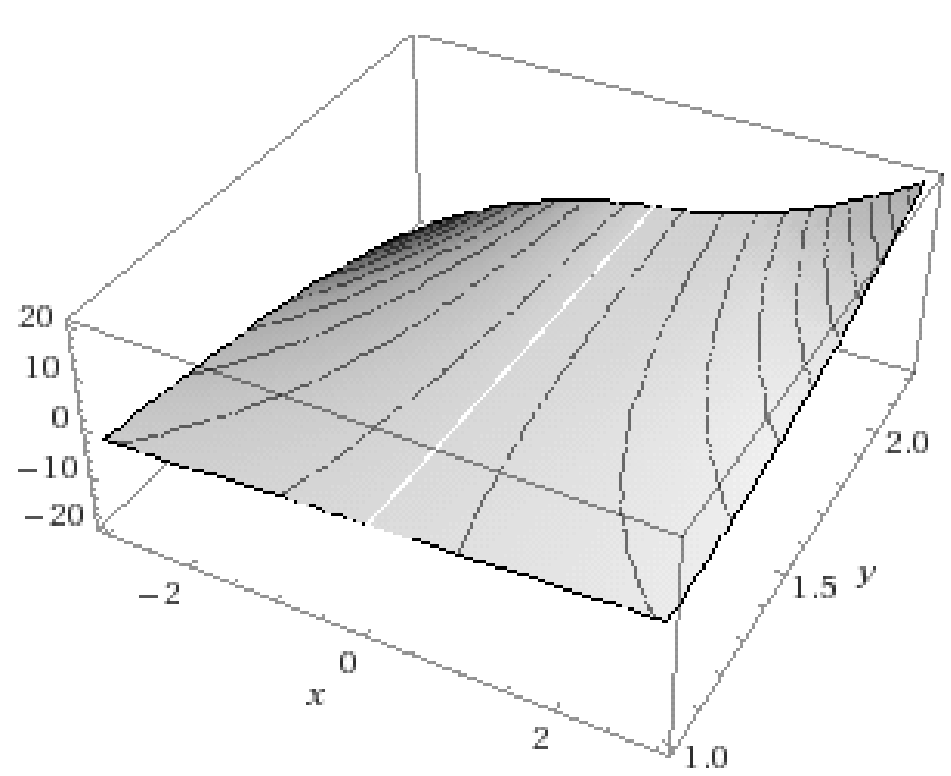

**Figure 9c. Plot of the PTVF**

$$\boldsymbol{\varphi}^{(\beth,\boldsymbol{\rho},\boldsymbol{\rho}^{(\beth)})}(\boldsymbol{x}) = \boldsymbol{F}^{(\boldsymbol{inv})}_{\boldsymbol{DPareto}(\boldsymbol{\rho}^{(\beth)})}\left(\boldsymbol{F}_{\boldsymbol{DPareto}(\boldsymbol{\rho})}(\boldsymbol{x})\right) =$$

$$= \begin{cases} \mathbf{1}-(\mathbf{1}-\boldsymbol{x})^{\frac{\mathbf{1}-\boldsymbol{\rho}}{\mathbf{1}-\boldsymbol{\rho}^{(\beth)}}}, -\mathbf{3} < \boldsymbol{x} < \mathbf{0} \\ (\mathbf{1}+\boldsymbol{x})^{\frac{\mathbf{1}-\boldsymbol{\rho}}{\mathbf{1}-\boldsymbol{\rho}^{(\beth)}}} - \mathbf{1},\ \ \mathbf{0} \le \boldsymbol{x} < \mathbf{3}, \boldsymbol{y} = \dfrac{\mathbf{1}-\boldsymbol{\rho}}{\mathbf{1}-\boldsymbol{\rho}^{(\beth)}} \in (\mathbf{1}, \mathbf{2.2}) \end{cases}$$

Again, the concave-convex shape of $\varphi^{(\beth,b,\rho)}(x), x \in \mathcal{R},$ is drastically different from the suggested PTVF given by (1), which again, as shown in this example, is meaningless from the point of view of the RDAPT.

We conclude this section with following the following two observations. First, PTVF given by (1) makes no sense within the framework RDAPT. Second, there is no universal parametric class of PTVFs. The choice of the parametric class of PTVFs is governed by the prior and posterior infinitely divisible return distribution ב is considering.

**3. Generalized Prospect Theory Weighing Function and Option Pricing with Logistic -Lévy Asset Return Process**

As we discussed in the previous section, the PTWF violates first-order stochastic dominance. To overcome this "weakness" of PT, Tversky and Kahneman (1992) (T&K hereafter) introduced the CPT, where the positive return (gains) and negative returns (losses) generated by financial assets are viewed differently as a result of the general "fear" disposition of traders, but the weighing function is defined on the space of cdfs of asset returns. Let again $\mathfrak{R}$ denote the return of a risky asset and $F^{(\mathfrak{R})}(x) = \mathbb{P}(\mathfrak{R} \leq x), x \geq 0$ refer to the cdf of $\mathfrak{R}$. Then T&K suggested that $\beth$ will overweight the loss $\mathfrak{R}Ind\{\mathfrak{R} < 0\}$ and underweight the gain $\mathfrak{R}Ind\{\mathfrak{R} < 0\}$. To quantify this assertion T&K introduced a continuous strictly increasing PWF $w^{(\beth)}(u), u \in [0,1], w^{(\beth)}(0) = 0, w^{(\beth)}(1) = 1,$ tempering the shape of $F^{(\mathfrak{R})}(x), x \in \mathcal{R}$. As a result, $\beth$ does not use $F^{(\mathfrak{R})}(x), x \in R,$ but instead $\beth$ will use as the cdf for $\mathfrak{R}$ the following penalized cdf:

$$F^{(\beth,\mathfrak{R})}(x) \coloneqq w^{(\beth)}\left(F^{(\mathfrak{R})}(x)\right). \tag{12}$$

In the sub-sections to follow we will study various parametric classes for the PWF $w^{(\beth)}(u), u \in [0,1]$ that have been suggested in the literature and we shall introduce new ones.

3.1 Tversky and Kahneman's Probability Weighting Function

Tversky and Kahneman (1992) introduced the following PWF:

$$w^{(\beth)}(u) \coloneqq \omega^{(\beth,\gamma)}(u) := \frac{u^{\gamma}}{[u^{\gamma}+(1-u)^{\gamma}]^{\frac{1}{\gamma}}}, u \in (0,1], \gamma \in [0,1]. \tag{13}$$

Figures 10a shows a plot of this shape.

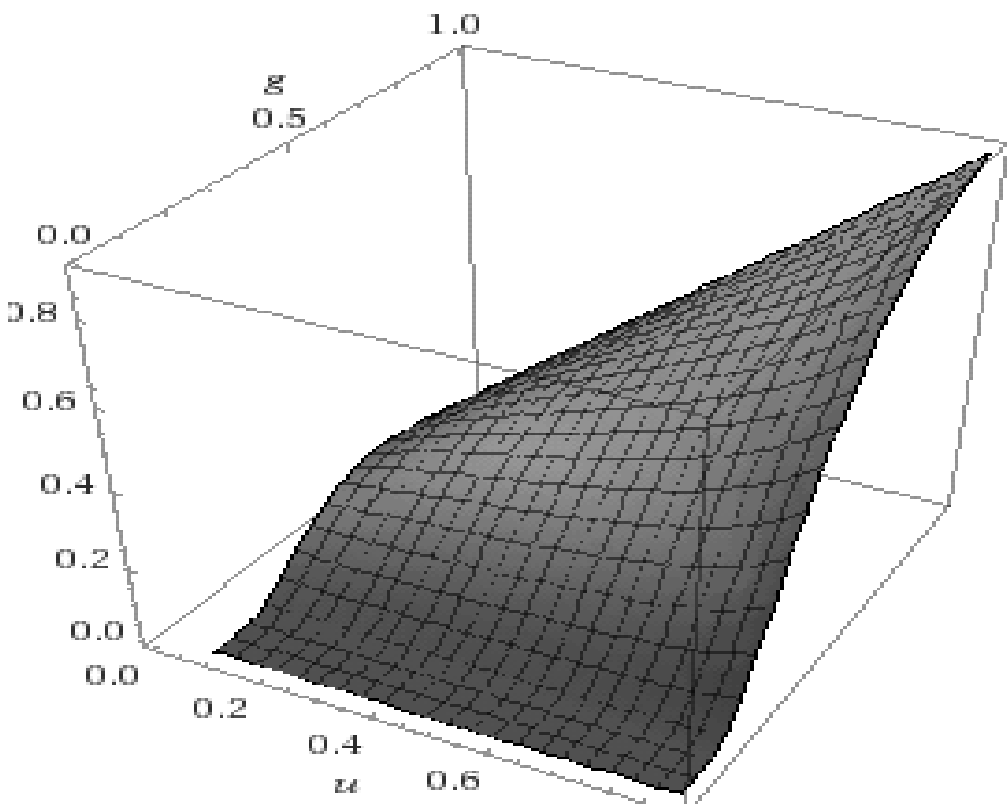

**Figure 10a: Plot of T&K's wpf for a fearful investor.**

**This figure gives the plot of** $\boldsymbol{w^{(\text{ב},\gamma)}(u) := \frac{u^{\gamma}}{[u^{\gamma}+(1-u)^{\gamma}]^{\frac{1}{\gamma}}}, u \in [0,1], 0 < \gamma < 1}$ where[17] it is clear that the closer parameter $\gamma$ is to 0.5, the more fearful ב is. Figure 10b shows the plot of the shape of the first derivative.

[17] Golstein and Einhorn (1987) and Gonzalez and Wu (1999) consider the slightly more general case by adding one more parameter:

$$\boldsymbol{w^{(\text{ב},\gamma,a)}(u) := \frac{au^{\gamma}}{[au^{\gamma}+(1-u)^{\gamma}]^{\frac{1}{\gamma}}}, u \in [0,1], 1 \leq \gamma \leq 2, a > 1.}$$

parameter $\gamma$ changes the curvature of $w^{(\text{ב},\gamma,a)}$, the parameter $a$ controls the height of $w^{(\text{ב},\gamma,a)}$. See also the reviews by Al-Nowaihi and Dhami (2010), Ackert and Deaves (2010, Chapter 3), and Takemura and Murakami (2016).

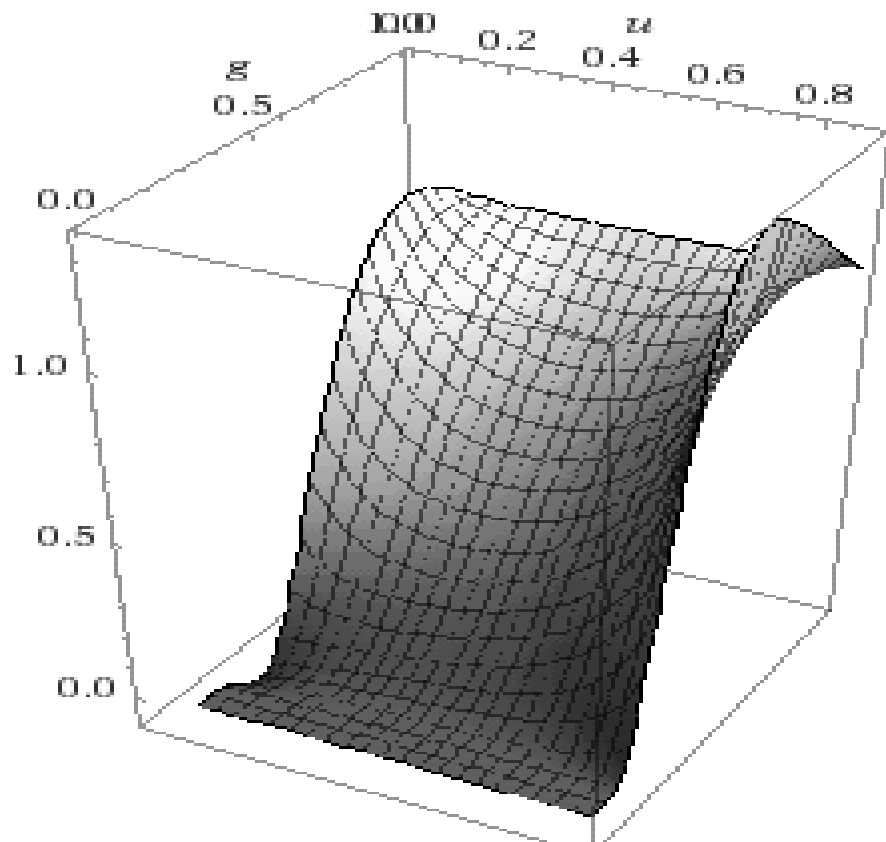

. **Figure 10b: Plot of the first derivative** $\frac{\partial w^{(\text{ב},\gamma)}(u)}{\partial u}(u) := \frac{u^{\gamma}}{[u^{\gamma}+(1-u)^{\gamma}]^{\frac{1}{\gamma}}}, u \in [0,1], 0<\gamma<1$ **of the T&K's wpf in the case of fearful investor.**

When $\gamma = 1, w^{(\text{ב},1)}(u) := u$, and $F^{(\text{ב},\Re)}(x) = F^{(\Re)}(x), x \in R$. Indeed, $lim_{\gamma \downarrow 0}\omega^{(\text{ב},\gamma)}(u) = 0$. As seen from Figures 11a and 11b, when $\boldsymbol{\gamma > 1}$, $\text{ב}'s$ disposition is "greedy", meaning that ב will overweight the gain $\Re Ind\{\Re > 0\}$ and underweight the loss $\Re Ind\{\Re > 0\}$.[18]

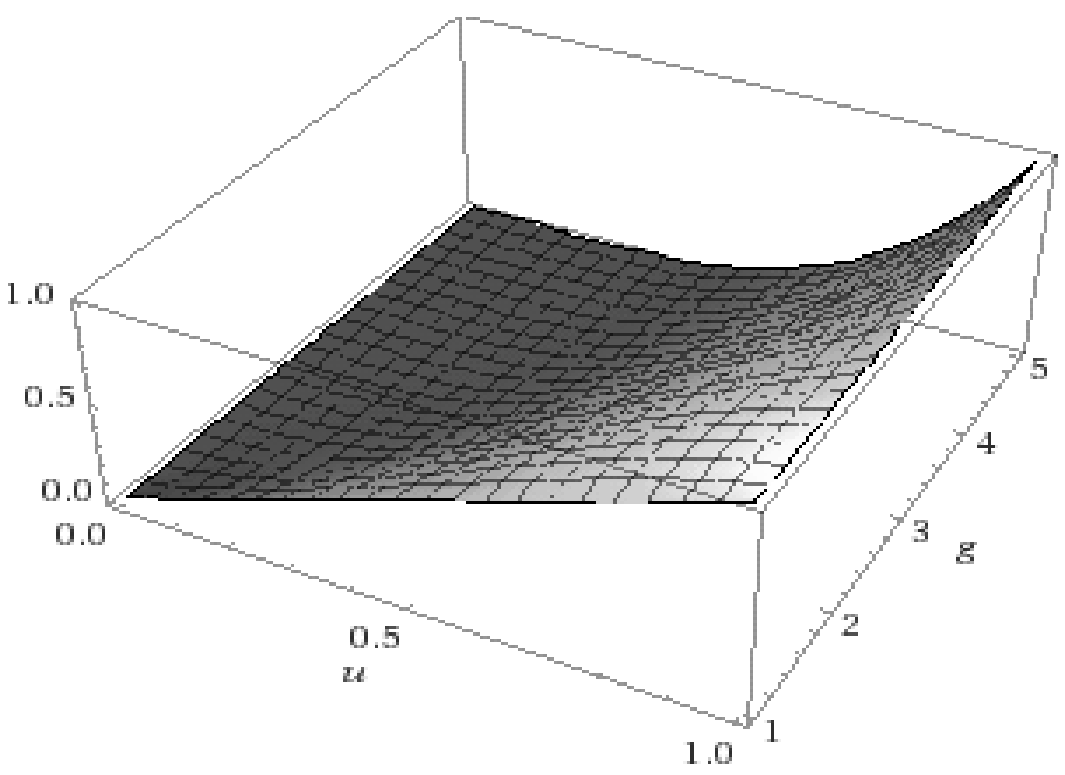

**Figure 11a: Plot of T&K's wpf** $w^{(\text{ב},\gamma)}(u) := \frac{u^{\gamma}}{[u^{\gamma}+(1-u)^{\gamma}]^{\frac{1}{\gamma}}}, u \in [0,1], 1<\gamma<5$ **for a greedy investor.**

---

[18] T&K assume that all traders are fearful. We do not know that, and would like to test whether the majority of market participants are fearful or greedy in a particular timeframe. That is why we assume that $\gamma > 0$ allowing for ב to be a greedy or fearful trader.

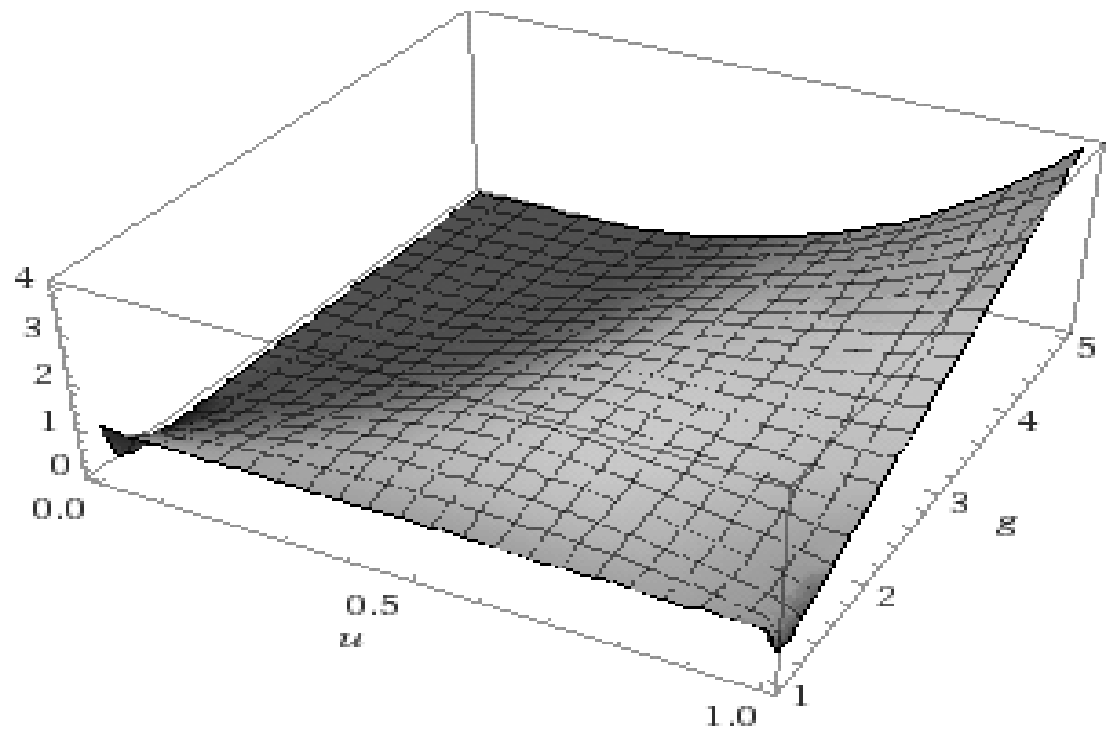

**Figure 11b: Plot of the first derivative T&K's wpf in the case of greedy investor.**

Note that from the point of view of risk-averse investor ב, the T&K-PWF $\omega^{(\beth,\gamma)}(u), 0 < u < 1$, makes contradictory sense. As a matter of fact, by applying (13), the investor with zero prior information on the stock return, becomes quite bullish on the stock. To see that, suppose ב has no information about the stock return, so ב's prior stock-return distribution is given by $\mathcal{U} \triangleq U[-1,1]$[19], a uniform rv on $[-1,1]$. Being most fearful investor ב decides to use as posterior return $\mathfrak{R}^{\left(\beth,\frac{1}{2}\right)} := \omega^{\left(\beth,\frac{1}{2}\right)}(\mathcal{U})$, as suggested in (13). Then the cdf and the pdf of $\mathfrak{R}^{\left(\beth,\frac{1}{2}\right)}$ have the form posterior return $\mathfrak{R}^{\left(\beth,\frac{1}{2}\right)} := \omega^{\left(\beth,\frac{1}{2}\right)}(\mathcal{U})$, as suggested in (13). Then the cdf and the pdf of $\mathfrak{R}^{\left(\beth,\frac{1}{2}\right)}$ have the form

$$F^{\left(\beth,\frac{1}{2}\right)}(x) = \mathbb{P}\left(\mathfrak{R}^{\left(\beth,\frac{1}{2}\right)} \le x\right) = \frac{\sqrt{2x+2}}{\left[\sqrt{x+1}+\sqrt{1-x}\right]^2}, -1 < x < 1$$

[19] $\mathcal{U} \triangleq U[-1,1]$ is not infinitely divisible and should not be used as a prior distribution with the RDAPT. However, we use here just as an illustration of the meaningless of using (13) even within the framework of BF only.

$$f^{\left(\beth,\frac{1}{2}\right)}(x)=\frac{\partial F^{\left(\beth,\frac{1}{2}\right)}(x)}{\partial x}=\frac{3+x-\sqrt{1-x^2}}{\sqrt{2-2x^2}\left[\sqrt{x+1}+\sqrt{1-x}\right]^3},-1<x<1,$$

see Figures 12a and 12b.

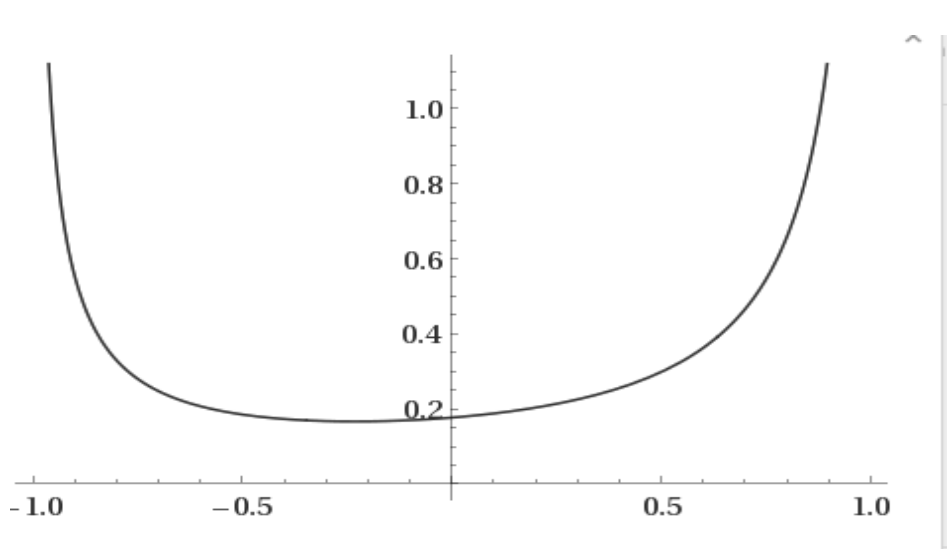

**Figure 12a: The pdf of $\mathfrak{R}^{\left(\beth,\frac{1}{2}\right)}$**

**Plot of** $\boldsymbol{f^{\left(\beth,\frac{1}{2}\right)}(x)=\frac{\partial F^{\left(\beth,\frac{1}{2}\right)}(x)}{\partial x}=\frac{3+x-\sqrt{1-x^2}}{\sqrt{2-2x^2}\left[\sqrt{x+1}+\sqrt{1-x}\right]^3},-1<x<1}$.

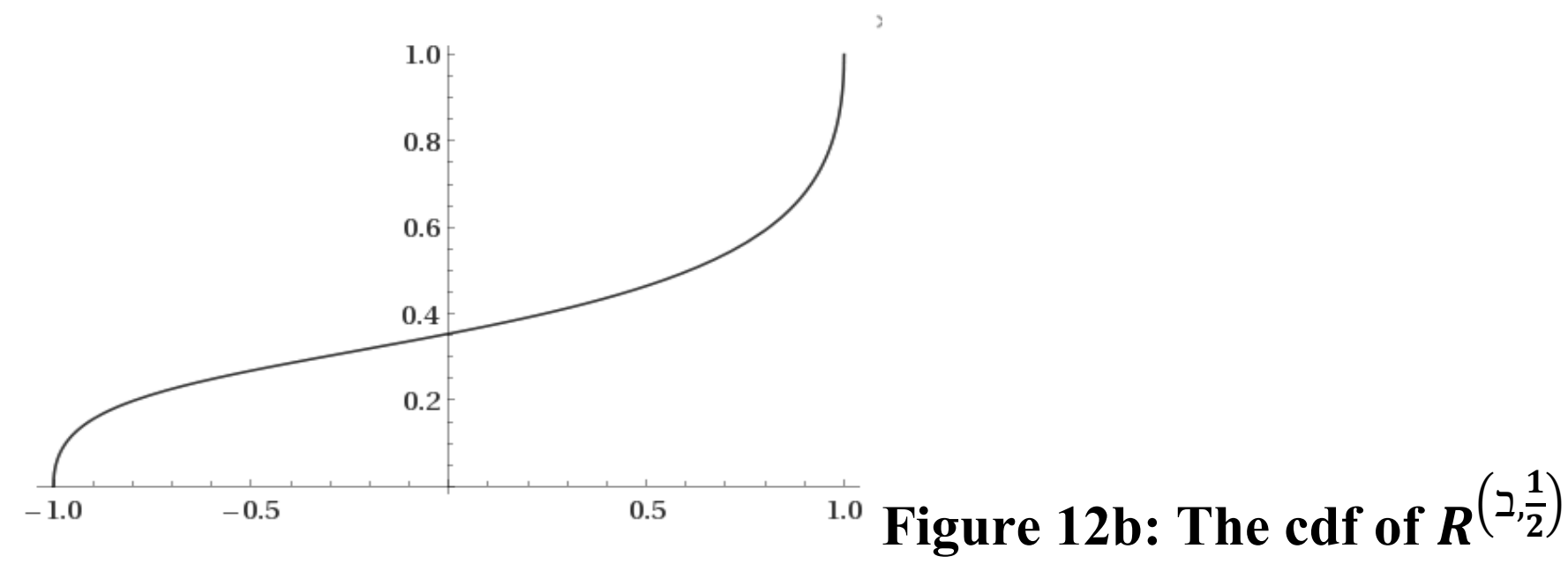

**Figure 12b: The cdf of $\boldsymbol{R^{\left(\beth,\frac{1}{2}\right)}}$**

**Plot of** $\boldsymbol{F^{\left(\beth,\frac{1}{2}\right)}(x)=\mathbb{P}\left(\mathfrak{R}^{\left(\beth,\frac{1}{2}\right)}\leq x\right)=\frac{\sqrt{2x+2}}{\left[\sqrt{x+1}+\sqrt{1-x}\right]^2}\ ,-1<x<1}$

The mean of $\mathfrak{R}^{\left(\beth,\frac{1}{2}\right)}$ is not anymore zero, rather it is positive $\mathbb{E}\mathfrak{R}^{\left(\beth,\frac{1}{2}\right)}=0.24645$. The variance is $var\mathfrak{R}^{\left(\beth,\frac{1}{2}\right)}=0.5917314$, and standard deviation $\sqrt{var\mathfrak{R}^{\left(\beth,\frac{1}{2}\right)}}=0.7692408$ . The information ratio with benchmark return being zero, is

$$\mathbb{I}\left(\mathfrak{R}^{\left(\beth,\frac{1}{2}\right)}\right)=\frac{\mathbb{E}\mathfrak{R}^{\left(\beth,\frac{1}{2}\right)}}{\sqrt{var\mathfrak{R}^{\left(\beth,\frac{1}{2}\right)}}}=0.3203808.$$

In the financial industry, it is typically accepted that an information ratio that falls within the range of 0.40 to 0.60 is “quite good”.[20] As a matter of fact, by changing $\beth's$ prior uniform distribution (for which the information ratio is indeed zero), $\beth$ becomes quite bullish on this particular stock.

As a second illustration, suppose $\beth's$ prior stock-return distribution is standard normal $\mathcal{N} \triangleq N[0,1]$, and again $\beth$ decides to use as the posterior return $\Re^{\left(\mathcal{N},\beth,\frac{1}{2}\right)} := \omega^{\left(\beth,\frac{1}{2}\right)}(\mathcal{N})$, as suggested in (13). The cdf and the pdf of $\Re^{\left(\mathcal{N},\beth,\frac{1}{2}\right)}$ do not have easily tractable analytical representation. Their graphs are given in Figures 13a and 13b.

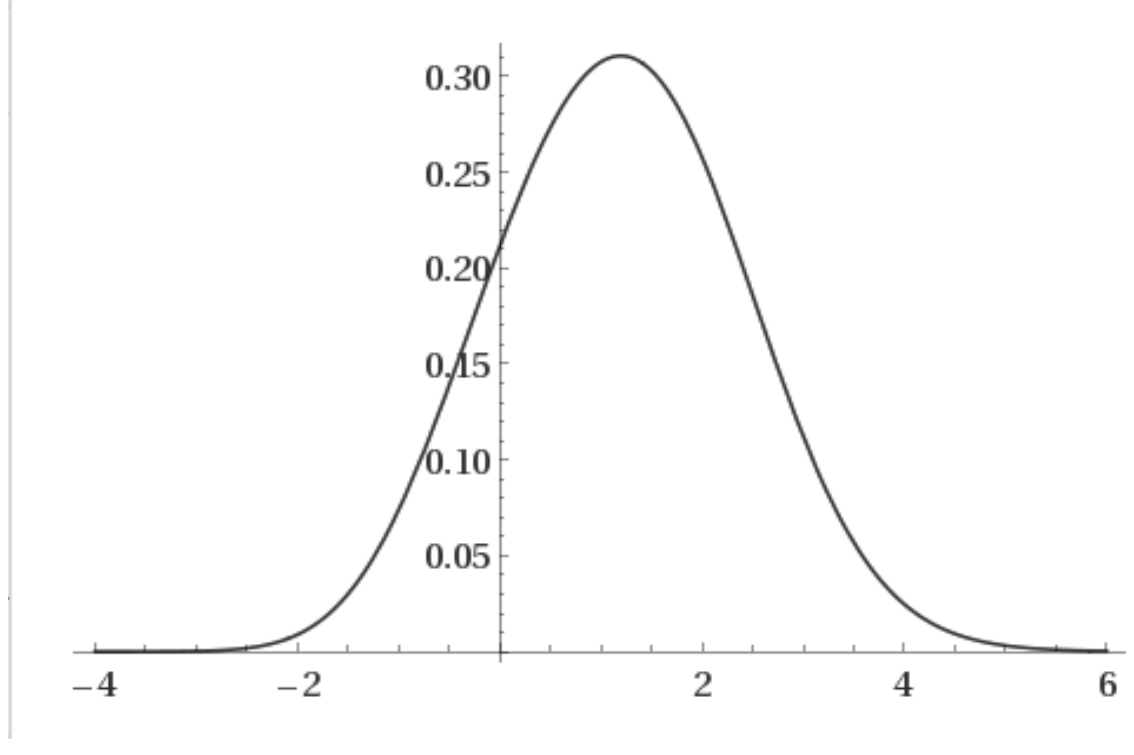

**Figure 13a: The pdf of $\Re^{\left(\mathcal{N},\beth,\frac{1}{2}\right)} := \omega^{\left(\beth,\frac{1}{2}\right)}(\mathcal{N})$.Plot of $f^{\left(\beth,\frac{1}{2}\right)}(x) = \frac{\partial F^{\left(\beth,\frac{1}{2}\right)}(x)}{\partial x} = \frac{3+x-\sqrt{1-x^2}}{\sqrt{2-2x^2}\left[\sqrt{x+1}+\sqrt{1-x}\right]^3}, -1 < x < 1$.**

[20] See the Blog of Zephur-Informa Investments Solutions, Informa Business Intelligence Inc., 2017. http://www.styleadvisor.com/resources/statfacts/information-ratio, 2017

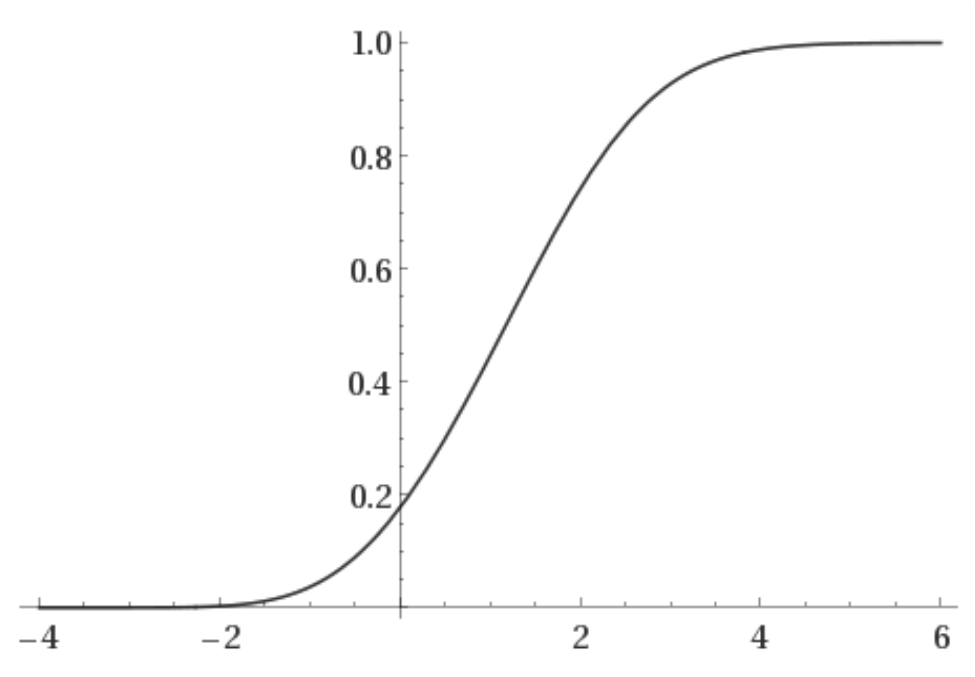

**Figure 13b: The cdf of $\boldsymbol{\mathfrak{R}^{\left(\mathcal{N},\text{ב},\frac{1}{2}\right)} := \omega^{\left(\text{ב},\frac{1}{2}\right)}(\mathcal{N})}$**

The mean of $\mathfrak{R}^{\left(\mathcal{N},\text{ב},\frac{1}{2}\right)}$ is again positive $\mathbb{E}\mathfrak{R}^{\left(\mathcal{N},\text{ב},\frac{1}{2}\right)} = 2.945167$. The variance is $var\mathfrak{R}^{\left(\mathcal{N},\text{ב},\frac{1}{2}\right)} = 4.12624$ and standard deviation $\sqrt{var\mathfrak{R}^{\left(\mathcal{N},\text{ב},\frac{1}{2}\right)}} = 2.031315$. The information ratio with benchmark return being zero, is

$$\mathbb{I}\left(\mathfrak{R}^{\left(\text{ב},\frac{1}{2}\right)}\right) = \frac{\mathbb{E}\mathfrak{R}^{\left(\mathcal{N},\text{ב},\frac{1}{2}\right)}}{var\mathfrak{R}^{\left(\mathcal{N},\text{ב},\frac{1}{2}\right)}} = 1.449882 > 1$$ [21]

In this case, ב becomes extremely bullish on this particular stock.

3.2 Prelec Probability Weighting Function and Option Pricing

Prelec (1998) [22] introduced the following WPF (designated as **PWPF**) :

---

[21] " Information ratios of 1.00 for long periods of time are rare. According to Informa Business Intelligence Inc., 2017.

[22] Prelec (1998) introduced also exponential power wpf, $w^{(\text{ב},\gamma,\eta)}(u) := \exp\left\{-\frac{\eta}{\gamma}(1-u^{\gamma})\right\}, 0 < u < 1, \gamma > 0, \eta > 0$, and hyperbolic-logarithmic $W^{(\text{ב},\gamma,\eta)}(u) = (1+\gamma lnp)^{-\frac{\eta}{\gamma}}, 0 < u < 1, \gamma > 0, \eta > 0$. Luce (2001) introduced the following WPT $\omega^{(\text{ב},\alpha,\beta)}(u) = \exp\left(-\beta\left(\frac{1-p}{p}\right)^{\alpha}\right), 0 < u <$

$$w^{(\text{ב},\delta,\rho)}(u) := \exp\{-(-\delta lnu)^{\rho}\}, u \in [0,1], 0 < \rho < 1, \delta > 0. \quad (14)$$

$(Pi)$ for $0 < \rho < 1$, and fixed $\delta > 0$, ב is fearful, see Figure 14a;

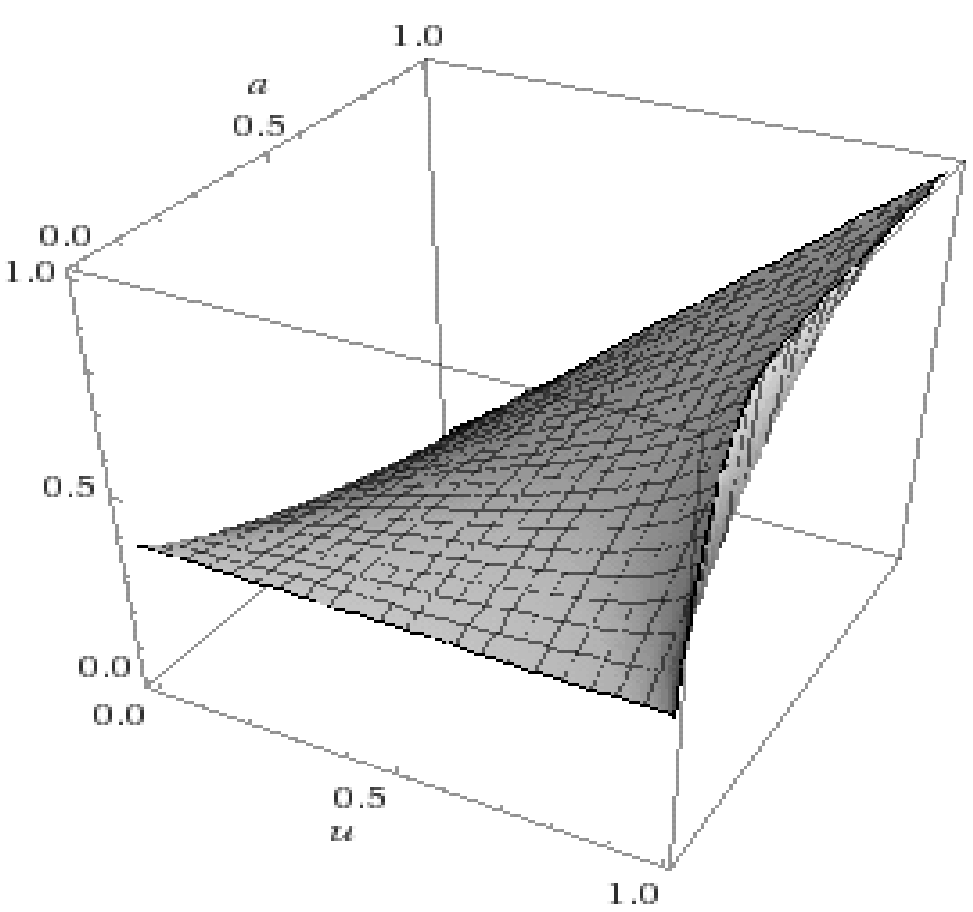

**Figure 14a.Plot of Prelec pwf with fixed scale parameter:**

$$w^{(\text{ב},1,\rho)}(u) := \exp\{-(-lnu)^{\rho}\}, u \in [0,1], 0 < \rho < 1.$$

$(Pii\ )$ for fixed $\rho$, $\delta$ controls the height of the wpf, see Figure 14b;

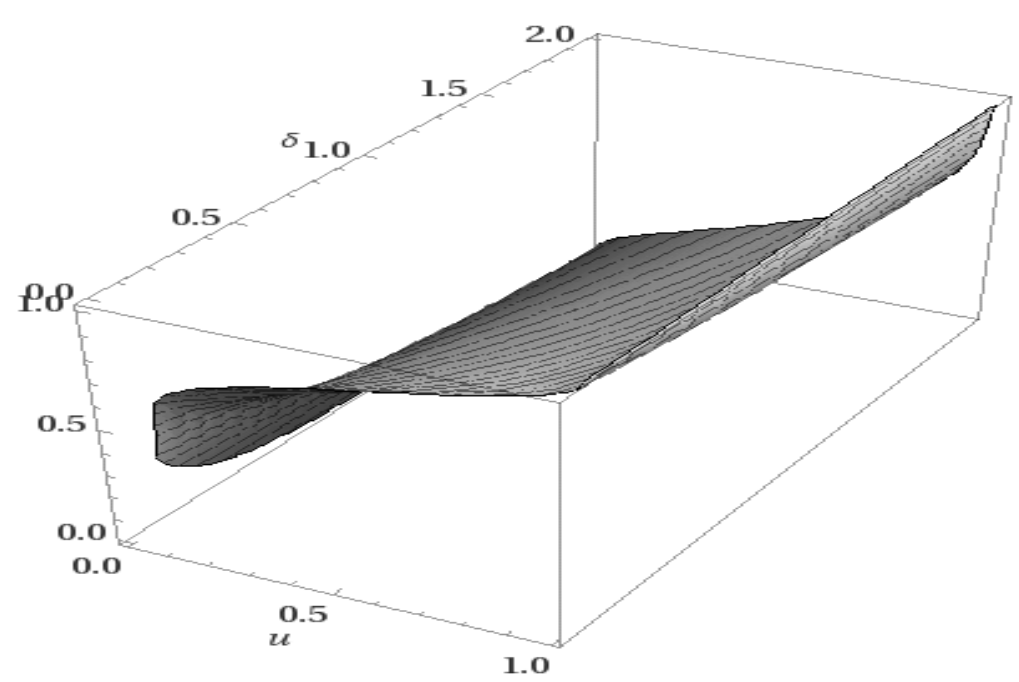

**Figure 14b.Plot of Prelec function with fixed shape parameter:** $\boldsymbol{w^{\left(\text{ב},\frac{1}{2},\delta\right)}(u) := \exp\left\{-\sqrt{-\delta lnu}\right\}, u \in [0,1], \delta \in [0,2]}$

---

$1, \alpha > 0, \beta > 0$. We were not able to extend the results in section 3.2 to those WPF, and it seems that this is not possible.

$(Piii)$ for $\rho > 1$, $\delta > 0$, $\beth$ becomes greedy, and the larger $\rho$ is the greedier $\beth$ become, see Figure 14c.

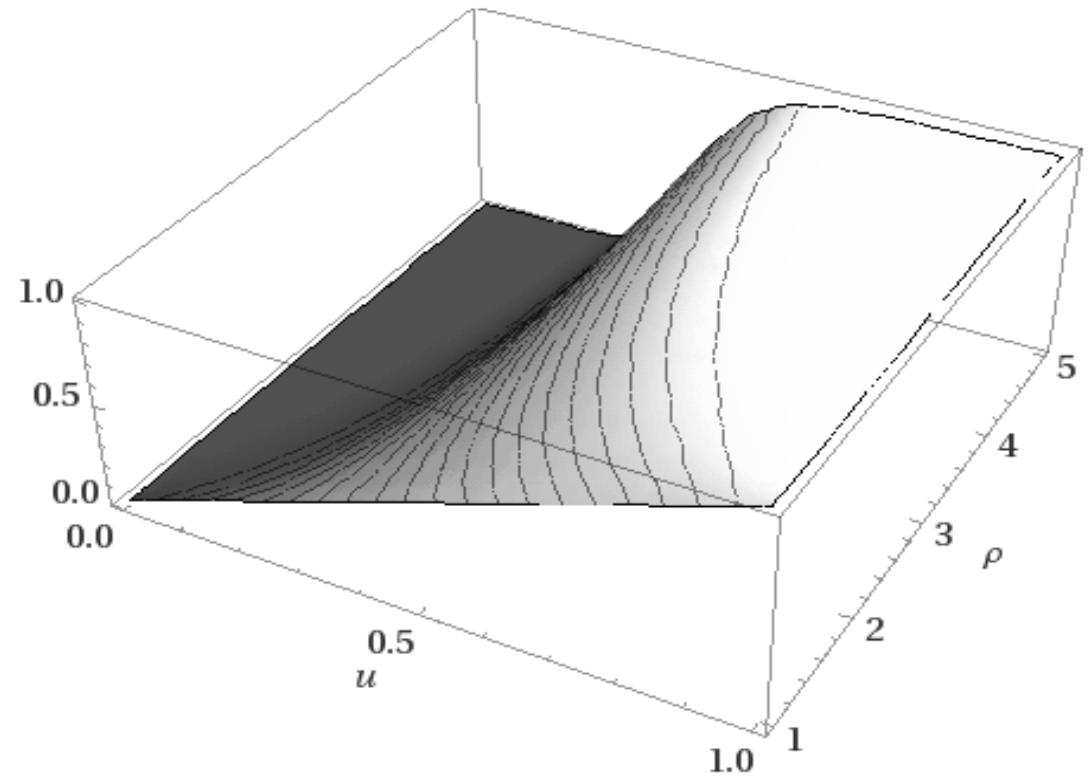

**Figure 14c: Plot of Prelec wpf for greedy investor $\boldsymbol{w^{(\beth,1,\rho)}(u) := \exp\{-(-ln u)^{\rho}\}, u \in [0,1], \rho \in [1,5]}$.**

To derive an option pricing formula similar to the one in the previous section we need to use the following minor modification of PWPF, designated as **MPWPF**, and having the representation:

$$W^{(\beth,\delta,\rho)}(u) = 1 - w^{(\beth,\delta,\rho)}(1-u) =$$

$$= 1 - \exp\{-(-\delta \ln(1-u))^{\rho}\}, u \in [0,1], \rho > 0, \delta > 0. \quad (15)$$

Similar to (14), suppose $\beth$ modifies her views on the asset (strictly increasing continuous) cdf $F^{(\Re)}(x), x \in \mathcal{R}$ by applying $W^{(\beth,\delta,\rho)}(\cdot)$ to obtain

$$F^{(\beth,\Re)}(x) := W^{(\beth,\delta,\rho)}\left(F^{(\Re)}(x)\right) \quad (16)$$

a posterior cdf $F^{(\beth,\Re)}(x)$. Then $P(i), P(ii)$, and $P(iii)$ hold, see Figures 15a,15b, and 15c.

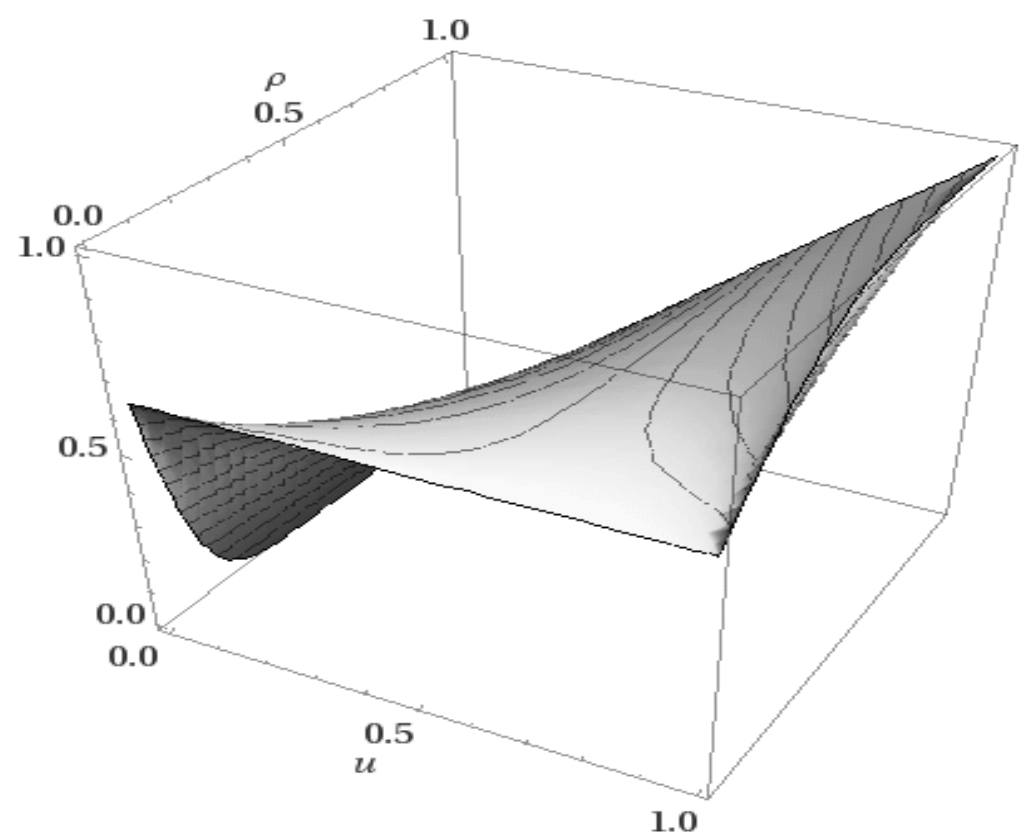

**Figure 15a:Plot of modified Prelec wpf with fixed scale parameter:** $W^{(\beth,1,\rho)}(u) = 1 - \exp\{-(-\ln(1-u))^{\rho}\}, u \in [0,1], \rho \in (0,1)$.

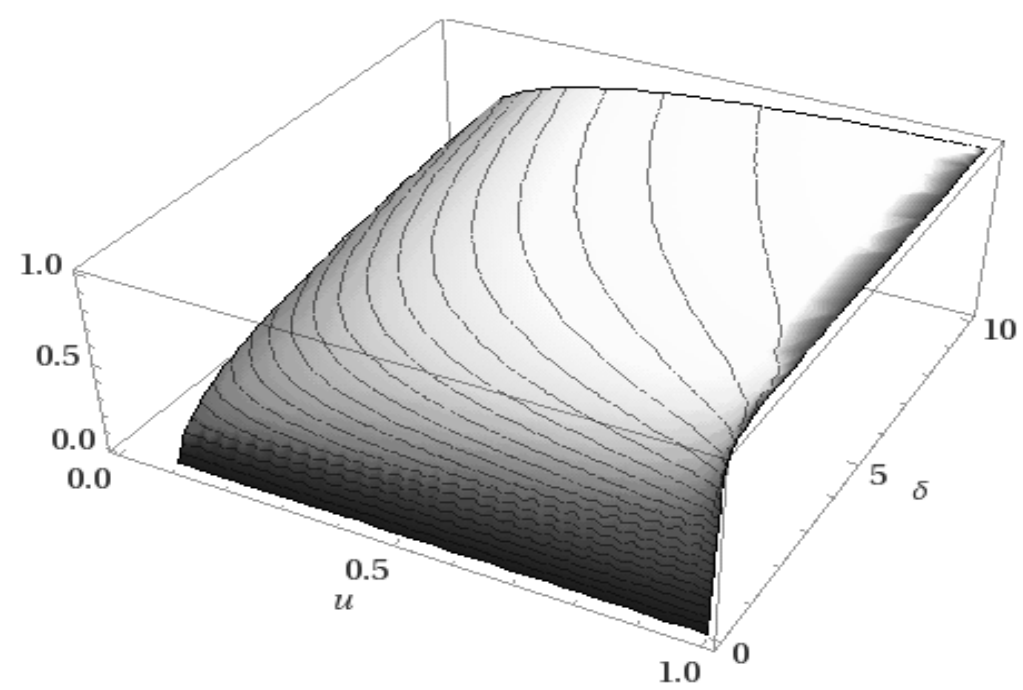

**Figure 15b:Plot of the modified Prelec wpf with fixed shape parameter:** $W^{\left(\beth,\delta,\frac{1}{2}\right)}(u) = 1 - \exp\left\{-\sqrt{-\delta\ln(1-u)}\right\}, u \in [0,1], \delta \in [0,10]$.

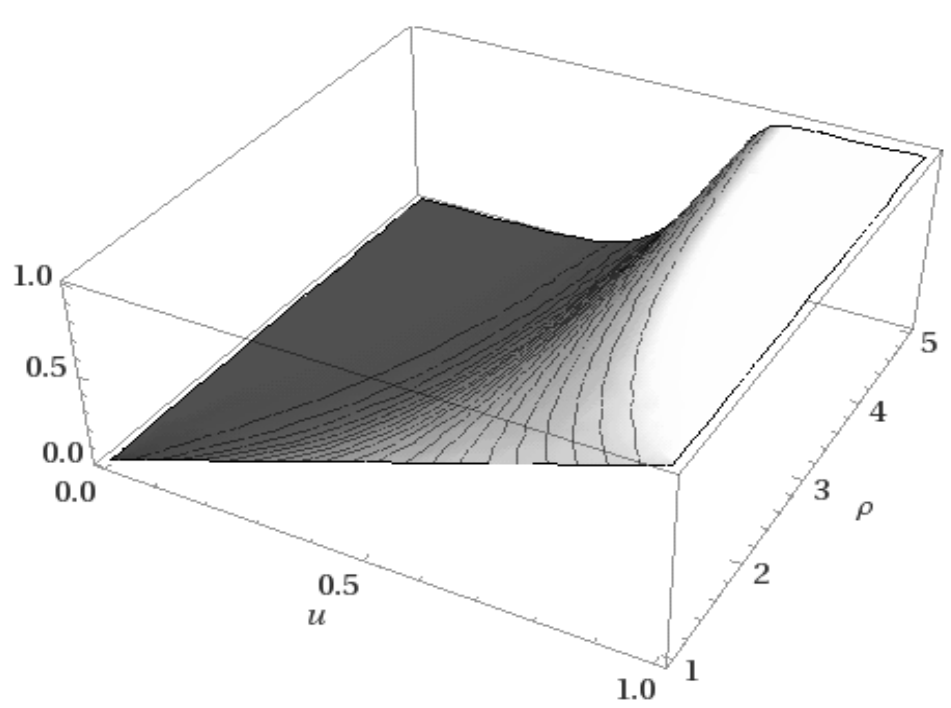

**Figure 15c:Plot of modified Prelec wpf for greedy investor:** $W^{(\beth,1,\rho)}(u) = 1 - \exp\{-(-\ln(1-u))^{\rho}\}, u \in [0,1], \rho \in (1,5]$.

Note however that while the shapes of PWPD and MPWPD are similar they are quite different, see Figure 15d.

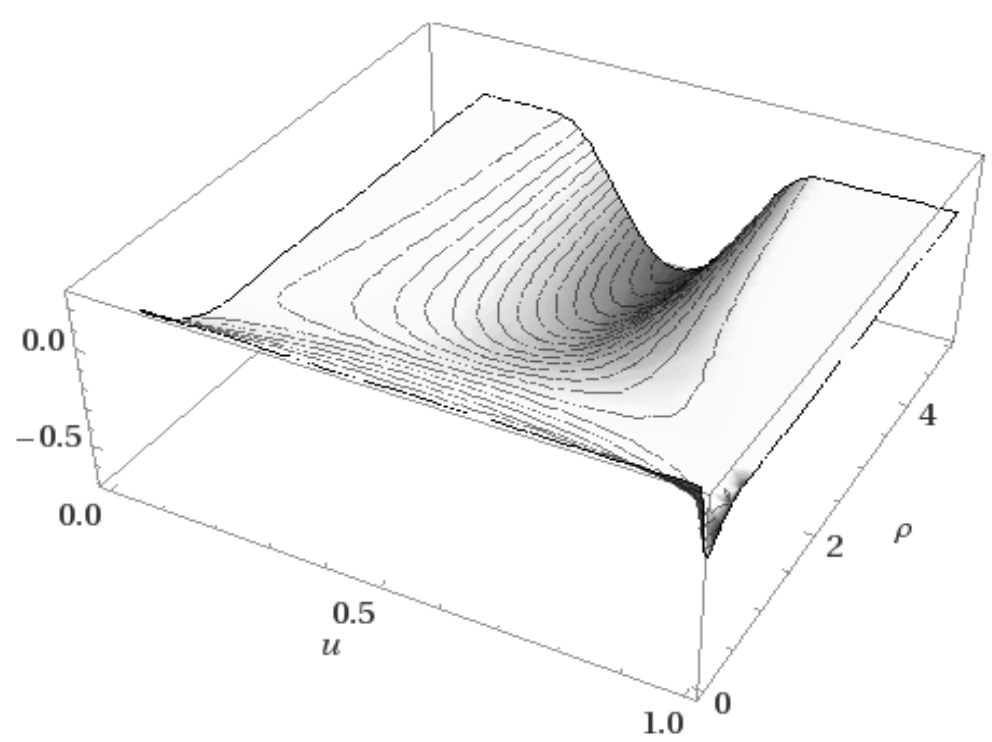

**Figure 15d: Plot of the difference between of modified Prelec wpf and Preelec wpf.:** $\boldsymbol{W^{(\beth,1,\rho)}(u) - w^{(\beth,\delta,\rho)}(u) = 1 - \exp\{-(-\ln(1-u))^{\rho}\} - \exp\{-(-\delta lnu)^{\rho}\}, u \in [0,1], \rho \in (0,5]}$.

Let $\mathcal{U} \triangleq U[0,1]$ be a uniformly distributed on $[0,1]$ rv. Let $F^{(\text{inv},\mathfrak{R})}(u) := \sup\{x : F^{(\mathfrak{R})}(x) \leq u\}, u \in [0,1]$ be the inverse of the cdf $F^{(\mathfrak{R})}(x), x \in \mathcal{R}$. Then $F^{(\text{inv},\mathfrak{R})}(\mathcal{U}) \triangleq \mathfrak{R}$. Set $\mathfrak{R}^{(\beth)} := F^{(\text{inv},\beth,\mathfrak{R})}(\mathcal{U})$, where $F^{(\text{inv},\beth,\mathfrak{R})}(u), u \in [0,1]$ is the inverse of $F^{(\beth,\mathfrak{R})}(x), x \in \mathcal{R}$. Then $F^{(\beth,\mathfrak{R})}$ is the cdf for $\mathfrak{R}^{(\beth)}$. Denote by $W^{(\text{inv}\beth)}(u), u \in [0,1]$, the inverse function of $W^{(\beth)}(u), u \in [0,1]$. Thus,

$$\mathfrak{R}^{(\beth)} \triangleq F^{(\text{inv},\mathfrak{R})}\left(W^{(\text{inv}\beth,\delta,\rho)}\left(F^{(\mathfrak{R})}(\mathfrak{R})\right)\right) \tag{17}$$

From (15), (16), and (17), it follows that a natural choice for $F^{(\Re)}(x), x \in \mathcal{R}$, is the negative Gumbel distribution, $\Re \triangleq NG(\mu, \varrho), \mu \in \mathcal{R},\ \varrho > 0,$ see Figures 5a and 5b, with $F^{(\Re)}(x) = F^{(NG(\mu,\rho))}(x) = 1 - \exp\left(-e^{\frac{x-\mu}{\varrho}}\right), x \in \mathcal{R}.$[23] With $\Re \triangleq NG(\mu, \varrho)$, it follows from (17), that

$\Re^{(\beth)} \triangleq NG\left(\mu^{(\beth)}, \varrho^{(\beth)}\right)$, where $\mu^{(\beth)} = \mu - \varrho \ln \delta$, and $\varrho^{(\beth)} = \frac{\varrho}{\rho}$. Thus remarkably, the MPWPF keeps the prior cdf $F^{(\Re)}$ and posterior cdf $F^{(\beth,\Re)}$ the same negative Gumbel distributional class.

The relation between the prior information ratio $\mathbb{I}(\Re) = \frac{\mu - \varrho\gamma^{(E-M)}}{\frac{\pi}{\sqrt{6}}\varrho}$ and posterior information ratio $\mathbb{I}\left(\Re^{(\beth)}\right) = \frac{\mu - 2\varrho\gamma^{(E-M)}}{\frac{2\pi}{\sqrt{6}}\varrho}$ is now totally dependent of the MPWPF parameter $\rho > 0$, prior mean return $\mathbb{E}\Re = \mu - \varrho\gamma^{(E-M)}$and posterior mean return $\mathbb{E}\Re^{(\beth)}$,

$$\mathbb{I}\left(\Re^{(\beth)}\right) - \mathbb{I}(\Re) = \frac{\sqrt{6}}{\pi}\left(\frac{\rho - 1}{\varrho}\mu - \rho \ln \delta\right) \qquad (18)$$

In contrast to the case with WPF given by (13), the relation given by (18) provides a flexible reasonable structure of the posterior information ratio in comparison with the prior information ratio.

Now our goal is to see whether the option market can give us some clues about the overall "market value" of the "greed-fear" parameter $\rho > 0$ in the MPWPF given by (15). Consider then

---

[23] This is because the negative Gumbel distribution is infinitely divisible, skewed to the left and heavy tailed and Gumbel Lévy process have been used in finance literature to model asset returns, see Bell (2006) and Markose and Alentorn (2011). As shown in Leadbetter, Lindgren,. and Rootzén, (1983*)*, Theorem 1.5.3, the asymptotic distribution of (properly scaled and shifted) minimum of iid standard normal distribution is a negative Gumbel distribution.

the negative-Gumbel Lévy market consisting of a risky asset $\mathcal{S}^{(\beth)}$ and a riskless asset (bond) $\mathcal{B}$.[24] The price dynamics of $\mathcal{S}^{(\beth)}$ are given by $S(t)^{(\beth)} = S(0)^{(\beth)}\mathrm{e}^{\mathbb{NG}(\mathrm{t})}, t \geq 0, S(0)^{(\beth)} > 0$, where $\mathbb{NG}(\mathrm{t}), t \geq 0$, is a negative Gumbel-Lévy process; that is, $\mathbb{NG}(\mathrm{t}), \mathrm{t} \geq 0$, is a Lévy process with unit increment $\mathbb{NG}(1) \triangleq NG\left(\mu^{(\beth)}, \varrho^{(\beth)}\right)$. By Sato's Theorem[25] on equivalent martingale measure $\mathbb{Q} \sim \mathbb{P}$ for Lévy processes, the risk-neutral dynamics of $S(t)^{(\beth)}$is given by $S(t)^{(\beth,\mathbb{Q})} = S(0)^{(\beth)} \exp\left\{\mathbb{NG}^{(\mathbb{Q})}(\mathrm{t})\right\}, t \geq 0$, where $\mathbb{NG}^{(\mathbb{Q})}(\mathrm{t}), t \geq 0$ is again a negative Gumbel-Lévy process, but with $(i)$ $\mathbb{NG}^{(\mathbb{Q})}(1) \triangleq NG\left(\mu^{(\beth,\mathbb{Q})}, \varrho^{(\beth)}\right)$, and $(ii)$ $\mu^{(\beth,\mathbb{Q})}$ satisfies the *martingale condition*:

$r = \ln\left(\mathbb{E}\left(e^{\mathbb{NG}^{(\mathbb{Q})}(1)}\right)\right) = \ln\left(e^{\mu^{(\beth,\mathbb{Q})}}\Gamma\left(1+\varrho^{(\beth)}\right)\right)$, where $r > 0$ is the riskless rate.[26] Thus, $\mu^{(\beth,\mathbb{Q})} = r - \ln\left(\Gamma\left(1+\varrho^{(\beth)}\right)\right)$. The characteristic function $\varphi^{\left(\mathbb{NG}^{(\mathbb{Q})}(\mathrm{t})\right)}(\theta), \theta \in \mathcal{R}$ of the $\mathbb{NG}^{(\mathbb{Q})}(\mathrm{t}), t \geq 0$, is given by $\varphi^{\left(\mathbb{NG}^{(\mathbb{Q})}(\mathrm{t})\right)}(\theta) = \mathbb{E}e^{\mathbb{NG}^{(\mathbb{Q})}(\mathrm{t})} = e^{i\theta\mu^{(\beth,\mathbb{Q})}t}\left(\Gamma\left(1+i\varrho^{(\beth)}\theta\right)\right)^{t}, \theta \in \mathcal{R}, t \geq 0$. Hence, the pdf $f^{\left(\mathbb{NG}^{(\mathbb{Q})}(\mathrm{t})\right)}(x), x \in \mathcal{R}$, of the $\mathbb{NG}^{(\mathbb{Q})}(\mathrm{t})$ is given by the inversion formula: $f^{\left(\mathbb{NG}^{(\mathbb{Q})}(\mathrm{t})\right)}(x) = \frac{1}{2\pi}\int_{-\infty}^{\infty} e^{-i\theta x}\,\varphi^{\left(\mathbb{NG}^{(\mathbb{Q})}(\mathrm{t})\right)}(\theta)d\theta$.

Consider a European contingent claim (ECC), $\mathcal{C}$, with price process $C(t), t \in [0,T]$ and final payoff is $C(T) = g\left(S(T)^{(\beth)}\right)$. Then the value of $\mathcal{C}$ at $t = 0$

$$C(0) = e^{-rT}\mathbb{E}g\left(S(0)^{(\beth)}e^{\mathbb{NG}^{(\mathbb{Q})}(\mathrm{t})}\right) = e^{-rT}\int_{-\infty}^{\infty} g\left(S(0)^{(\beth)}e^{x}\right) f^{\left(\mathbb{NG}^{(\mathbb{Q})}(\mathrm{t})\right)}(x)dx. \quad (19)$$

---

[24] See also Schoutens (2003, Section 6.2), Carr and Wu (2004), Bell (2006), and Tankov (2011).

[25] Sato (1999, p. 218), Theorem 33.1. See also Kyprianou, Schoutens, and Wilmott ((2005) Section 8.5.1, and Bell (2006).

[26] We assume that the price dynamics of the bond $\mathcal{B}$ is given by $e^{rt}, t \geq 0$.

To evaluate the fear-greedy profile of the financial market based on MPWPF one needs to calibrate the option pricing formula (19) to market option prices.

3.3 General Form of WPF Consistent with The Rational Dynamic Asset Pricing Theory

We start with the following observation regarding the nature of Prelec's WPF given by (14). Suppose $\beth's$ prior return distribution is a Gumbel distribution: $\Re \triangleq G(\mu,\varrho), \mu \in \mathcal{R}, \varrho > 0, F^{(G(\mu,\rho))}(x) = \exp\left(-e^{-\frac{x-\mu}{\varrho}}\right), x \in \mathcal{R}$, and $\beth's$ goal is to find WPF $w^{(\beth,\mu,\varrho,\mu^{(\beth)},\varrho^{(\beth)})}(u), u \in (0,1)$ such that the posterior return distribution $\Re^{(\beth)} \triangleq G\left(\mu^{(\beth)},\varrho^{(\beth)}\right), \mu^{(\beth)} \in \mathcal{R}, \varrho^{(\beth)} > 0$, with cdf $F^{(\Re^{(\beth)})}(x) = w^{(\beth,\mu,\varrho,\mu^{(\beth)},\varrho^{(\beth)})} \circ F^{(\Re)}(x)$. Then $w^{(\beth,\mu,\varrho,\mu^{(\beth)},\varrho^{(\beth)})}(u) = \exp\left(-\left(-a^{(\beth)}\ln(u)\right)^{b^{(\beth)}}\right)$, where $a^{(\beth)} = e^{\frac{\mu-\mu^{(\beth)}}{\rho}}$ >0 and $b^{(\beth)} = \frac{\rho}{\rho^{(\beth)}} > 0$, which is Pwpf $w^{(\beth,a^{(\beth)},b^{(\beth)})}(u), u \in (0,1)$, see (14).

Similarly, if the distribution is $\Re \triangleq NG(\mu,\varrho), \mu \in \mathcal{R}, \varrho > 0, F^{(NG(\mu,\rho))}(x) = 1 - \exp\left(-e^{\frac{x-\mu}{\varrho}}\right), x \in \mathcal{R}$, and $\beth's$ goal is to find WPF $W^{(\beth,\mu,\varrho,\mu^{(\beth)},\varrho^{(\beth)})}(u), u \in (0,1)$ such that posterior return distribution $\Re^{(\beth)} \triangleq NG\left(\mu^{(\beth)},\varrho^{(\beth)}\right), \mu^{(\beth)} \in \mathcal{R}, \varrho^{(\beth)} > 0$, with cdf $F^{(\Re^{(\beth)})}(x) = W^{(\beth,\mu,\varrho,\mu^{(\beth)},\varrho^{(\beth)})} \circ F^{(\Re)}(x)$. Then $W^{(\beth,\mu,\varrho,\mu^{(\beth)},\varrho^{(\beth)})}(u) = 1 - \exp\left(-\left(-a^{(\beth)}\ln(1-u)\right)^{b^{(\beth)}}\right)$ which is the MPWPF given by (15).

These observations lead to the following general form of WPFs consistent with the rational asset pricing theory. Let $\Re$ and $\Re^{(\beth)}$ be infinitely divisible random variables, with cdfs $F^{(\Re)}(x), x \in \mathcal{R}$ and $F^{(\Re^{(\beth)})}(x), x \in \mathcal{R}$, and let the mgf $\mathcal{M}^{(\Re^{(\beth)})}(s) = \mathbb{E}e^{s\Re^{(\beth)}} < \infty, 0 < s < s^{(0)}$. Define the *general form of WPF* , $w^{(\beth,\Re,\Re^{(\beth)})}(u), u \in (0,1)$, *consistent with RDAPT* as the solution to the following equation:

$$F^{(\Re^{(\beth)})}\ (x) = w^{(\beth,\Re,\Re^{(\beth)})} \circ F^{(\Re)}(x), x \in \mathcal{R}.$$

Next, we provide some illustrative examples.

**Example 3. (Logistic WPF)** Let $\Re \triangleq Logistic(\mu, \rho), F^{(\Re)}(x) = \frac{1}{1+\exp\left(-\frac{x-\mu}{\rho}\right)}, x \in \mathcal{R}, \mu \in \mathcal{R},$

$\rho > 0,$ and $\Re^{(\beth)} \triangleq Logistic\left(\mu^{(\beth)}, \varrho^{(\beth)}\right)$[27]. Then $w^{(\beth,\Re,\Re^{(\beth)})} = \frac{1}{1+c^{(\beth)}\left(\frac{1}{u}-1\right)^{b^{(\beth)}}}, u \in (0,1), c^{(\beth)} =$

$\mathrm{e}^{-\frac{\mu-\mu^{(\beth)}}{\varrho^{(\beth)}}} > 0, b^{(\beth)} = \frac{\rho}{\varrho^{(\beth)}} > 0.$ For $b^{(\beth)} \in (0,1)$, $w^{(\beth,\Re,\Re^{(\beth)})}$ has a concave-convex shape ($\beth$ "fearful" ), while for $b^{(\beth)} > 0, w^{(\beth,\Re,\Re^{(\beth)})}$ has a convex-convex shape ($\beth$ "greedy" ). Parameter $c^{(\beth)} > 0$ controls the height of the $w^{(\beth,\Re,\Re^{(\beth)})}$, see Figures 16a and 16b.

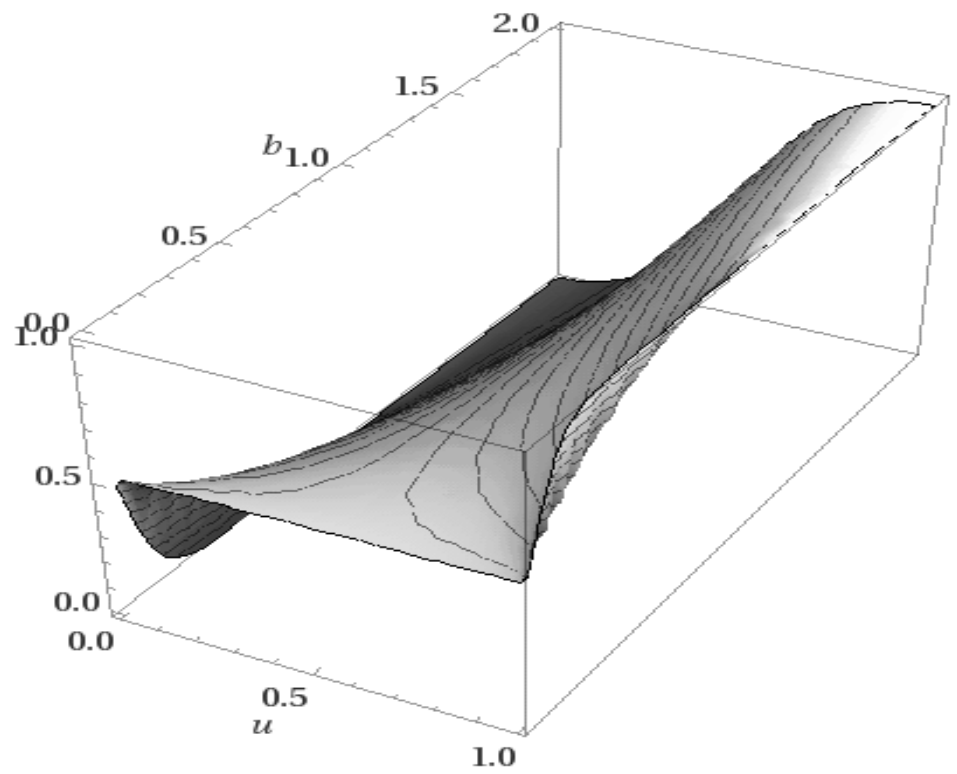

**Figure16a. Plot of Logistic-Logistic wpf**

$$\boldsymbol{w^{(\beth,\Re,\Re^{(\beth)})} = \frac{1}{1+\left(\frac{1}{u}-1\right)^{b^{(\beth)}}}, u \in (0,1), 0.01 < b^{(\beth)} < 2}$$

[27] Logistic distribution is infinitely divisible, see Steutel and van Harn (2004), Appendix B2.

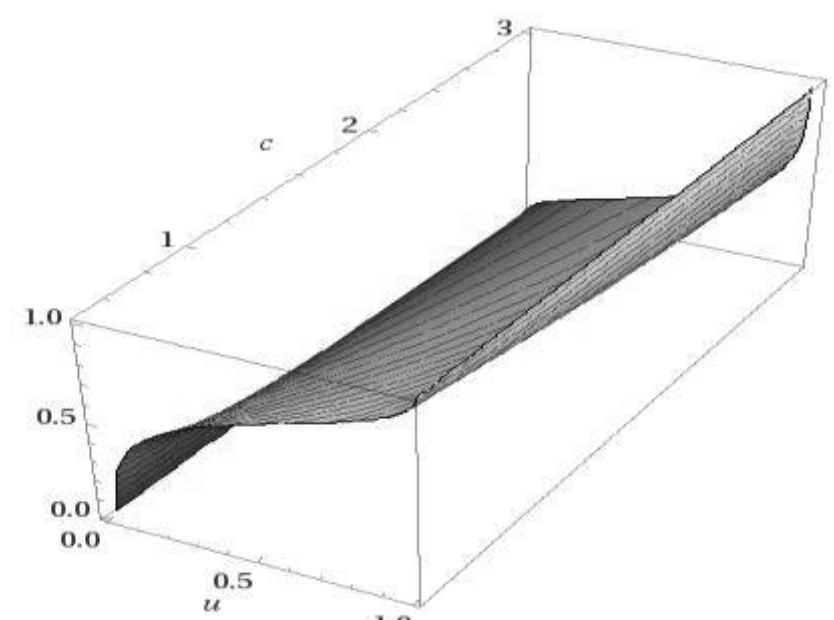

**Figure16b. Plot of Logistic-Logistic wpf**$\boldsymbol{w^{(\beth,\Re,\Re^{(\beth)})} = \frac{1}{1+c^{(\beth)}\sqrt{\left(\frac{1}{u}-1\right)}}, u \in (0,1), 0.5 < c^{(\beth)} < 3}$

**Example 4. (Gumbel-Logistic WPF)** Let $\Re \triangleq Logistic(\mu,\rho), \mu \in \mathcal{R}, \rho > 0$ and $\Re^{(\beth)} \triangleq G\left(\mu^{(\beth)},\varrho^{(\beth)}\right), \mu^{(\beth)} \in \mathcal{R}, \varrho^{(\beth)} > 0.$ Then $w^{(\beth,\Re,\Re^{(\beth)})}(u) = \exp\left(-c^{(\beth)}\left(\frac{1}{u}-1\right)^{b^{(\beth)}}\right), 0 < u < 1,$ see Figures 17a and 17b. This is indeed a version of PWPF (14), where $w^{(\beth,\delta,\rho)}(y) := \exp\{-(-\delta ln y)^{\rho}\}, y \in [0,1],$ with $y = \frac{1}{u} - 1$ .

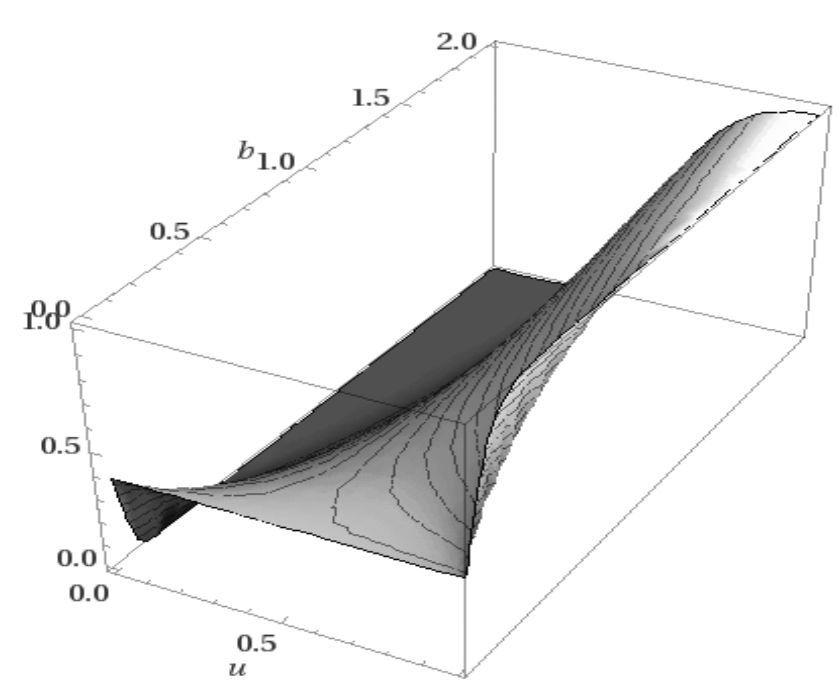

**Figure 17a. Plot of Gumbel-Logistic wpf** $\boldsymbol{w^{(\beth,\Re,\Re^{(\beth)})} = \exp\left(-\left(\frac{1}{u}-1\right)^{b^{(\beth)}}\right), 0 < u < 1, 0 < b < 2}$

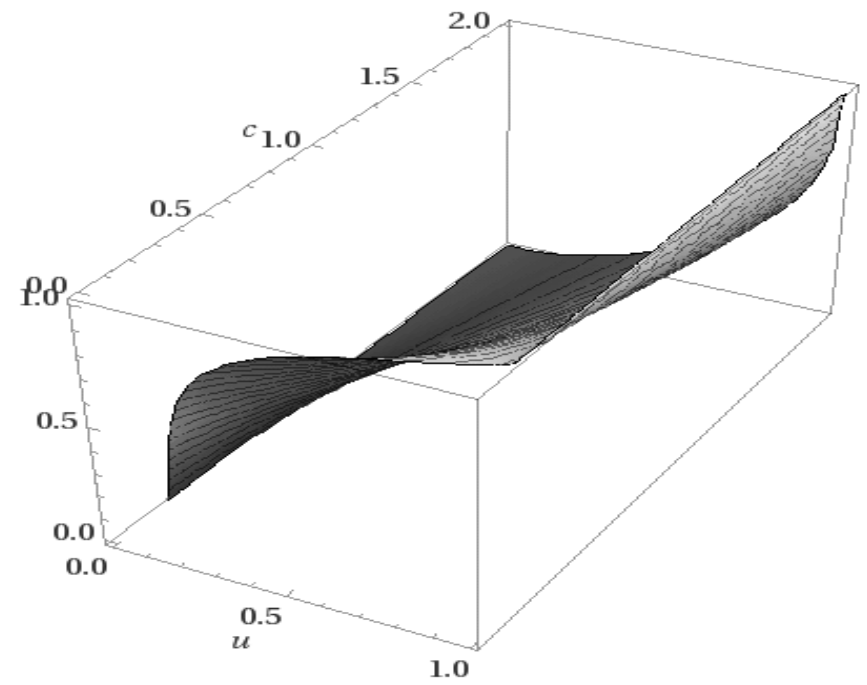

**Figure 17b. Plot of Gumbel-Logistic wpf** $\boldsymbol{w^{(\text{ב},\Re,\Re^{(\text{ב})})} = \exp\left(-c^{(\text{ב})}\sqrt{\frac{1}{u}-1}\right), 0 < u < 1, 0.2 < b < 2}$

**Example 5. (Double Pareto WPF)** Let

$\Re \triangleq DPareto(\rho), \rho > 1$, and $\Re^{(\text{ב})} \triangleq DPareto\left(\varrho^{(\text{ב})}\right), \varrho^{(\text{ב})} > 1$.

Then

$$w^{(\text{ב},\Re,\Re^{(\text{ב})})}(u) = \begin{cases} \frac{1}{2}(2u)^{\gamma^{(\text{ב})}}, 0 < u < \frac{1}{2} \\ 1 - \frac{1}{2}\left(2(1-u)\right)^{\gamma^{(\text{ב})}}, \frac{1}{2} < u < 1 \end{cases}, \gamma^{(\text{ב})} = \frac{1-\varrho^{(\text{ב})}}{1-\rho}.$$

For $\gamma^{(\text{ב})} \in (0,1)$, ב “fearful” while $\gamma^{(\text{ב})} > 1$, ב is “greedy”, see Figure 18.

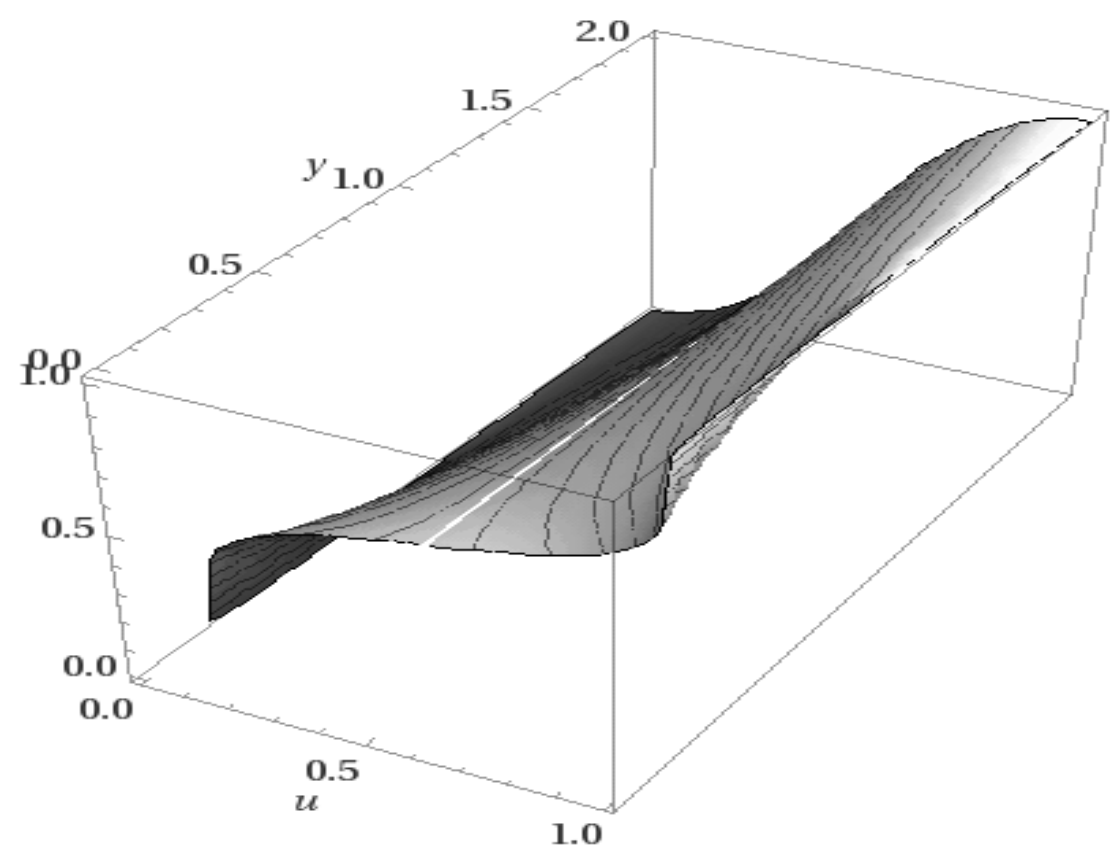

**Figure 18. Plot of Double Pareto wpf**

$$\mathbf{w}^{(\text{ב},\mathfrak{R},\mathfrak{R}^{(\text{ב})})}(\mathbf{u}) = \begin{cases} \frac{1}{2}(2\mathbf{u})^{\gamma^{(\text{ב})}}, \mathbf{0} < \mathbf{u} < \frac{1}{2} \\ \mathbf{1} - \frac{1}{2}\left(\mathbf{2}(\mathbf{1}-\mathbf{u})\right)^{\gamma^{(\text{ב})}}, \frac{1}{2} < \mathbf{u} < \mathbf{1} \end{cases}, \boldsymbol{\gamma}^{(\text{ב})} \in (\mathbf{0}, \mathbf{2})$$

**Example 6. (Double Pareto-Laplace WPF).** Let $\mathfrak{R} \triangleq DPareto(\rho)\ \rho > 1$ and $\mathfrak{R}^{(\text{ב})} \triangleq Laplace\left(b^{(\text{ב})}\right), b^{(\text{ב})} > 0$. Then

$$w^{(\text{ב},\mathfrak{R},\mathfrak{R}^{(\text{ב})})}(u) = \begin{cases} \dfrac{1}{2} e^{\frac{1-(2u)^{\frac{1}{1-\rho}}}{b^{(\text{ב})}}}, 0 < u < \dfrac{1}{2} \\ 1 - \dfrac{1}{2} e^{\frac{1-(2-2u)^{\frac{1}{1-\rho}}}{b^{(\text{ב})}}}, \dfrac{1}{2} < u < 1, \quad \rho > 1, b > 0 \end{cases}$$

Applying **Double Pareto-Laplace WPF, ב** passes from prior heavy-tailed distribution with power tails to a posterior distribution with thin (exponential) tails, see Figures 19 a,19b,19c and 19d.

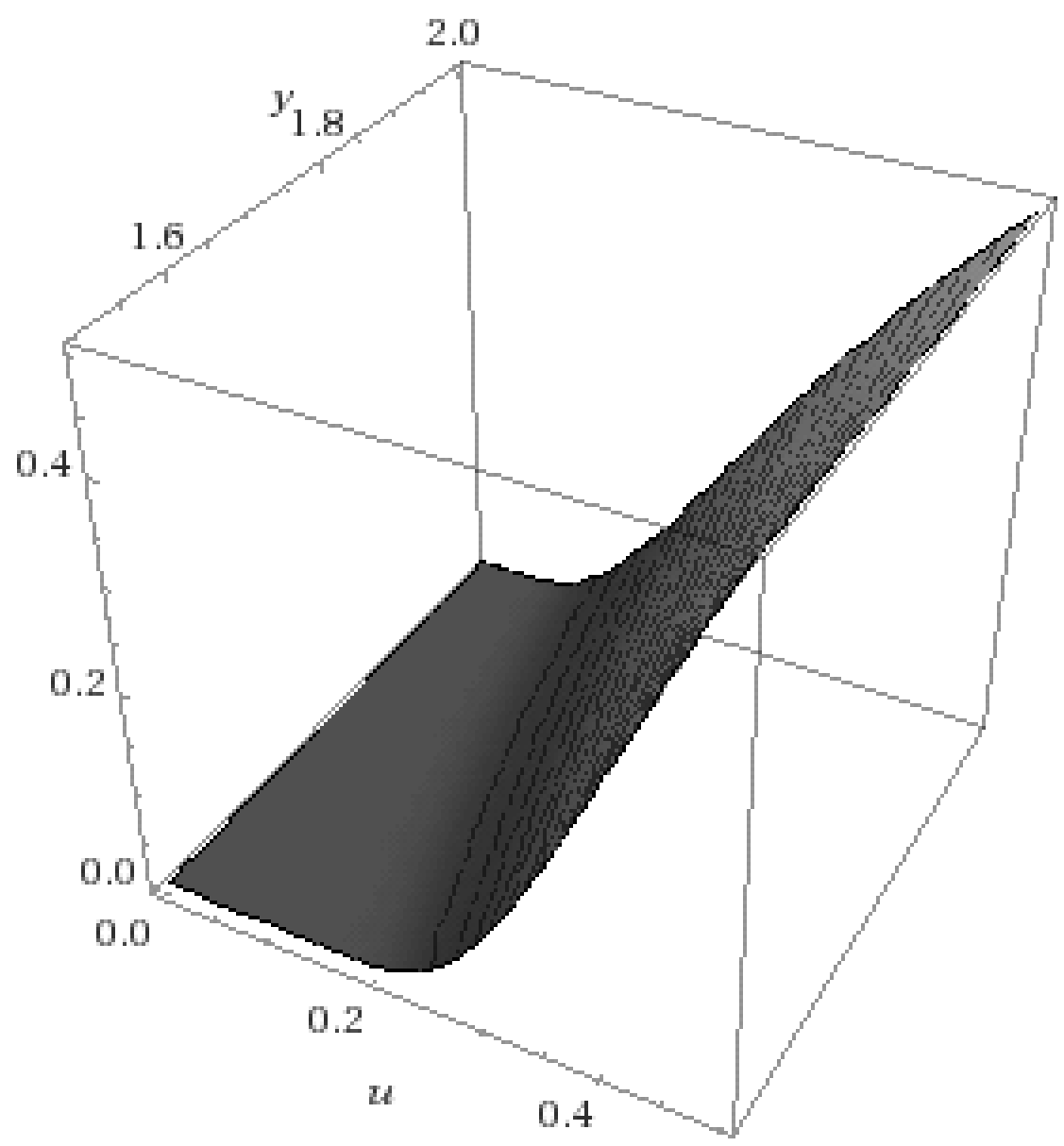

**Figure 19a. Plot of Double Pareto-Laplace WPF $w^{\left(\beth,\mathfrak{R},\mathfrak{R}^{(\beth)}\right)}(u) = \frac{1}{2}e^{\frac{1-(2u)^{\frac{1}{1-\rho}}}{b^{(\beth)}}}, 0 < u < \frac{1}{2}, 1.5 < \rho < 2, b = 1.$**

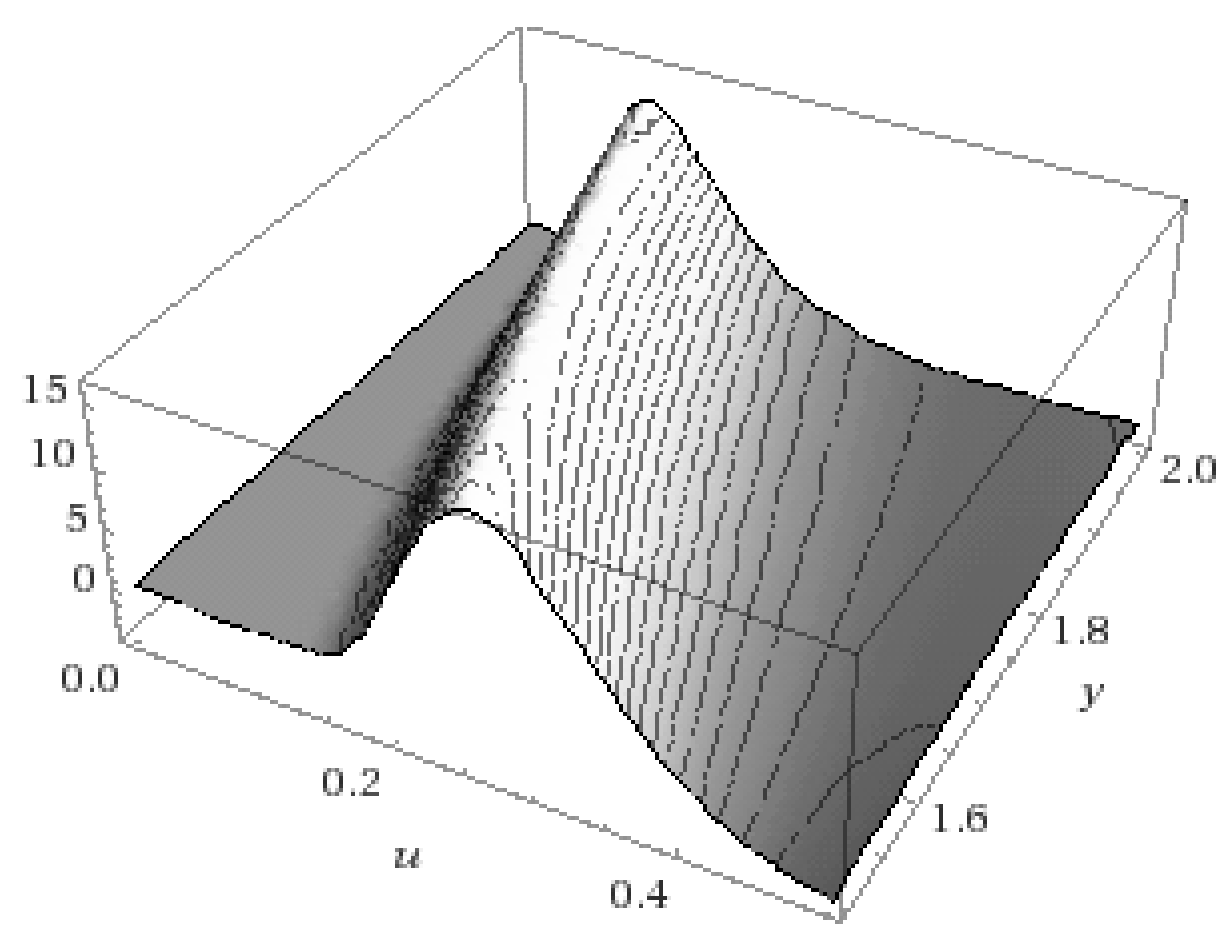

**Figure 19b. Plot of Double Pareto-Laplace WPF second derivative $\frac{\partial^2 w^{\left(\beth,\mathfrak{R},\mathfrak{R}^{(\beth)}\right)}(u)}{\partial u^2} = (\rho - 1)^{-2} 2^{\frac{\rho}{1-\rho}} u^{\frac{1}{1-\rho}-2}\left((2u)^{\frac{1}{1-\rho}} - \rho\right) exp\left(1 - (2u)^{\frac{1}{1-\rho}}\right) > 0$ showing the convexity of $w^{\left(\beth,\mathfrak{R},\mathfrak{R}^{(\beth)}\right)}(u), 0 < u < \frac{1}{2} < \rho < 2, b = 1$**

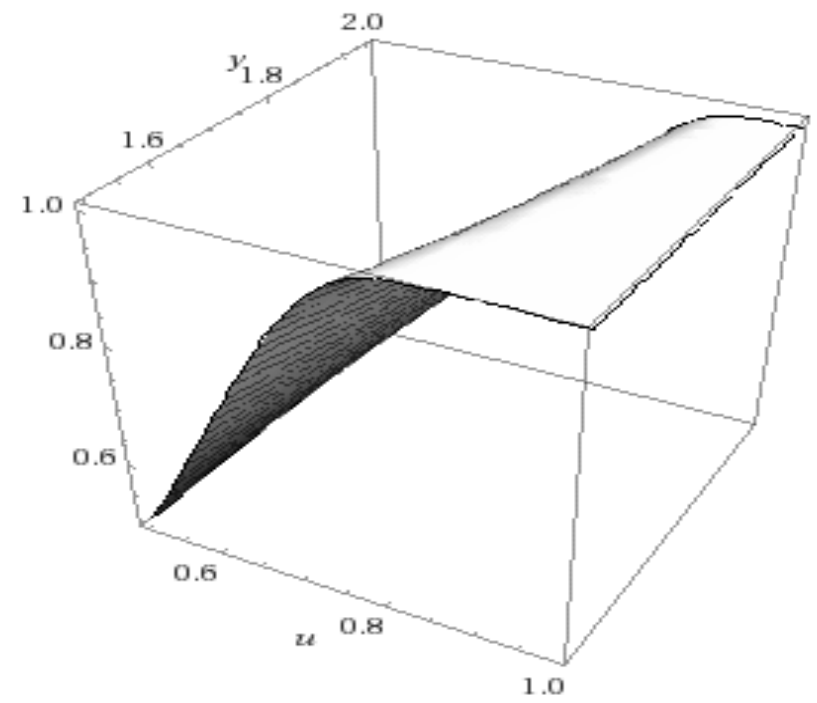

**Figure 19c. Plot of Double Pareto-Laplace WPF $w^{(\beth,\Re,\Re^{(\beth)})}(u) = 1 - \frac{1}{2}e^{\frac{1-(2-2u)^{\frac{1}{1-\rho}}}{b^{(\beth)}}}, \frac{1}{2} \le u < 1, 1.5 < \rho < 2, b = 1$**

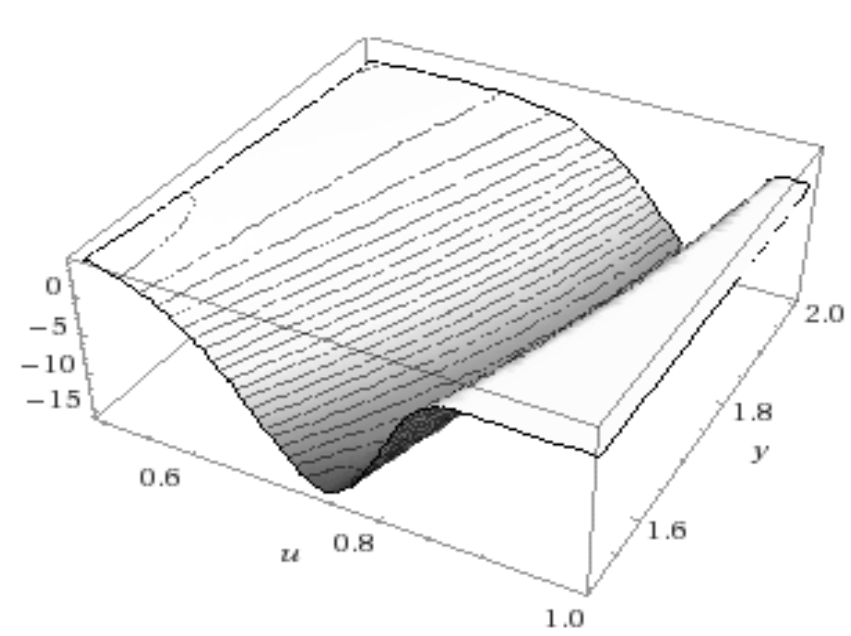

**Figure 19d. Plot of Double Pareto-Laplace WPF second derivative $\frac{\partial^2 w^{(\beth,\Re,\Re^{(\beth)})}(u)}{\partial u^2} = -(\rho-1)^{-2}2^{-\frac{\rho}{1-\rho}}(1-u)^{\frac{1}{1-\rho}-2}\left((2-2u)^{\frac{1}{1-\rho}} - \rho\right) exp\left(1-(2-2u)^{\frac{1}{1-\rho}}\right) < 0$ showing the concavity of $w^{(\beth,\Re,\Re^{(\beth)})}(u), \frac{1}{2} < u < 1, 1.5 < \rho < 2, b = 1$.**

**Example 7. (Cauchy WPF).** Let $\Re \triangleq Cauchy(c), c > 0$ and $\Re^{(\beth)} \triangleq Cauchy\left(c^{(\beth)}\right)$, The pdf $f^{(\Re)}(x) = f_{Cauchy(c)}(x), x \in \mathcal{R}$, and the cdf cdf $F^{(\Re)}(x) = F_{Cauchy(c)}(x), x \in \mathcal{R}$ and are given by

$$f_{Cauchy(c)}(x) = \frac{1}{\pi}\frac{c}{c^2 + x^2}, x \in \mathcal{R}, c > 0, F_{Cauchy(c)}(x) = \frac{1}{2} + \frac{1}{\pi}\arctan\left(\frac{x}{c}\right),$$

see Figures, 20a and 20b.

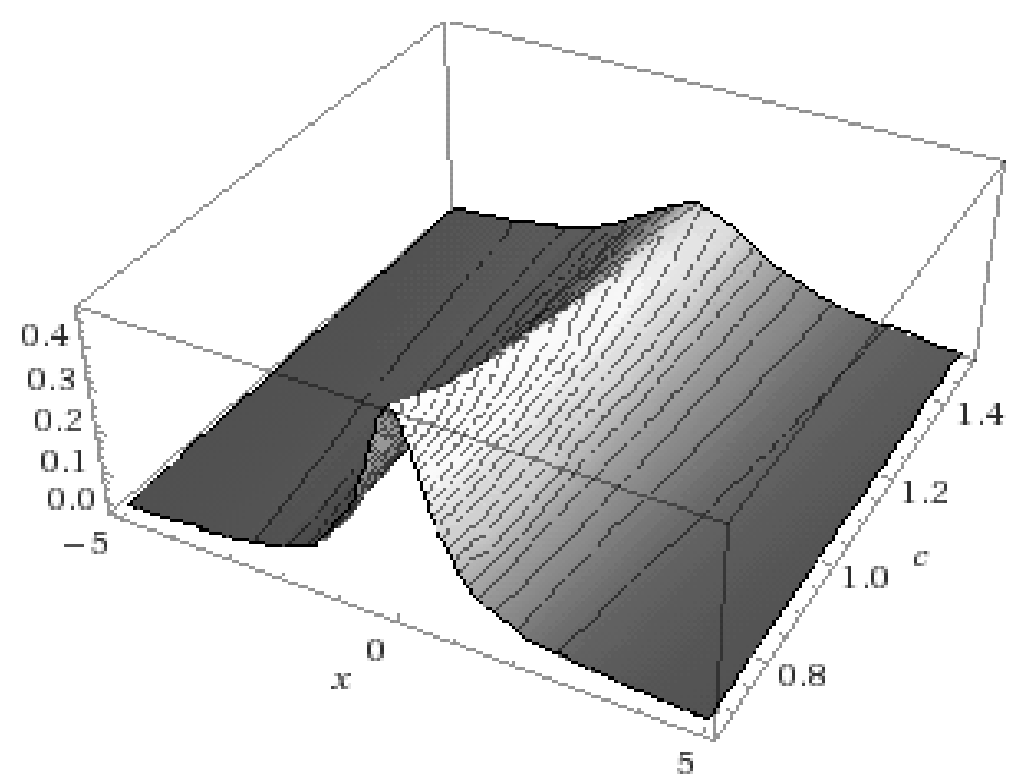

**Figure 20a Plot of the pdf** ,$\boldsymbol{f_{Cauchy(c)}(x) = \frac{1}{\pi}\frac{c}{c^2+x^2}, x \in \mathcal{R}, c > 0 + \frac{1}{\pi}\arctan\left(\frac{x}{c}\right), -5 < x < 5, \frac{2}{3} < c < 3/2}$

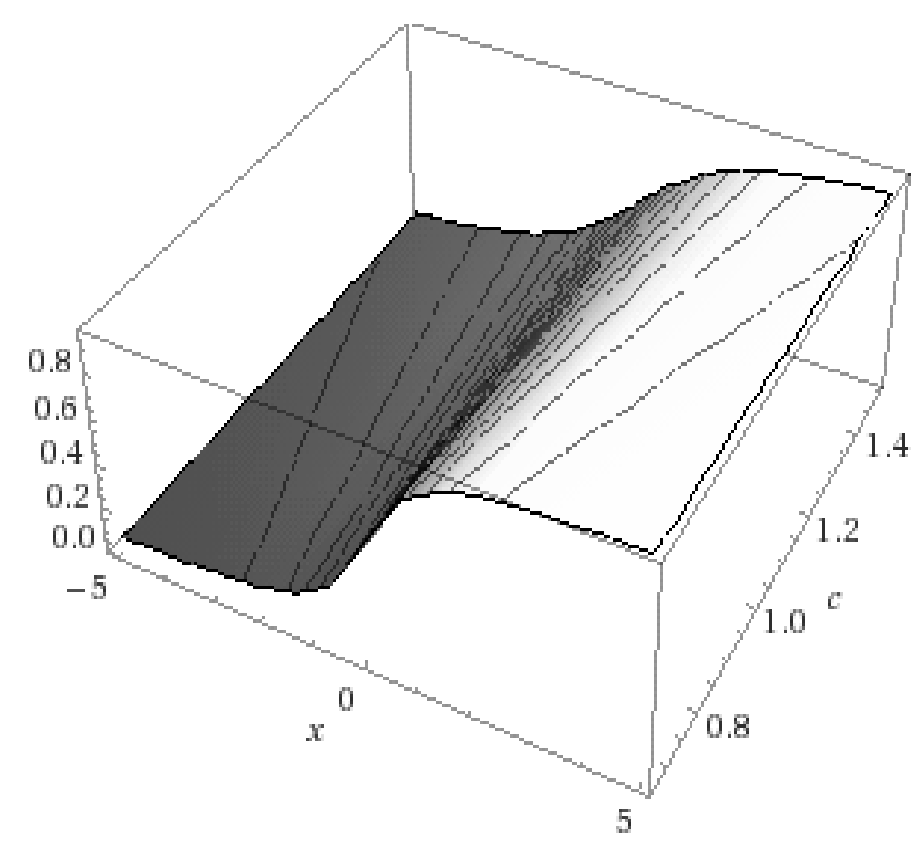

**Figure 20b Plot of the cdf** $\boldsymbol{F_{Cauchy(c)}(x) = \frac{1}{2} + \frac{1}{\pi}\arctan\left(\frac{x}{c}\right), -5 < x < 5, 2/3 < c < 3/2}$

Then $w^{(\beth,\mathfrak{R},\mathfrak{R}^{(\beth)})}(u) = \frac{1}{2} + \frac{1}{\pi}\arctan\left(\frac{c\tan\left(\pi u - \frac{\pi}{2}\right)}{c^{(\beth)}}\right)$,see Figure 21a. When $a^{(\beth)} := \frac{c}{c^{(\beth)}} \in (0,1)$, ב is "fearful", while when $a^{(\beth)} > 1$ , ב is "greedy", see the plot of the second derivative of $w^{(\beth,\mathfrak{R},\mathfrak{R}^{(\beth)})}(u), 0 < u < 1$ in Figure 21.b.

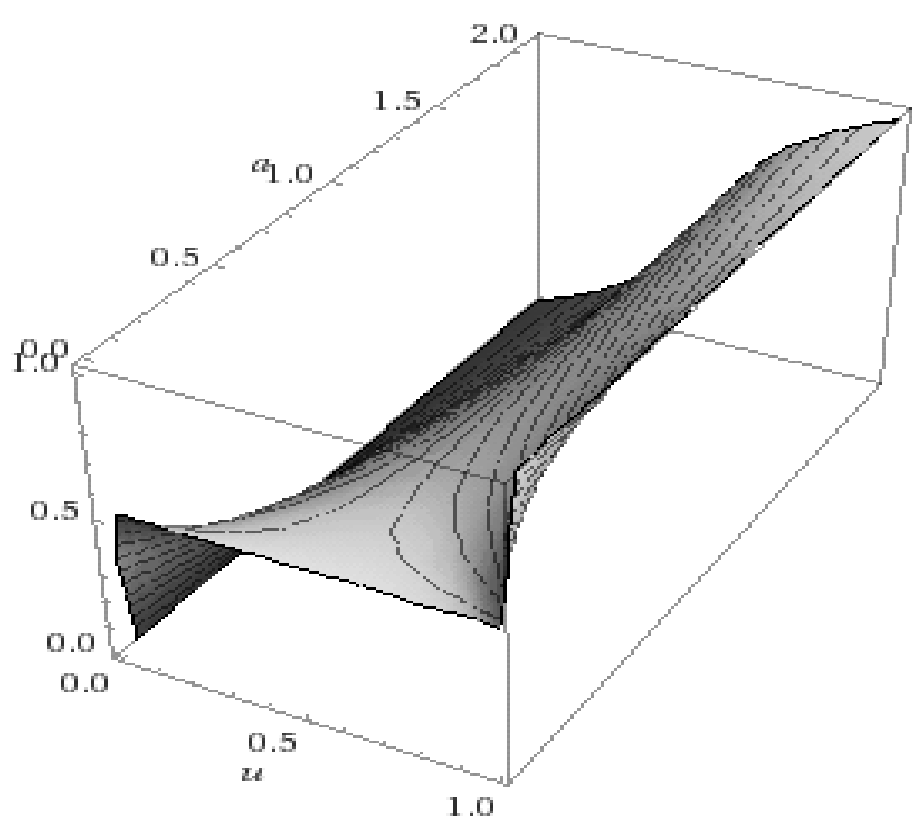

**Figure 21a. Plot of Cauchy WPF**

$$w^{(\text{ב},\Re,\Re^{(\text{ב})})}(u)=\frac{1}{2}+\frac{1}{\pi}\arctan\left(a^{(\text{ב})}\tan\left(\pi u-\frac{\pi}{2}\right)\right), 0<u<1, 0<a^{(\text{ב})}<2.$$

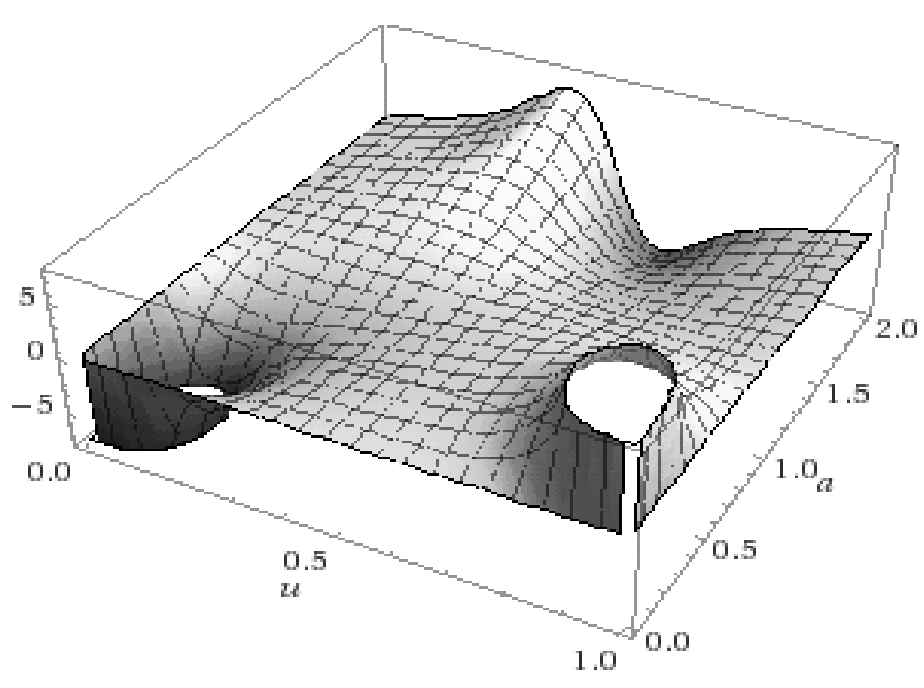

**Figure 21a. Plot of the second derivative of Cauchy WPF** $\frac{\partial^2 w^{(\text{ב},\Re,\Re^{(\text{ב})})}(u)}{\partial u^2}=\frac{4\pi a^{(\text{ב})}\left({a^{(\text{ב})}}^2-1\right)\sin(2\pi u)}{\left({a^{(\text{ב})}}^2(\cos(2\pi u)+1)-\cos(2\pi u)+1\right)^2}, 0<u<1, 0<a^{(\text{ב})}<2.$

**Example 8. (Cauchy - Gumbel WPF).** Let $\Re \triangleq Cauchy(c), c>0$ and $\Re^{(\text{ב})} \triangleq G\left(\mu^{(\text{ב})},\varrho^{(\text{ב})}\right), \mu^{(\text{ב})}\in\mathcal{R}, \varrho^{(\text{ב})}>0$. Then $w^{(\text{ב},\Re,\Re^{(\text{ב})})}(u)=\exp\left(-e^{-\frac{c\tan\left(\pi u-\frac{\pi}{2}\right)-\mu^{(\text{ב})}}{\varrho^{(\text{ב})}}}\right)$, see Figure 22a. When $a^{(\text{ב})}:=\frac{c}{\rho^{(\text{ב})}}\in(0,1)$, ב is “fearful”, while when $a^{(\text{ב})}>1$ , ב is “greedy”, see the plot of the second derivative of $w^{(\text{ב},\Re,\Re^{(\text{ב})})}(u), 0<u<1$ in Figure 22.b.

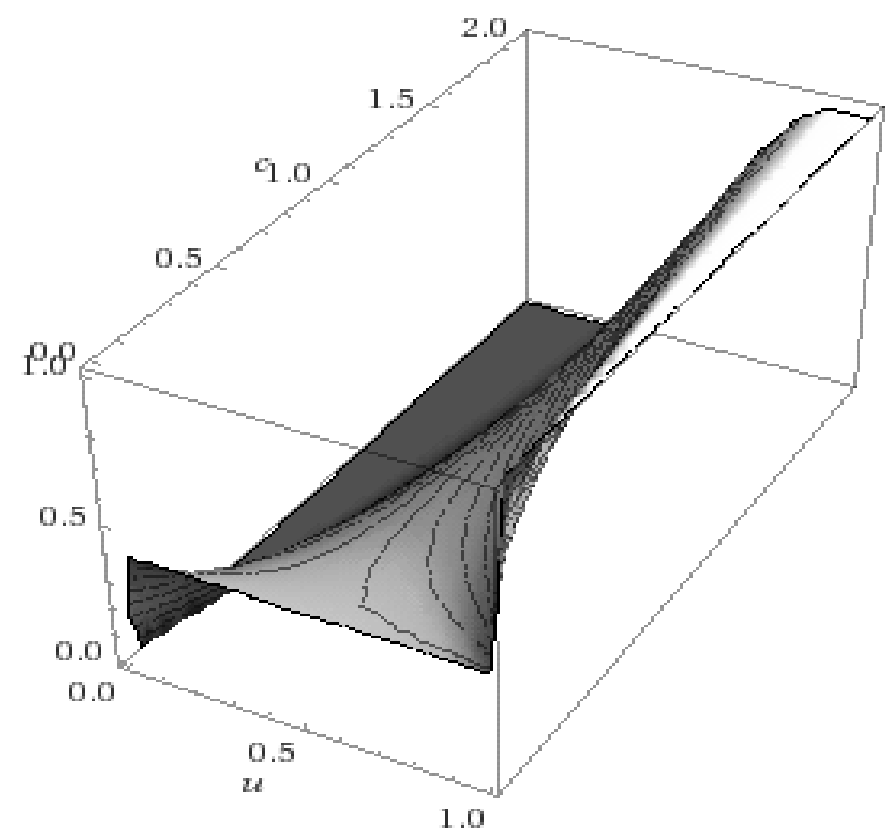

**Figure 22a. Plot of Cauchy-Gumbel WPF** $w^{(\text{ב},\Re,\Re^{(\text{ב})})}(u) = \exp\left(-e^{-c\cdot\tan\left(\pi u - \frac{\pi}{2}\right)}\right)$ $\mathbf{0 < u < 1, 0 < c < 2}, \mu^{(\text{ב})} = 0, \rho^{(\text{ב})} = 1$

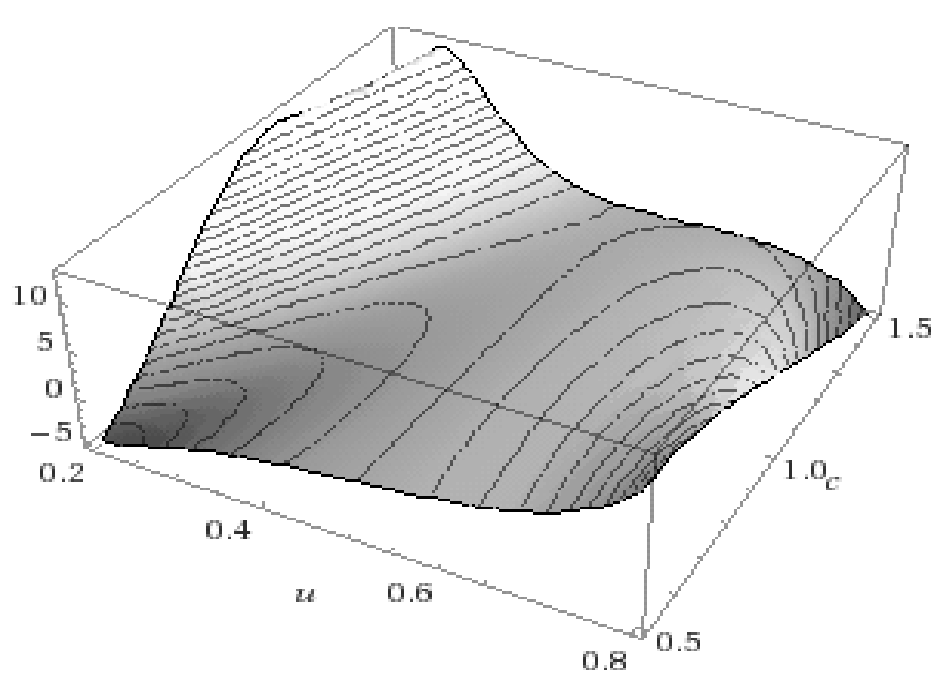

**Figure 22b. Plot of the second derivative of Cauchy WPF** $\frac{\partial^2 w^{(\text{ב},\Re,\Re^{(\text{ב})})}(u)}{\partial u^2} = -c\pi^2 csc^4(\pi u)\left(c - ce^{c\cdot cot(\pi u)} + sin(2\pi u)\right) e^{c\cdot cot(\pi u) - e^{c\cdot cot(\pi u)}}$, $\mathbf{0 < u < 1, 0 < c < 2}, \mu^{(\text{ב})} = 0, \rho^{(\text{ב})} = 1$

**Example 9. (Laplace WPF).** Let $\Re \triangleq Laplace(b), > 0$ and $\Re^{(\text{ב})} \triangleq Laplace\left(b^{(\text{ב})}\right), b^{(\text{ב})} > 0$. Then

$$w^{(\text{ב},\Re,\Re^{(\text{ב})})}(u) = \begin{cases} \frac{1}{2}(2u)^{\frac{b}{b^{(\text{ב})}}}, & 0 < u < \frac{1}{2} \\ 1 - \frac{1}{2}(2-2u)^{\frac{b}{b^{(\text{ב})}}}, & \frac{1}{2} \leq u < 1, \end{cases}$$

see Figure 23. When $\frac{b}{b^{(\beth)}} \in (0,1)$, $\beth$ is “fearful”, while when $\frac{b}{b^{(\beth)}} > 1$ , $\beth$ is “greedy”, see the plot of the second derivative of $w^{(\beth,\Re,\Re^{(\beth)})}(u), 0 < u < 1$.

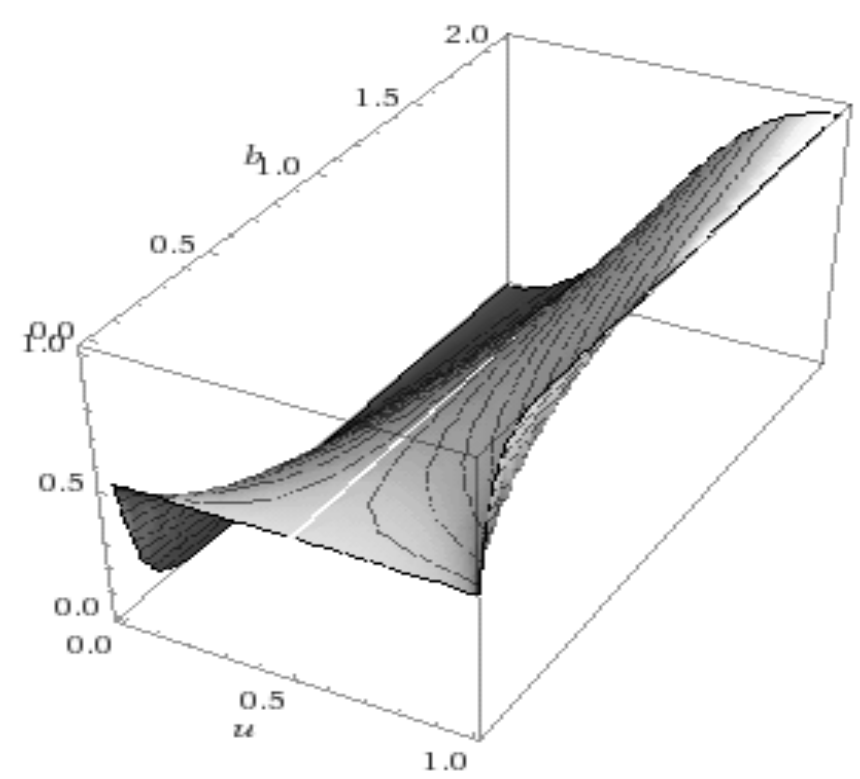

**Figure 23. Plot of Laplace WPF**

$$\boldsymbol{w^{(\beth,\Re,\Re^{(\beth)})}(u)} = \begin{cases} \frac{1}{2}(2u)^{b}, & 0 < u < \frac{1}{2} \\ 1 - \frac{1}{2}(2-2u)^{b}, & \frac{1}{2} \le u < 1 \end{cases}, 0 < b < 2, b^{(\beth)} = 1$$

**Example 10. (Gaussian WPF).** Let $\Re \triangleq \mathcal{N}(\mu, \sigma^2), \mu \in \mathcal{R}, \sigma > 0$, and $\Re^{(\beth)} \triangleq \mathcal{N}\left(\mu^{(\beth)}, {\sigma^{(\beth)}}^2\right), \mu^{(\beth)} \in \mathcal{R}, \sigma^{(\beth)} > 0$. Then

$$w^{(\beth,\Re,\Re^{(\beth)})}(u) = F_{\mathcal{N}\left(\mu^{(\beth)},{\sigma^{(\beth)}}^2\right)} \circ F^{inv}_{\mathcal{N}(\mu,\sigma^2)}(u) = \frac{1}{2} erfc\left(\frac{1}{\sqrt{2}} \frac{\mu^{(\beth)} - F^{inv}_{\mathcal{N}(\mu,\sigma^2)}(u)}{\sigma^{(\beth)}}\right) =$$

$$= \frac{1}{2} erfc\left(\frac{1}{\sqrt{2}} \frac{\mu^{(\beth)} - \mu + \sqrt{2}\sigma erfc^{(Inv)}(2u)}{\sigma^{(\beth)}}\right),$$

see Figures 24a and 24b. When $\frac{\sigma}{\sigma^{(\beth)}} \in (0,1)$, $\beth$ is “fearful”, while when $\frac{\sigma}{\sigma^{(\beth)}} > 1$ , $\beth$ is “greedy”.

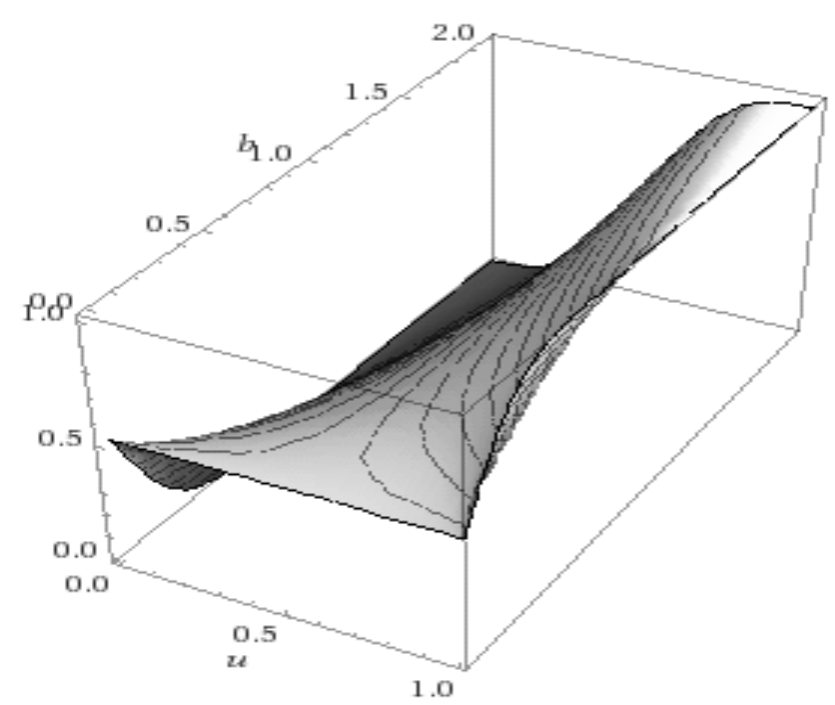

**Figure 24a. Plot of Gaussian WPF** $\boldsymbol{w}^{(\text{ב},\mathfrak{R},\mathfrak{R}^{(\text{ב})})}(\boldsymbol{u}) = \frac{1}{2}erfc\left(\frac{1}{\sqrt{2}}\frac{\mu^{(\text{ב})}-\mu+\sqrt{2}\sigma erfc^{(Inv)}(2u)}{\sigma^{(\text{ב})}}\right), 0 < u < 1,\ \mu^{(\text{ב})} = \mu, b = \frac{\sigma}{\sigma^{(\text{ב})}} \in (0,2).$

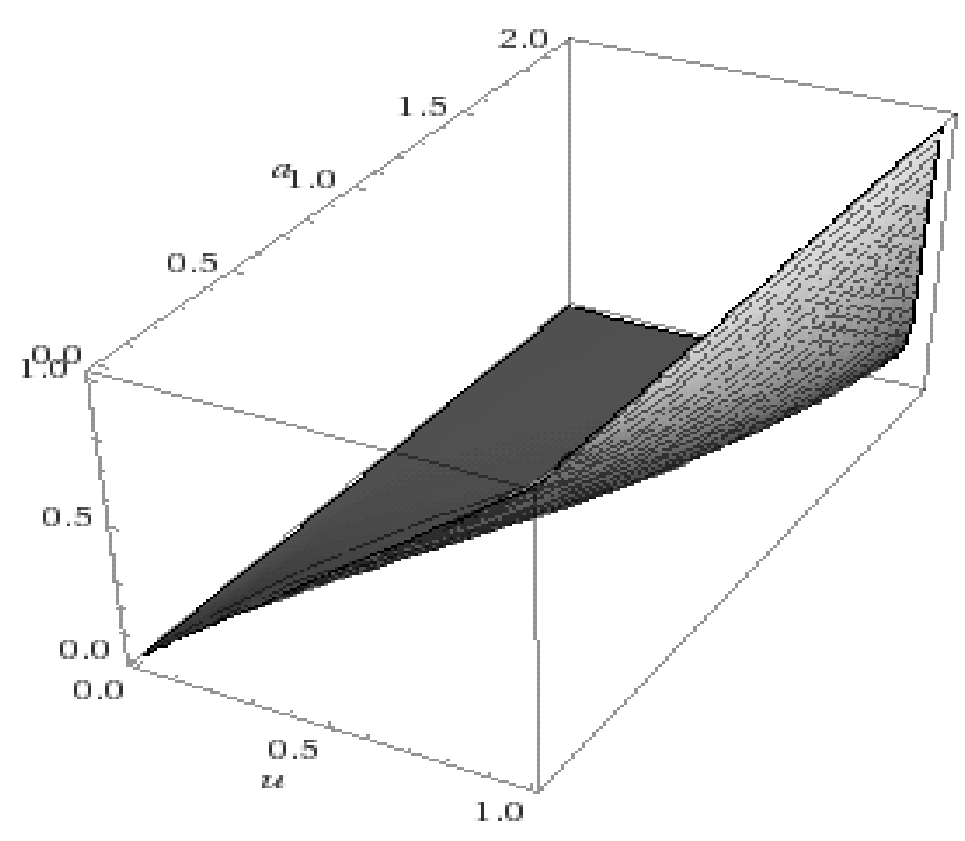

**Figure 24b. Plot of Gaussian WPF** $\boldsymbol{w}^{(\text{ב},\mathfrak{R},\mathfrak{R}^{(\text{ב})})}(\boldsymbol{u}) = \frac{1}{2}erfc\left(\frac{1}{\sqrt{2}}\frac{\mu^{(\text{ב})}-\mu+\sqrt{2}\sigma erfc^{(Inv)}(2u)}{\sigma^{(\text{ב})}}\right), 0 < u < 1,\ a = \mu^{(\text{ב})} - \mu \in (0,2), b = 2.$

**Example 11. (Gaussian -Negative Gumbel WPF).** Let $\mathfrak{R} \triangleq \mathcal{N}(\mu,\sigma^2)$ and $\mathfrak{R}^{(\text{ב})} \triangleq NG\left(\mu^{(\text{ב})},\varrho^{(\text{ב})}\right), \mu^{(\text{ב})} \in \mathcal{R}, \varrho^{(\text{ב})} > 0$. Then

$$w^{(\text{ב},\mathfrak{R},\mathfrak{R}^{(\text{ב})})}(u) = F_{NG\left(\mu^{(\text{ב})},\varrho^{(\text{ב})}\right)} {}^{\circ}F^{inv}_{\mathcal{N}(\mu,\sigma^2)}(u) =$$

$$= 1 - \exp\left(-e^{\frac{\mu-\mu^{(\text{ב})}-\sqrt{2}\sigma erfc^{(Inv)}(2u)}{\rho^{(\text{ב})}}}\right), 0 < u < 1.$$

see Figures 25a and 25b. When $\frac{\boldsymbol{\sigma}}{\boldsymbol{\rho}^{(\text{ב})}} \in (0,1)$, ב is “fearful”, while when $\frac{\boldsymbol{\sigma}}{\boldsymbol{\rho}^{(\text{ב})}} > 1$ , ב is “greedy”.

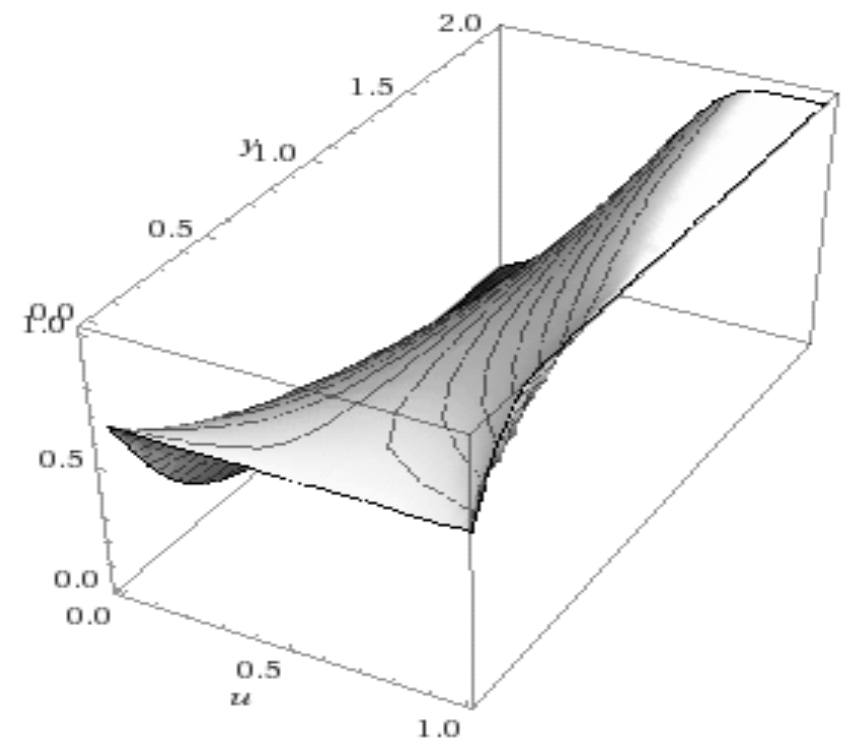

**Figure 25a. Plot of Gaussian -Negative Gumbel WPF :** $w^{(\beth,\Re,\Re^{(\beth)})}(u) = F_{NG(\mu^{(\beth)},\varrho^{(\beth)})} \circ F^{inv}_{\mathcal{N}(\mu,\sigma^2)}(u) =$

$$= 1 - \exp\left(-e^{\frac{\mu-\mu^{(\beth)}-\sqrt{2}\sigma erfc^{(Inv)}(2u)}{\rho^{(\beth)}}}\right), 0 < u < 1, \mu^{(\beth)} = \mu, y = \frac{\sigma}{\rho^{(\beth)}} \in (0, 2).$$

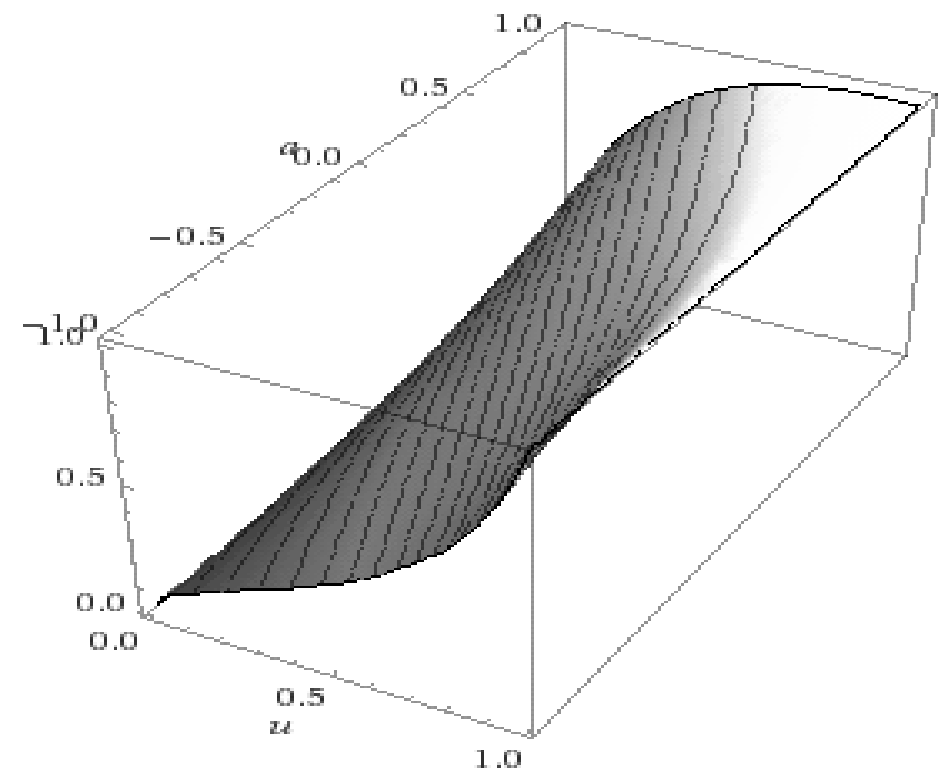

**Figure 25b. Plot of Gaussian -Negative Gumbel WPF:** $w^{(\beth,\Re,\Re^{(\beth)})}(u) = F_{NG(\mu^{(\beth)},\varrho^{(\beth)})} \circ F^{inv}_{\mathcal{N}(\mu,\sigma^2)}(u) =$

$$= 1 - \exp\left(-e^{\frac{\mu-\mu^{(\beth)}-\sqrt{2}\sigma erfc^{(Inv)}(2u)}{\rho^{(\beth)}}}\right), 0 < u < 1, a = \mu^{(\beth)} - \mu \in (-1, 1), \frac{\sigma}{\rho^{(\beth)}} = 1 .$$

**Example 12. (Gaussian -Logistic WPF).** Let $\Re \triangleq \mathcal{N}(\mu, \sigma^2)$ and $\Re^{(\beth)} \triangleq Logist(m^{(\beth)}, \rho^{(\beth)}), m^{(\beth)} \in \mathcal{R}, \rho^{(\beth)} > 0$. Then

$$w^{(\text{ב},\mathfrak{R},\mathfrak{R}^{(\text{ב})})}(u) = F_{Logist(m^{(\text{ב})},\rho^{(\text{ב})})} \circ F^{inv}_{\mathcal{N}(\mu,\sigma^2)}(u) =$$

$$= \frac{1}{1+exp\left(\frac{m^{(\text{ב})} - \mu + \sqrt{2}\sigma erfc^{(Inv)}(2u)}{\rho^{(\text{ב})}}\right)}, x \in \mathcal{R}$$

When $\frac{\sigma}{\rho^{(\text{ב})}} \in (0,1)$, ב is "fearful", while when $\frac{\sigma}{\rho^{(\text{ב})}} > 1$ , ב is "greedy", Figure 26a. The difference $m^{(\text{ב})} - \mu$ controls the height of the wpf, see Figures 26b.

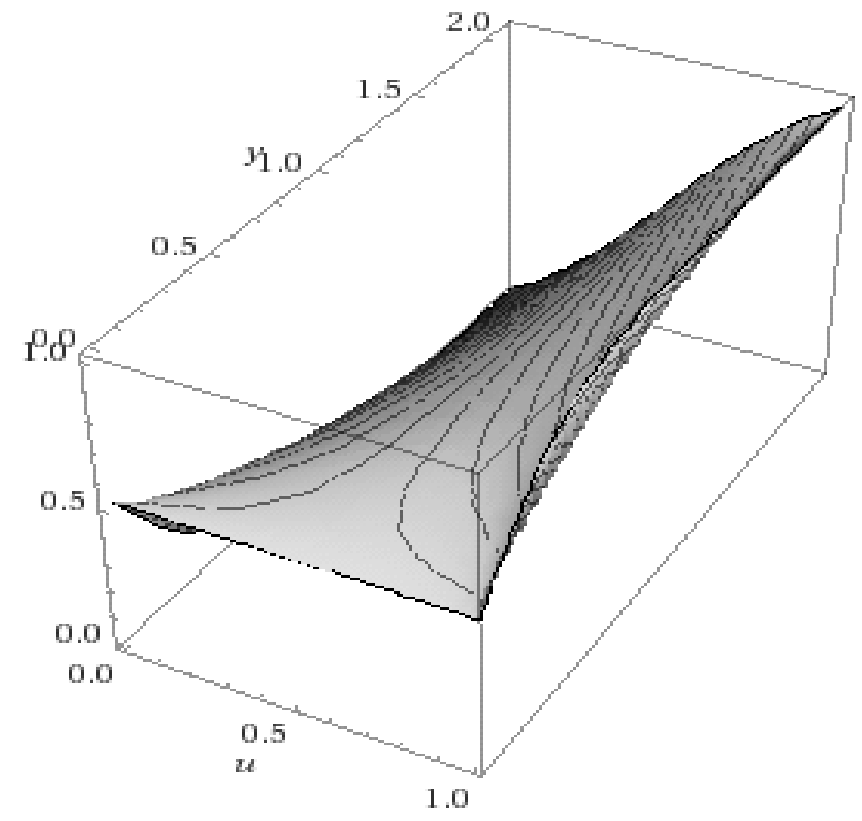

**Figure 26a. Plot of Gaussian -Logistic WPF :** $\boldsymbol{w^{(\text{ב},\mathfrak{R},\mathfrak{R}^{(\text{ב})})}(u) = F_{Logist(m^{(\text{ב})},\rho^{(\text{ב})})} \circ F^{inv}_{\mathcal{N}(\mu,\sigma^2)}(u) =}$

$\boldsymbol{F^{(\mathfrak{R},m,\rho)}(x) = \frac{1}{1+exp\left(\frac{m^{(\text{ב})} - \mu + \sqrt{2}\sigma erfc^{(Inv)}(2u)}{\rho^{(\text{ב})}}\right)}, x \in \mathcal{R}, \mu^{(\text{ב})} = \mu, y = \frac{\sigma}{\rho^{(\text{ב})}} \in (0,2).}$

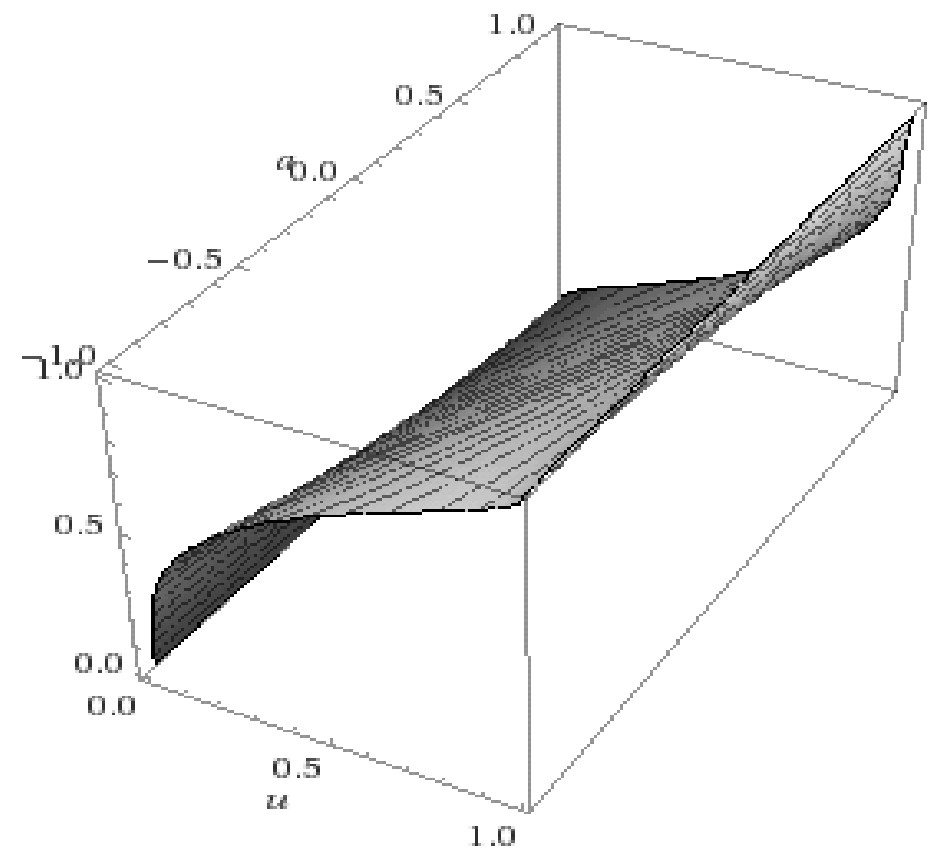

**Figure 26a. Plot of Gaussian -Logistic WPF :** $\boldsymbol{w^{(\beth,\Re,\Re^{(\beth)})}(u) = F_{Logist(m^{(\beth)},\rho^{(\beth)})} \circ F^{inv}_{\mathcal{N}(\mu,\sigma^2)}(u) =}$

$$\boldsymbol{= \frac{1}{1+exp\left(\frac{m^{(\beth)}-\mu+\sqrt{2}\sigma erfc^{(Inv)}(2u)}{\rho^{(\beth)}}\right)}, x \in \mathcal{R}, a = \mu^{(\beth)} - \mu \in (-1,1), y = \frac{\sigma}{\rho^{(\beth)}} = 1}$$

## 4. **Option Pricing with Greed and Fear Factor and the General Itô Processes**

Consider the Black-Scholes market for a risky asset (stock) $\mathcal{S}$ and a riskless asset (bond) $\mathcal{B}$. The stock price follows the dynamics of an Itô process:

$$\frac{dS(t)}{S(t)} = \mu(t, S(t))dt + \sigma(t, S(t)dB(t), t \geq 0, S(0) > 0, \mu\big(t, S(t)\big) > 0, \sigma(t, S(t) > 0, \qquad (20)$$

where $B(t), t \geq 0$, is a Brownian motion generating a stochastic basis $(\Omega, \mathcal{F}, \mathbb{F} = (\mathcal{F}_t, t \geq 0), \mathbb{P})$ and the instantaneous mean return $\mu\big(t, S(t)\big), t \geq 0$ and $\sigma(t, S(t), t \geq 0$, satisfy the usual regularity conditions.[28] The bond dynamics is given by

$$\frac{d\beta(t)}{\beta(t)} = r(t, S(t)dt, \beta(0) > 0, \qquad (21)$$

---

[28] See Duffie (2001), Chapter 5, Section C.

where $r\big(t,S(t)\big)>0, t\geq 0,$ is the riskless rate, which is $\mathbb{F}$-adapted, and $\sup_{t\geq 0}\left\{r\big(t,S(t)\big)+\frac{1}{r\left(t,S(t)\right)}\right\}<\infty$, $\mathbb{P}$-a.s.

Consider a ECC $\mathcal{C}$ with price process $C(t)=f\big(t,S(t)\big)$, where $f(t,x)>0, t\geq 0, x>0,$ has continuous derivatives $\frac{\partial f\left(t,S(t)\right)}{\partial t}$, and $\frac{\partial^2 f\left(t,S(t)\right)}{\partial x^2}, t\geq 0, v>0$. The terminal time for $\mathcal{C}$ is $>0$ , and the final payoff is $C(T)=f\big(T,S(T)\big)=g\big(S(T)\big)$, for some continuous $g(x), x>0$. Then by the Itô formula,

$$df\big(t,S(t)\big)=\left(\frac{\partial f\left(t,S(t)\right)}{\partial t}+\mu\big(t,S(t)\big)S(t)\frac{\partial f\left(t,S(t)\right)}{\partial x}+\frac{1}{2}\sigma\big(t,S(t)\big)^2 S(t)^2\ \frac{\partial^2 f\left(t,S(t)\right)}{\partial x^2}\right)dt+$$

$$+\sigma\big(t,S(t)\big)S(t)\frac{\partial f\left(t,S(t)\right)}{\partial x}B(t), t\geq 0. \qquad (22)$$

Suppose $\beth$ is taking a short position in the $\mathcal{C}$ contract. $\beth$ hedges the short position under a certain level of "greed & fear", which we quantify as follows. When $\beth$ trades the stock $\mathcal{S}$, the stock dynamics $S(t), t\geq 0$ is different from (8) due to $\beth's$ superior or inferior trading performance.[29] As a result, $\beth$ trades $\mathcal{S}$ under the following price dynamics:

$$\frac{dS(t)}{S(t)}=\mu^{(\tau)}(t,S(t))dt+\sigma^{(\tau)}(t,S(t)dB(t), t\geq 0, S(0)>0, \qquad (23)$$

for some $\mu^{(\tau)}\big(t,S(t)\big)>0, \sigma^{(\tau)}(t,S(t)>0$. $\beth$ chooses a $\mathbb{F}$-adapted trading strategy $a(t), b(t), t\geq 0,$

[29] This could be attributable to transaction costs, liquidity constraints, trading the stock in a better or worse trading frequency, and the like. Another possible view could be that $\beth$ has different estimates for the drift and diffusion parameters than the publicly available ones.

$$f\big(t,S(t)\big)\Big(1+\mathcal{G}\big(t,S(t)\big)\Big)=P\big(t,S(t)\big)\coloneqq a(t)S(t)+\beta(t)\,b(t),t\geq 0\,, \qquad (24)$$

where $\mathcal{G}\big(t,S(t)\big),t\geq 0$, is the "greed & fear" functional. If $\mathcal{G}\big(t,S(t)\big)>0$, ב is in a "greedy" hedging disposition, believing that $P\big(t,S(t)\big)$ cannot only cover ב's short position in $\mathcal{C}$ , but also generates some dividend stream. This is due to ב$'s$ belief that following the price dynamics given (23) has a superior trading dynamics over the publicly available trading dynamics given by (20). If $\mathcal{G}\big(t,S(t)\big)<0$, ב is in a "fear" hedging disposition, believing that $P\big(t,S(t)\big)$ will not be able to cover ב$'$s short position in $\mathcal{C}$ due to ב$'s$ belief that the trading dynamics given by (20) is inferior. If $\mathcal{G}\big(t,S(t)\big)=0$, ב has taken the standard hedge position. Note that $\mathcal{G}\big(t,S(t)\big),t\geq 0$ changes dynamically over time and can oscillate, describing the fact that ב might dynamically change his or her greed and fear disposition.

Next, suppose that ב's choses the dynamics of the self-financing portfolio $P\big(t,S(t)\big)$ to be

$$dP\big(t,S(t)\big)=$$

$$=\Big(a(t)S(t)\mu^{(\tau)}\big(t,S(t)\big)+b(t)r\big(t,S(t)\big)\beta(t)\Big)\,dt+a(t)S(t)\sigma^{(\tau)}(t,S(t)dB(t), \qquad (25)$$

where $a(t)$ and $b(t)$ are chosen so that the following utility function $\mathcal{U}^{(\text{ב})}(t,t+dt),t\geq 0$, is maximized:

$$\mathcal{U}^{(\text{ב})}(t,t+dt)\coloneqq\mathcal{G}\big(t,S(t)\big)\mathbb{E}_t\Big(dP\big(t,S(t)\big)\Big)-var_t\Big(df\big(t,S(t)\big)-dP\big(t,S(t)\big)\Big)=$$

$$= \mathcal{G}\big(t,S(t)\big)\begin{pmatrix} a(t)S(t)\left(\mu^{(\tau)}\big(t,S(t)\big) - r\big(t,S(t)\big)\right) + \\ +r\big(t,S(t)\big)\, f\big(t,S(t)\big)\big(1-\mathcal{G}\big(t,S(t)\big)\big) \end{pmatrix} dt -$$

$$-S(t)^2\left(\sigma\big(t,S(t)\big)\frac{\partial f\big(t,S(t)\big)}{\partial x} - a(t)\sigma^{(\tau)}\big(t,S(t)\big)\right)^2 dt \qquad (26)$$

Again, if ב is in a "greedy" disposition, $\mathcal{G}\big(t,S(t)\big) > 0$, then (25) implies that ב is seeking an extra return from the hedged portfolio, foregoing the perfect replication. If ב is in a "fearful" disposition, $\mathcal{G}\big(t,S(t)\big) < 0$, then (25) implies that ב is willing to cut some of the return from the hedged portfolio to add to the hedge. To find $a(t)$ and $b(t)$, ב solves for $a(t)$ the equation

$$\frac{\partial \mathcal{U}^{(ב)}(t,t+dt)}{\partial a(t)} = 0.$$

Because $\frac{\partial^2 \mathcal{U}^{(ב)}(t,t+dt)}{\partial^2 a(t)} = -S(t)^2\sigma^{(\tau)}\big(t,S(t)\big)^2 dt,$ the optimal $a(t)$ is given by

$$a(t) = \frac{\sigma^{(\tau)}\big(t,S(t)\big)}{\sigma\big(t,S(t)\big)}\frac{\partial f\big(t,S(t)\big)}{\partial x} + \frac{\mathcal{G}\big(t,S(t)\big)\left(\mu^{(\tau)}\big(t,S(t)\big) - r\big(t,S(t)\big)\right)}{\sigma^{(\tau)}\big(t,S(t)\big)^2 S(t)} \qquad (26)$$

and then,

$$b(t) = \frac{1}{\beta(t)}\left\{\begin{matrix} f\big(t,S(t)\big)\left(1+\mathcal{G}\big(t,S(t)\big)\right) - \\ -\frac{\sigma^{(\tau)}\big(t,S(t)\big)}{\sigma\big(t,S(t)\big)}\frac{\partial f\big(t,S(t)\big)}{\partial x}S(t) - \frac{\mathcal{G}\big(t,S(t)\big)\left(\mu^{(\tau)}\big(t,S(t)\big) - r\big(t,S(t)\big)\right)}{\sigma^{(\tau)}\big(t,S(t)\big)^2} \end{matrix}\right\}. \qquad (27)$$

As a best replication strategy, ב chooses the mean dynamics given by:

$$\frac{\partial f\big(t,S(t)\big)}{\partial t} + \mu\big(t,S(t)\big)S(t)\frac{\partial f\big(t,S(t)\big)}{\partial x} + \frac{1}{2}\sigma\big(t,S(t)\big)^2 S(t)^2\,\frac{\partial^2 f\big(t,S(t)\big)}{\partial x^2} =$$

$$= a(t)S(t)\mu^{(\tau)}\big(t,S(t)\big) + b(t)r\big(t,S(t)\big)\beta(t) \qquad (28)$$

Next, we use the following notations:

$(i)\ r^{(\text{ב})}\big(t,S(t)\big) \coloneqq r\big(t,S(t)\big)\left(1+\mathcal{G}\big(t,S(t)\big)\right)$ is the specific for $\text{ב}'s$ discount rate based on $\text{ב}'s$ level of "greed & fear";

$(ii)\ \ \theta\big(t,S(t)\big) = \frac{\mu(t,S(t))-r(t,S(t))}{\sigma(t,S(t))}$ is the Sharpe ratio for the publicly traded stock $\mathcal{S}$ ;

$(iii)\ \ \theta^{(\tau)}\big(t,S(t)\big) = \frac{\mu^{(\tau)}(t,S(t))-r(t,S(t))}{\sigma^{(\tau)}(t,S(t))}$ is the Sharpe ratio for stock $\mathcal{S}$ when traded by ב;

$(iv)\quad D_y^{(\tau)}\big(t,S(t)\big)$

$$\coloneqq \frac{\theta^{(\tau)}\big(t,S(t)\big)\sigma^{(\tau)}\big(t,S(t)\big)^2 - \theta\big(t,S(t)\big)\sigma^2\big(t,S(t)\big)}{\sigma\big(t,S(t)\big)} + \mathcal{G}\big(t,S(t)\big)r\big(t,S(t)\big)$$

is the yield (positive or negative) accumulated byב while trading $\mathcal{S}$;

$(v)\quad R^{(\text{ב})}\big(t,S(t)\big) \coloneqq r^{(\text{ב})}\big(t,S(t)\big) - D_y^{(\tau)}\big(t,S(t)\big)$ is reduced by the dividend yield $D_y^{(\tau)}\big(t,S(t)\big)$ , $\text{ב}'s$ discount rate $r^{(\text{ב})}\big(t,S(t)\big)$;

$(iv)\ \ h^{(\tau)}(t,S(t)) := \theta^{(\tau)}(t,S(t))^2\mathcal{G}(t,S(t))$ is "$\text{ב}'s$ running trading reward" .

Then, (26), (27) and (28) lead to the following partial differential equation for$f(t,x),\ t \in [0,T), x > 0$,

$$\frac{\partial f(t,x)}{\partial t} + R^{(\tau)}(t,x)x\frac{\partial f(t,x)}{\partial x} - r^{(\text{ב})}(t,x)f(t,x) + \frac{1}{2}\sigma(t,x)^2x^2\ \frac{\partial^2 f(t,x)}{\partial x^2} - h^{(\tau)}(t,x) = 0, \quad (29)$$

with boundary condition $f(T,x) = g(x), x > 0.$ The partial differential equation given by (29) admits the Feynman-Kac solution:[30]

[30] See Duffie (2001), Appendix E.

$$f(t,x) = \mathbb{E}\left\{\varphi^{(\beth)}(t,T)g(X(T) - \int_t^T \varphi^{(\beth)}(t,s)h^{(\tau)}(s,X(s))\,ds\right\} \tag{30}$$

where $\varphi^{(\beth)}(t,s) := \exp\{-\int_t^s r^{(\beth)}(t,x)\}$, and $X(s), s \geq t, X(t) = x$ is the Itô process

$$\frac{dX(s)}{X(s)} = R^{(\tau)}(s,X(s))\,ds + \sigma\big(s,X(s)\big)dB(s). \tag{31}$$

Now (30) and (31) provide the following risk-neutral valuation of $\beth's$ trading activities. While hedging, ב is trading $\mathcal{S}$ with the risk-neutral dynamics given by (31), viewing the stock as paying dividend yield $D_y^{(\tau)}\big(t,S(t)\big)$. ב trades at a discount rate $\varphi^{(\beth)}(0,t), t \geq 0$. Finally, while trading, ב enjoys the running trading reward $h^{(\tau)}\big(t,S(t)\big)$.

Consider the following special case: $\mathcal{G}\big(t,S(t)\big) = \mathcal{G} \in \mathcal{R}, \mu\big(t,S(t)\big) = \mu > r = r\big(t,S(t)\big) > 0$, $\sigma^{(\tau)}\big(t,S(t)\big) = \sigma\big(t,S(t)\big) = \sigma > 0, \mu^{(\tau)}(t,x) = \mu^{(\tau)} = (1+\mathcal{G})\mu$, $g(x) = \max(0, x-K)$. Then the call option formula is given by

$$C(t) = f\big(t,S(t)\big) = C\big(t,S(t),X,T,r,\sigma,D_y^{\beth}\big) =$$

$$= e^{-(\mathcal{G}(\mu-r))(T-t)}S(t)\Phi\big(D^{(1)}(t)\big) - Ke^{-r(T-t)}\Phi\left(D^{(2)}(t)\right) - \left(\frac{(1+\mathcal{G})\mu - r}{\sigma}\right)^2 \frac{\mathcal{G}}{r(1+\mathcal{G})} \tag{32}$$

where $\Phi(\mathrm{x}), \mathrm{x} \in \mathcal{R}$, is the standard normal cumulative distribution function and

$$D^{(1)}(t) = \frac{\ln\left(\frac{S(t)}{X}\right) + \left(r - \mathcal{G}(\mu - r) + \frac{1}{2}\sigma^2\right)(T-t)}{\sigma(T-t)}, D^{(2)}(t) = D^{(1)}(t) - \sigma(T-t).$$

### 5. Option Pricing with Greed & Fear Factor When Stock Price Dynamics Follows a Binomial Tree

There is no unique way in which the "greed & fear" factor can manifest in the trading activities of a trader. We illustrate that fact in the following alternative "greed & fear" trading model.

Consider again the Black-Scholes market for a risky asset (stock) $\mathcal{S}$ and a riskless asset (bond) $\mathcal{B}$. The stock price follows the dynamics of a geometric Brownian motion:

$$\frac{dS(t)}{S(t)} = \mu dt + \sigma dB(t), t \geq 0, S(0) > 0, \mu > 0, \sigma > 0 \tag{33}$$

where $B(t), t \geq 0$, is a Brownian motion generating a stochastic basis $(\Omega, \mathcal{F}, \mathbb{F} = (\mathcal{F}_t, t \geq 0), \mathbb{P})$. Then the corresponding binomial pricing tree:

$$S((k+1)\Delta t) = \begin{cases} S\big((k+1)\Delta t\big)^{(up)} = S(k\Delta t)\big(1 + \mu\Delta t + \sigma\sqrt{\Delta t}\,\big) \;\; w.p. 1/2 \\ S\big((k+1)\Delta t\big)^{(down)} = S(k\Delta t)\big(1 + \mu\Delta t - \sigma\sqrt{\Delta t}\,\big) \;\; w.p. 1/2 \end{cases} \tag{34}$$

where $k = 0,1, \dots, n-1, n\Delta t = T$, generates a right continuous with left limits process converging weakly in Skorokhod $D([0,T])$ – topology to $S(t), t \in [0,T]$.[31] The bond dynamics is given by

$$\frac{d\beta(t)}{\beta(t)} = rdt, \beta(0) > 0, r \in (0, \mu), \tag{35}$$

where $r$ is the riskless rate.

Consider an ECC $\mathcal{C}$ with price process $C(t) = f\big(t, S(t)\big)$. Suppose a trader ב is taking a short position in the $\mathcal{G}$-contract. ב hedges the short position under a certain level of "greed & fear", which we quantify as follows. At time $t^{(k)} = k\Delta t, k = 0, \dots, n-1,$ ב forms the hedge portfolio $P\left(t^{(k)}\right)$ of one short $\mathcal{C}$ and $a(t^{(k)})$- stock shares:

$$P\left(t^{(k)}\right) = -C\big(t^{(k)}\big) + a\big(t^{(k)}\big)S\big(t^{(k)}\big). \tag{36}$$

At $t^{(k+1)} = (k+1)\Delta t$, the hedge portfolio has values:

[31] See Kim et al (2016).

$$P\left(t^{(k+1)}\right) =$$

$$\begin{cases} P\left(t^{(k+1)}\right)^{(up)} = -C\left(t^{(k+1)}\right)^{(up)} + a\left(t^{(k)}\right)S\left(t^{(k+1)}\right)^{(up)} & w.p.\frac{1}{2} \\ P\left(t^{(k+1)}\right)^{(down)} = -C\left(t^{(k+1)}\right)^{(down)} + a\left(t^{(k)}\right)S\left(t^{(k+1)}\right)^{(down)} & w.p.\frac{1}{2} \end{cases}$$

ב selects $a\left(t^{(k)}\right)$ so that the following “greed & fear” functional $\mathfrak{G}(t^{(k)})$ is minimized:

$$\mathfrak{G}\left(t^{(k)}\right) = var\left(P\left(t^{(k+1)}\right)\right) - \mathcal{G}^{(\text{ב})}\,\mathbb{E}\left(P\left(t^{(k+1)}\right)\right), \tag{37}$$

where

$$\mathcal{G}^{(\text{ב})} := \mathcal{A}^{(\text{ב})}\sigma\left(C\left(t^{(k+1)}\right)^{(up)} - C\left(t^{(k+1)}\right)^{(down)}\right)\sqrt{\Delta t} \tag{38}$$

and $\mathcal{A}^{(\text{ב})} \in \mathcal{R} = (-\infty, +\infty)$ is $\text{ב}'s$ greed & fear coefficient. If $\mathcal{A}^{(\text{ב})} < 0$ (resp. $\mathcal{A}^{(\text{ב})} < 0$ ) $\text{ב}'s$ hedge decision is based on a certain level of greed (resp. fear). If $\mathcal{A}^{(\text{ב})} = 0$, $\text{ב}'s$ hedge decision is not influenced by fear or greed, leading to the standard risk-neutral binomial option pricing hedge. ב determines $a\left(t^{(k)}\right)$ such that $\mathfrak{G}\left(t^{(k)}\right) \Longrightarrow min$. Then, ב obtains

$$a\left(t^{(k)}\right) = \frac{C\left(t^{(k+1)}\right)^{(up)} - C\left(t^{(k+1)}\right)^{(down)}}{S\left(t^{(k+1)}\right)^{(up)} - S\left(t^{(k+1)}\right)^{(down)}} + \mathcal{G}^{(\text{ב})}\frac{S\left(t^{(k+1)}\right)^{(up)} + S\left(t^{(k+1)}\right)^{(down)}}{\left(S\left(t^{(k+1)}\right)^{(up)} - S\left(t^{(k+1)}\right)^{(down)}\right)^2}. \tag{39}$$

Next, ב choses

$$\mathbb{E}\left(P\left(t^{(k+1)}\right)\right) =$$

$$= \frac{1}{2}\left(a\left(t^{(k)}\right)\left(S\left(t^{(k+1)}\right)^{(up)} + S\left(t^{(k+1)}\right)^{(down)}\right) - \left(C\left(t^{(k+1)}\right)^{(up)} + C\left(t^{(k+1)}\right)^{(down)}\right)\right)$$

as the desirable value of the hedged portfolio at $t^{(k+1)}$. ב uses $e^{-r\Delta t}\mathbb{E}\left(P\left(t^{(k+1)}\right)\right)$ as a proxy for the $P\left(t^{(k)}\right)$ and computes the option value $C\left(t^{(k)}\right)$ by solving the equation

$$e^{-r\Delta t}\mathbb{E}\left(P\left(t^{(k+1)}\right)\right) = -C\left(t^{(k)}\right) + a\left(t^{(k)}\right)S\left(t^{(k)}\right), \tag{40}$$

with $a\left(t^{(k)}\right)$ given by (49). Applying (44) and (50), ב obtains the following option value at $t^{(k)}$:

$$C\left(t^{(k)}\right) = \left\{\frac{1}{2} - \frac{1}{2}\theta\sqrt{\Delta t}\right\} C\left(t^{(k+1)}\right)^{(up)} + \left\{\frac{1}{2} + \frac{1}{2}\theta\sqrt{\Delta t}\right\} C\left(t^{(k+1)}\right)^{(down)} - \mathcal{G}^{(\text{ב})}\frac{\theta}{2\sigma} \tag{41}$$

where $\theta = \frac{\mu - r}{\sigma}$ is the Sharpe ratio. When $\mathcal{G}^{(\text{ב})} = 0$, $C\left(t^{(k)}\right)$ is the binomial option pricing formula.[32] From (41) and (38),

$$C\left(t^{(k)}\right) = \left\{\frac{1}{2} - \frac{1}{2}\,\theta^{(\text{ב})}\sqrt{\Delta t}\right\} C\left(t^{(k+1)}\right)^{(up)} + \left\{\frac{1}{2} + \frac{1}{2}\,\theta^{(\text{ב})}\sqrt{\Delta t}\right\} C\left(t^{(k+1)}\right)^{(down)} \tag{42}$$

where $\theta^{(\text{ב})} = \frac{\mu + D_y^{\text{ב}} - r}{\sigma}$ is the Sharpe ratio for a stock-paying dividend yield $D_y^{\text{ב}} = (\mu - r)\,\mathcal{A}^{(\text{ב})}$.

Thus, ב$'s$ hedge decision based on minimizing $\mathfrak{G}\left(t^{(k)}\right)$ in (25) is equivalent to a risk-neutral hedge, when ב$'s$ risk-neutral strategy hedging is based on trading the stock with dividend yield $D_y^{\text{ב}}$. However, in reality, ב trades the stock with no dividends, and thus, if ב is "greedy" (that is, $D_y^{\text{ב}} = (\mu - r)\,\mathcal{A}^{(\text{ב})} > 0$ ), then the value of ב$'s$ hedge portfolio will be smaller because some risk has not been hedged. In contrast, if ב is "fearful" (that is, $D_y^{\text{ב}} = (\mu - r)\,\mathcal{A}^{(\text{ב})} < 0$ ), then the value of the hedge portfolio will be larger, due to the fact that ב$'s$ portfolio is in fact over-hedged.

From (52) the corresponding Black-Scholes equation is then

$$\frac{\partial f(t,x)}{\partial t} + \left(r - D_y^{\text{ב}}\right)x\frac{\partial f(t,x)}{\partial x} + rf(t,x) + \frac{1}{2}\sigma^2 x^2\frac{\partial^2 f(t,x)}{\partial x^2} = 0. \tag{43}$$

---

[32] See Kim et al (2016, p. 6), Section 3.2.

If $\mathcal{C}$ is a European call option with maturity $T$ and strike , that is, $f(t,x)=\max(0,x-X)$, then

$$C(t)=f\big(t,S(t)\big)=C\big(t,S(t),X,T,r,\sigma,D_y^{\beth}\big)=$$

$$= \; e^{-D_y^{\beth}(T-t)}S(t)\Phi\big(D^{(1)}(t)\big)-Xe^{-r(T-t)}\Phi\left(D^{(2)}(t)\right), \tag{44}$$

where $\Phi(\mathrm{x}), \mathrm{x}\in\mathcal{R}$, is the standard normal cumulative distribution function and

$$D^{(1)}(t)=\frac{\ln\left(\frac{S(t)}{X}\right)+\left(r-D_y^{\beth}+\frac{1}{2}\sigma^2\right)(T-t)}{\sigma(T-t)}, D^{(2)}(t)=D^{(1)}(t)-\sigma(T-t).$$

Calibration of the implied dividend based on option data follows a standard technique.[33] The optimization problem is: Given market data for European calls $C^{(marmket)}\big(t,S(t),X^{(i)},T^{(i)}\big), i=1,\dots,N$, find $C\left(t,S(t),X,T,r,\sigma^{(impl)},D_y^{\beth^{(impl)}}\right)$ solving

$$\min_{\sigma>0,D_y^{\beth}\in\mathcal{R}}\sum_{i=1}^{N}\left(\frac{C^{(marmket)}\big(t,S(t),X^{(i)},T^{(i)}\big)-C\big(t,S(t),X,T,r,\sigma,D_y^{\beth}\big)}{C^{(marmket)}(t,S(t),X^{(i)},T^{(i)})}\right)^2$$

## 6. Conclusion

In this paper we attempt to imbed basic notions and facts of behavioral finance within the realm of rational finance. The goal is to show within a few important areas with behavioral finance theory, such as prospect theory and cumulative prospect theory, that after natural adaptation both theories can be placed within the framework of dynamic asset pricing theory. We also show that rational finance can benefit by extending dynamic pricing theory to accommodate traders "greed and fear" factors. We provide option pricing formulas within PT and CPT, allowing one to initiate empirical work on estimating financial markets level of "greed and fear" from market option pricing data. We have provided new prospect theory value functions and weighting probability functions showing possible extensions of the classical PT and CPT. While our study is limited in scope, covering only a few of the most important concept in behavioral finance, we hope that it shows the general direction of placing behavioral finance theory into the solid quantitative framework of rational finance theory. At the same time we show the natural extension of the rational dynamic asset pricing theory to accommodate important concepts and findings of reported by the behavioral finance camp.

[33] See, for example, Cao (2005), Niburg (2009), and Bilson, Kang and Luo (2015).